\documentclass[10pt]{article}
\usepackage[square,authoryear]{natbib}
\usepackage{graphicx}
\usepackage{marsden_article}
\usepackage{stackrel}
\usepackage{ulem}

\usepackage{eucal}
\usepackage{stmaryrd}
\usepackage{mathrsfs}

\usepackage{epstopdf,framed}
\usepackage{pdfsync}
\usepackage{framed}

\DeclareFontFamily{OT1}{pzc}{}
\DeclareFontShape{OT1}{pzc}{m}{it}{<-> s * [1.100] pzcmi7t}{}
\DeclareMathAlphabet{\mathpzc}{OT1}{pzc}{m}{it}

\begin{document}
\newtheorem{remark}[theorem]{Remark}


\title{A Lagrangian formalism for nonequilibrium thermodynamics}
\vspace{-0.2in}

\newcommand{\todoFGB}[1]{\vspace{5 mm}\par \noindent
\framebox{\begin{minipage}[c]{0.95 \textwidth} \color{red}FGB: \tt #1
\end{minipage}}\vspace{5 mm}\par}


\author{\hspace{-1cm}
\begin{tabular}{cc}
Fran\c{c}ois Gay-Balmaz &
Hiroaki Yoshimura
\\ CNRS - LMD - IPSL  & School of Science and Engineering
\\ Ecole Normale Sup\'erieure de Paris & Waseda University
\\  24 Rue Lhomond 75005 Paris, France & Okubo, Shinjuku, Tokyo 169-8555, Japan \\ francois.gay-balmaz@lmd.ens.fr & yoshimura@waseda.jp\\
\end{tabular}\\\\
}

\maketitle
\vspace{-0.3in}

\begin{center}
\abstract{In this paper, we present a Lagrangian formalism for nonequilibrium thermodynamics.
This formalism is an extension of the Hamilton principle in classical mechanics that allows the inclusion of irreversible phenomena in both discrete and continuum systems  (i.e., systems with finite and infinite degrees of freedom).
The irreversibility is encoded into a nonlinear nonholonomic constraint given by the expression of  entropy production associated to all the irreversible processes involved.
Hence from a mathematical point of view, our variational formalism may be regarded as a generalization of the Lagrange-d'Alembert principle used in nonholonomic mechanics. In order to formulate the nonholonomic constraint, we associate to each irreversible process a variable called the thermodynamic displacement. This allows the definition of a corresponding variational constraint. 
Our theory is illustrated with various examples of discrete systems such as mechanical systems with friction, matter transfer, electric circuits, chemical reactions, and diffusion across membranes. 
For the continuum case, the variational formalism is naturally extended to the setting of infinite dimensional nonholonomic Lagrangian systems
and is expressed in material representation, while its spatial version is obtained via a nonholonomic Lagrangian reduction by symmetry. In the continuum case, our theory is systematically illustrated by the example of a multicomponent viscous heat conducting fluid with chemical reactions and mass transfer.}
\vspace{2mm}
\end{center}
\tableofcontents

\section{Introduction}

\subsection{Motivations and aims of the paper}\label{Int_mot}
\paragraph{Some backgrounds and history.}
Thermodynamics is a phenomenological theory which aims to {\it identify} and {\it describe} the relations between the observed {\it macroscopic} properties of a physical system with the help of fundamental laws, without aiming to explain the microscopic origin of these properties. The field of thermodynamics naturally includes macroscopic disciplines such as classical mechanics, fluid dynamics, and electromagnetism. In such a theory it is assumed that the macroscopic physical system can be described by a "small number" of variables whose value can be measured exactly. It is the goal of {\it statistical physics} to justify the principles of thermodynamics via microscopic properties and to provide methods to obtain the phenomenological coefficients appearing in the thermodynamic theory.

Historically, thermodynamics was first developed to treat almost exclusively equilibrium states and transition from one equilibrium state to another, in which changes in temperature play an important role. In this context, thermodynamics appeared mainly as a theory of heat and is viewed today as a branch of {\it equilibrium thermodynamics}.
Such a classical theory, {\it that does not aim to describe the dynamic} evolution of the system, can be developed in a well established setting (\cite{Gibbs1902}), governed by the well-known first and second laws (in their standard formulation found in any textbook on the subject). It is worth noting that classical mechanics, fluid dynamics and electromagnetism, being essentially dynamical theories, {\it can not} been treated in the context of the {\it classical} equilibrium thermodynamics, whereas, according to the general definition of thermodynamics mentioned at the beginning of this introduction, they belong to the subject of thermodynamics, also named \textit{nonequilibrium thermodynamics}.

Much effort has been dedicated to the construction of a general setting for {\it nonequilibrium} thermodynamics. Although the groundwork was laid by the classical investigation of Clausius, Kelvin, Maxwell, and Rayleigh, the classical theory of nonequilibrium thermodynamics did not emerge until the work of Onsager (\cite{Onsager1931}) on reciprocal relations connecting the coefficients which occur in the linear phenomenological equations relating the irreversible fluxes and the thermodynamic forces. These reciprocal relations were derived from the time reversal invariance of the {\it microscopic} equations of motion. Onsager's approach was followed by the contributions of Meixner, Prigogine, Coleman, Truesdell, among others (see e.g. the books \cite{deGrootMazur1969}, \cite{Truesdell1969}, \cite{GlPr1971}, \cite{StSc1974}, \cite{Bi1975}, \cite{Woods1975}, \cite{Lavenda1978}, \cite{KoPr1998}).

While the theory of nonequilibrium thermodynamics is still today a very active subject of research, relevant with many disciplines of physics, chemistry, biology and engineering, one cannot say that it has reached a level of completeness.  One of the reasons lies in the lack of a general Lagrangian variational formalism for nonequilibrium thermodynamics that would reduce to the classical Lagrangian variational formalism of mechanics in absence of irreversible processes. Various variational approaches have been proposed in relation with nonequilibrium thermodynamics. At the heart of most of them, is the \textit{principle of least dissipation of energy} by \cite{Onsager1931}, later extended in \cite{OnMa1953}, \cite{MaOn1953}, that underlies the reciprocal relations in the linear case. Another principle was formulated by \cite{Prigogine1947}, \cite{GlPr1971} as a condition on steady state processes, known as the \textit{principle of minimum entropy production}. Onsager's approach was generalized in \cite{Ziegler1968} to the case of systems with nonlinear phenomenological laws. We refer to \cite{Gyarmati1970} for reviews and developments of Onsager's variational principles, and for a study of the relation between Onsager's and Prigogine's principles. In this direction, we also refer to e.g. \cite[\S6]{Lavenda1978} and \cite{Ichiyanagi1994} for overviews on variational approaches to irreversible processes. Another important work  was done by \cite{Bi1975, Bi1984} in conjunction with thermoelasticity, viscoelasticity and heat transfer, where a {\it principle of virtual dissipation} in a generalized form of d'Alembert principle was used with various applications to nonlinear irreversible thermodynamics.
It was noteworthy that this variational approach was restricted to {\it weakly irreversible systems} or {\it thermodynamically holonomic and quasi-holonomic} systems, which one can obtain by  the 
assumption of isothermal systems or quasi-isothermal systems, namely, the assumption that the temperature remains constant and uniform enables us to simplify the required constraints associated with the rate of entropy production to be holonomic or quasi-holonomic, although \cite{Bi1975} mentioned the relations between the rate of entropy production and state variables may be given as {\it nonholonomic constraints} (see, equation (8.7) on page 21).

More recently, \cite{FuFu2012} showed a variational formalism of viscoelastic fluids, in which the internal conversion of mechanical power into heat power due to frictional forces was written as a nonholonomic constraint.

\paragraph{Main features of our variational formalism.} 

The variational formalism for nonequilibrium thermodynamics developed in this present paper is distinct from the earlier variational approaches mentioned above, both in its physical meaning and in its mathematical structure, as well as in its goal. Roughly speaking, while most of the earlier variational approaches mainly underlie the equation for the rate of entropy production, aiming to justify the expression of the phenomenological laws governing the irreversible processes involved, our variational approach aims to underlie {\it the complete set of time evolution equations of the system\/}, in such a way that it extends the classical Lagrangian formalism for discrete and continuum systems in mechanics to systems including irreversible processes.

This is accomplished by constructing a generalization of the {\it Lagrange-d'Alembert principle\/} of nonholonomic mechanics, where the entropy production of the system, written as the sum of the contribution of each of the irreversible processes, is incorporated into {\it nonlinear nonholonomic constraints\/}. As a consequence, all the {\it phenomenological laws\/} are encoded in the nonholonomic constraint, to which we naturally associate a {\it variational constraint\/} on the allowed variations of the action functional. A natural definition of the variational constraint in terms of the phenomenological constraint is possible thanks to the introduction of the concept of a \textit{thermodynamic displacement} that generalizes the concept of a displacement associated to the temperature, called the thermal displacement. When irreversible processes are not taken into account, our variational formalism consistently recovers Hamilton's principle in Lagrangian mechanics.
\medskip

One of the essential ideas in our approach may be given as follows. Consider the entropy production written in the generic form
\[
I= \frac{1}{T} \sum_ \alpha {J} _\alpha {X} ^\alpha,
\]
where $X ^\alpha $ denotes the {\it thermodynamic affinity\/} associated to an irreversible process and $J _\alpha $ is the corresponding {\it irreversible flux\/}. Our approach consists in associating to each irreversible process the  rate $\dot \Lambda  ^\alpha $ of a ``new" quantity $ \Lambda  ^\alpha $, called a {\it thermodynamic displacement}, such that  $\dot\Lambda  ^\alpha =X ^\alpha $. This allows us to define a {\it virtual} quantity $ \delta \Lambda  ^\alpha $ for each of the irreversible process and to write the corresponding {\it admissible constraint\/} to be imposed on the variations of the action functional associated to the Lagrangian of the irreversible system. 
\medskip

Even in the simplest examples of discrete systems in which irreversible processes are included (such as pistons, chemical reactions, or electric circuits), our approach allows for an efficient way to derive the coupled evolution for the mechanical variables $(q,\dot{q})$ and the thermodynamic variable (entropy) $S$ by considering the following {\it nonholonomic Lagrangian variational formalism} for a Lagrangian $L(q, \dot q, S)$: 
\[
\delta \int_{ t _1 }^{ t _2 }L(q , \dot q , S)dt +\int_{ t _1 }^{ t _2 }\left\langle F^{\rm ext}(q, \dot q, S), \delta q\right\rangle  =0,
\]
where one imposes the {\it nonlinear nonholonomic constraints}: 
\begin{equation}\label{constraint_introduction} 
\frac{\partial L}{\partial S}(q, \dot q, S)\dot S  = \left\langle F^{\rm fr}(q, \dot q, S) , \dot q \right\rangle -  P_H^{\rm ext}, 
\end{equation} 
and with respect to the variations $ \delta q $ and $\delta S$ subject to the associated {\it variational constraint}
\[
\frac{\partial L}{\partial S}(q, \dot q, S)\delta S= \left\langle F^{\rm fr}(q , \dot q , S),\delta q \right\rangle.
\]
Here $P_H^{\rm ext}$ denotes the external heat power supply, $ F^{\rm fr}$ denotes the friction force, of phenomenological nature, and $T:= -\frac{\partial L}{\partial S}$ is the temperature. The constraint \eqref{constraint_introduction} is thus a \textit{phenomenological constraint} that encodes the entropy production of the system: $\dot S= I+ \frac{1}{T}P_H^{\rm ext}$, where $I=- \frac{1}{T}\left\langle F^{\rm fr}(q , \dot q , S),\dot q \right\rangle$ is the {\it internal entropy production}.

\medskip

Later, we will come back to this variational formalism and will show how the dynamics of various discrete (finite dimensional) systems such as mechanical systems with friction, matter transfer, chemical reactions and electric circuits can be derived from this variational formalism. Furthermore, we will extend the nonholonomic variational formalism to continuum systems, where  we will formulate the complete evolution equations for irreversible continuum mechanics in an infinite dimensional setting by first working with the material representation. This will be illustrated with the general example of a multicomponent fluid in which the irreversible processes associated to viscosity, heat and matter transfer, and chemical reactions are included.

\paragraph{Contributions of the paper.} The main contributions of this paper are as follows:
\begin{itemize}
\item 
Given the Lagrangian of the irreversible system, the phenomenological laws relating thermodynamic affinities and irreversible fluxes, and the power due to transfer of heat or matter between the system and the exterior, our variational formalism yields the time evolution equations for the nonequilibrium dynamics of this system in accordance with the two fundamental laws in Stueckelberg's axiomatic formulation of thermodynamics.
\item 
It is well-known that the evolution equations of \textit{classical mechanics\/} can be derived and studied by the Lagrangian formalism using the \textit{critical action principle of Hamilton\/}. Such a formalism is especially well suited to study symmetries and naturally extends to mechanical systems with nonholonomic constraints and/or subject to external forces, via the  \textit{Lagrange-d'Alembert principle}. Our variational formalism is an \textit{extension of this general setting to the field of nonequilibrium thermodynamics\/}.  In particular, it allows for the study of symmetries of the system and for the implementation of the reduction processes associated to these symmetries.
\item
In the context of {\it continuum systems}, our variational formalism for nonequilibrium thermodynamics can be employed both in the {\it material} and {\it spatial} (or {\it Eulerian}) representations. In spatial representation, the principle is more involved but it is naturally explained and justified by a Lagrangian reduction by symmetry of the principle in material representation which, in turn, is the natural extension of the same principle used for thermodynamics of discrete systems.
\item
The formalism unifies in a systematic way a wide class of examples appearing in nonequilibrium thermodynamics, ranging from electric circuits, matter transfer, and chemical reactions, to viscous heat conducting multicomponent reacting fluids. It therefore helps the understanding of the analogy between various examples, which is of primordial importance for the future developments of nonequilibrium thermodynamics, whose main difficulties is essentially due to its \textit{multiphysical character}.
\item
Being derived from a geometric point of view, the formalism automatically produces intrinsic (coordinate free) equations of motion and clearly keeps track of the various reference fields (e.g. Riemannian metrics) that are underlying the theory in the continuum systems, which play a crucial role in the understanding of the material covariance of the theory. In particular, it provides a geometrically meaningful setting for the derivation of phenomenological laws among irreversible fluxes and thermodynamic forces.
%
%
\item
In many areas of classical mechanics, especially continuum mechanics, Hamilton's principle has played an important role in deriving new models. Such models would have been very difficult, if not impossible, to obtain via the exclusive use of balance equations arising from Newton's laws. We hope that the formalism developed in this paper will be of similar utility for the more general case of nonequilibrium thermodynamics, especially for systems involving several physical areas.
\end{itemize}

Let us also mention that our variational formalism is potentially useful for the future derivation of variational numerical integrators for nonequilibrium thermodynamics, obtained by a discretization of the variational structure, and therefore extending the variational integrators for classical and continuum mechanics (\cite{MaWe2001}, \cite{LeMaOrWe2003}).

\subsection{Fundamental laws of thermodynamics}
We close this introduction by recalling the fundamental laws of nonequilibrium thermodynamics. We follow the axiomatic formulation of thermodynamics developed by Stueckelberg around 1960 (see, for instance, \cite{StSc1974}), which is extremely well suited for the study of this field as a general macroscopic dynamic theory that extends classical mechanics to account for irreversible processes. In his axiomatic approach, Stueckelberg introduced two state functions, the {\it energy} and the {\it entropy} obeying the two fundamental laws of thermodynamics, formulated as first order differential equations. The equations describing the dynamic evolution of the system can be derived by using exclusively the two fundamental laws in a systematic way. We refer to  e.g. \cite{Gr1999}, \cite{FeGr2010}, \cite{GrBr2011} for a systematic use of Stueckelberg's formalism in several examples.

\paragraph{Stueckelberg's axiomatic formulation of thermodynamics.} Let us denote by $ \boldsymbol{\Sigma}  $ a physical system and by $ \boldsymbol{\Sigma} ^{\rm ext}$ its exterior. The state of the system is defined by a set of mechanical variables and a set of thermal variables. State functions are functions of these variables.
\medskip

\noindent {\bf First law:} For every system $ \boldsymbol{\Sigma} $, there exists an extensive scalar state function $E$, called {\bfi energy}, which satisfies
\[
\frac{d}{dt} E(t) = P^{\rm ext}_W(t)+P^{\rm ext}_H(t)+P^{\rm ext}_M(t),
\]
where $t$ denotes {\it time}, $ P^{\rm ext}_W(t)$ is the {\it power due to external forces} acting on the mechanical variables of the system, $P^{\rm ext}_H(t)$ is the {\it power due to heat transfer}, and $P^{\rm ext}_M(t)$ is the {\it power due to matter transfer} between the system and the exterior.
%

\begin{framed}
\noindent\underline{\textsf{Thermodynamic systems:}}\\[3mm]
Given a {\it thermodynamic system}, the following terminology is generally adopted:
\begin{itemize}
\item A system is said to be {\bfi closed} if there is no exchange of matter, i.e.,  $P^{\rm ext}_M(t)=0$. 
\item 
A system is said to be {\bfi adiabatically closed} if it is closed and there is no heat exchanges, i.e., $P^{\rm ext}_M(t)=P^{\rm ext}_H(t)=0$. 
\item 
A system is said to be {\bfi isolated} if it is adiabatically closed and there is no mechanical power exchange, i.e., $P^{\rm ext}_M(t)=P^{\rm ext}_H(t)=P^{\rm ext}_W(t)=0$.
\end{itemize}
\end{framed}

From the first law, it follows that the {\it energy of an isolated system is constant}.
\medskip

\noindent {\bf Second law:} For every system $ \boldsymbol{\Sigma} $, there exists an extensive scalar state function $S$, called {\bfi entropy}, which obeys the following two conditions (see \cite{StSc1974}, p.23)
\begin{itemize}
\item[(a)]  Evolution part:\\
If the system is adiabatically closed, the entropy $S$ is a non-decreasing function with respect to time, i.e., 
\[
\frac{d}{dt} S(t)=I(t)\geq 0,
\]
where $I(t)$ is the {\it entropy production rate} of the system accounting for the irreversibility of internal processes.
\item[(b)] Equilibrium part:\\
If the system is isolated, as time tends to infinity the entropy tends towards a finite local maximum of the function $S$ over all the thermodynamic states $ \rho $ compatible with the system, i.e., 
\[
\lim_{t \rightarrow +\infty}S(t)= \max_{ \rho \; \text{compatible}}S[\rho ].
\]
\end{itemize}


By definition, the evolution of an isolated system is said to be {\it reversible} if $I(t)=0$, namely, the entropy is constant. In general, the evolution of a system $ \boldsymbol{\Sigma} $ is said to be reversible, if the evolution of the total isolated system with which $ \boldsymbol{\Sigma} $ interacts is reversible.
\medskip

The laws of thermodynamics are often formulated in terms of differentials, especially for equilibrium thermodynamics. Note that since $E$ is a state function, we can form its differential $ \mathbf{d} E$, which is an exact form on the state space. We can therefore write the left hand side of the first law of thermodynamics as 
$$
 \frac{d}{dt} E(t)= \left<\mathbf{d} E( \zeta (t)), \dot { \zeta }(t) \right>, 
$$
 where $ \zeta $ denotes the collection of all state variables and $\dot { \zeta }$ its time derivative. However, it is important to note that $P^{\rm ext}_W$, $P^{\rm ext}_H$ and $P^{\rm ext}_M$ are {\it not} necessarily given in terms of the paring between differential forms and vector fields in general. It turns out that in some particular situations they can be described in terms of such a paring, but this is {\it not always case} in general, nor required by the fundamental laws of thermodynamics.

\section{Nonequilibrium thermodynamics of discrete systems}\label{Section2} 
\paragraph{Discrete systems and simple systems.} A \textit{discrete system} $ \boldsymbol{\Sigma}$ is a collection $ \boldsymbol{\Sigma} =\cup_{A=1}^N \boldsymbol{\Sigma} _A $ of a finite number of interacting simple systems $ \boldsymbol{\Sigma} _A $.
By definition, following \cite{StSc1974}, a \textit{simple system}$^{1}$ $\boldsymbol{\Sigma}$ is a macroscopic system for which one (scalar) thermal variable $ \tau  $ and a finite set $( \xi ^1 , ..., \xi ^n )$ of mechanical variables are sufficient to describe entirely the state of the system.  From the second law of thermodynamics, \textit{we can always choose $ \tau $ as the entropy $S$}.

\addtocounter{footnote}{1}
\footnotetext{In \cite{StSc1974} they are called \textit{\'el\'ement de syst\`eme} (French). We choose to use the English terminology \textit{simple system} instead of \textit{system element}.}

We shall first present our Lagrangian formalism for the case of simple systems and we will illustrate it for four representative examples, namely, the case of mechanics coupled with one thermal equation, the case of chemical reactions, the case of matter transfer, and the case of a nonlinear RLC circuit. For simple systems, no internal heat transfer can occur.
Then we shall extend our approach to treat the case of an arbitrary discrete system $ \boldsymbol{\Sigma} =\cup_{A=1}^N \boldsymbol{\Sigma} _A $, in which internal heat exchanges can occur. The theory will be illustrated with the examples of the two-cylinder problem and the thermodynamics of electric circuits.

\paragraph{A review of the Lagrange-d'Alembert principle for nonholonomic mechanics.} 
As mentioned in the introduction, our variational formalism for nonequilibrium thermodynamics is based on an extension of the Lagrange-d'Alembert principle for nonholonomic mechanics. As a preparation, we first recall here this principle as it applies to the standard case of {\it linear nonholonomic constraints}.

Consider a mechanical system with an $n$ dimensional configuration manifold $Q$ and a Lagrangian function $L:TQ \rightarrow \mathbb{R}  $ defined on the tangent bundle (velocity phase space, or state space) of the manifold $Q$. We will use the local coordinates $(q, \dot q)$ for an element in $TQ$. We suppose that the motion is constrained by a regular distribution (i.e., a smooth vector subbundle) $ \Delta_{Q}=\{\dot{q} \in T_{q}Q \mid \phi^{a}(q,\dot{q})=\left< \omega^{a}(q), \dot{q}\right>=0\} \subset TQ$, where $\omega^{a}, a=1,...,r <n$ are given one-forms on $Q$ and the constraints are {\it linear} in velocity. We denote by $ \Delta _Q(q) \subset T_qQ$ the vector fiber of $ \Delta _Q$ at $q \in Q$. We recall that the constraint $ \Delta _Q$ is \textit{holonomic} if and only if for all $q \in Q$ there exists a submanifold $N \subset Q$ with $ q \in N$ and such that $ \Delta _Q|_N=TN$. Otherwise the constraint is said to be \textit{nonholonomic}. We also assume that the system is subject to an exterior force, given by a fiber preserving map $F^{\rm ext}:TQ \rightarrow T^*Q$, not necessarily linear on the fibers. Here $T^*Q$ denotes the cotangent bundle (momentum phase space) of $Q$. We shall denote by $ \left\langle \alpha _q , v _q \right\rangle $ the pairing between $ \alpha _q \in T^*_qQ$ and $v _q\in  T_qQ$.

The equations of motion for {\it nonholonomic mechanics} are obtained by application of the \textit{Lagrange-d'Alembert principle}
\begin{equation}\label{LdA_VP} 
\delta \int_{t_1 }^{t _2}L(q, \dot q) dt+\int_{t _1 }^{t _2 } \left\langle F^{\rm ext}(q, \dot q), \delta q \right\rangle dt=0,
\end{equation} 
where $\dot q \in \Delta _Q$ and the variations $ \delta q $ are such that $ \delta q\in \Delta _Q$ and with $ \delta q(t _1 )= \delta q( t _2 )=0$. The resulting equations are called the {\it Lagrange-d'Alembert equations}, which are given by 
\begin{equation}\label{LdA_eq} 
\frac{d}{dt} \frac{\partial L}{\partial \dot q } - \frac{\partial L}{\partial q}-F^{\rm ext}(q, \dot q) =F^{c},\qquad \dot q \in \Delta _Q.
\end{equation} 
In the above, $F^{c}$ denotes {\it constraint forces}, which is expressed by using Lagrange multipliers $\lambda_{a}$ as $F^{c}=\lambda_{a}\omega^{a}(q) \in \Delta_{Q}^{\circ}(q)$, where $ \Delta _Q^\circ=\{ \alpha _q\in T^*Q\mid \left\langle \alpha _q , v _q \right\rangle =0,\;\forall\; v_q \in \Delta _Q(q)\}$ is the annihilator of $ \Delta _Q$. For an extended discussion of the Lagrange-d'Alembert equations for nonholonomic mechanics as well as for several references on the subject, see, for instance, \cite{Bl2003}. Here it is important to note that the Lagrange-d'Alembert principle \eqref{LdA_VP} is only valid for the special case of \textit{linear} nonholonomic constraints.

As we will show later, our treatment of nonequilibrium thermodynamics involves {\it nonlinear nonholonomic constraints} and hence we {\it cannot} employ the conventional Lagrange-d'Alembert principle \eqref{LdA_VP} for our purpose of formulating nonequilibrium thermodynamics. For  systems with nonlinear nonholonomic constraints, several variational formalisms have been developed.

One can develop {\it Hamilton's variational principle} over the curves $q(t)$ satisfying the nonlinear nonholonomic constraints $C=\{(q,\dot{q}) \in TQ \mid \phi^{a}(q,\dot{q})=0\} \subset TQ$. Such a principle would use Lagrange multipliers $\lambda_{a}$ to construct an augmented Lagrangian $L(q,\dot{q}) +\lambda_{a} \phi^{a}(q,\dot{q})$ to yield Euler-Lagrange equations from the critical condition of the action integral. However, the resultant Euler-Lagrange equations {\it do not} describe the evolution equation for nonholonomic mechanics, but different dynamics called the {\it vakonomic mechanics} (see \cite{Arnold, JiYo2015}), which are useful in optimal control problems. Even for the case of linear nonholonomic systems in which there is no external force $F^{\rm ext}(q,\dot{q})$, the Euler-Lagrange equations developed from Hamilton's principle for the augmented Lagrangian $L(q,\dot{q}) +\lambda_{a} \left< \omega^{a}(q), \dot{q}\right>$ are essentially different from  the Lagrange-d'Alembert equations \eqref{LdA_eq} obtained from the Lagrange-d'Alembert principle \eqref{LdA_VP}. The difference between the two variational formalisms lie in the way of taking variations; namely, in the Lagrange-d'Alembert principle, constraints are only imposed on the velocity $\dot q(t)$, i.e., at $\varepsilon =0$, and on the variations $ \delta q(t)= \left.\frac{d}{d\varepsilon}\right|_{\varepsilon=0} q_ \varepsilon (t)$, while in the vakonomic formalism, the constraints are imposed on the velocity vectors of $q_ \varepsilon (t)$ for all $ \varepsilon$ in a neighborhood of $0$. We also note that in the Lagrange-d'Alembert principle, the constraint $ \Delta _Q$ has a double role. On one hand, it imposes \textit{kinematic constraints} on the velocity $\dot{q}(t)$. On the other hand, it imposes \textit{variational constraints} on the admissible variations $\delta q(t)$ of a curve $q(t)$.

Some generalization of the Lagrange-d'Alembert principle to nonlinear nonholonomic mechanics was developed in \cite{Ch1934}, see also \cite{Ap1911}, \cite{Pi1983}. In Chetaev's approach, the variational constraint is derived from the nonlinear constraint submanifold $C \subset TQ$. However, it has been pointed out in \cite{Ma1998} that this principle does not always lead us to the correct equations of motion for mechanical systems and in general one has to consider the nonlinear constraint on velocities and the variational constraint as independent notions. 
A general geometric approach to nonholonomic systems with nonlinear and (possibly) higher order constraints has been developed by \cite{CeIbdLdD2004}. It generalizes both the Lagrange-d'Alembert principle and Chetaev's approach. It is important to point out that for these generalizations, including Chetaev's approach, {\it energy may not be conserved} along the solution of the equations of motion. This implies that the this geometric approach includes the case of non-ideal constraints.
From a mathematical point of view, the variational formalism for nonequilibrium thermodynamics that we develop in this paper falls into this general setting. Consistently with the first law, in our setting the constraints are ideal if and only if the system is isolated.

\subsection{Variational formalism for nonequilibrium thermodynamics of simple systems}\label{system_elements}
In order to present our formalism, let us first consider a simple and closed system, so $P^{\rm ext}_M=0$. In this case, the Lagrangian of the system is a function $L=L(q, \dot q, S):TQ \times \mathbb{R}  \rightarrow \mathbb{R}  $, where, as mentioned in \S\ref{Int_mot}, we notice that $L$ includes a thermodynamic variable, the entropy $S$, in addition to the mechanical variables $(q, \dot q)$.
We assume that the system is subject to an exterior force $F^{\rm ext}:TQ\times \mathbb{R}   \rightarrow T^*Q$ so the associated mechanical power is $P_W^{\rm ext}=\left\langle F^{\rm ext}(q, \dot q, S), \dot q \right\rangle $. We also assume that there is a friction force $F^{\rm fr}:TQ\times \mathbb{R}   \rightarrow T^*Q$. The forces are assumed to be fiber preserving, that is, $F^{\rm ext}(q, \dot q, S) \in T_q^*Q$ and $F^{\rm fr}(q, \dot q, S) \in T_q^*Q$, for all $(q, \dot q)\in TQ$, for all $ S \in \mathbb{R}  $.

As before, we denote by $P_H^{\rm ext}(t) $ the power due to heat transfer between the system and the exterior. Note that $P_H^{\rm ext}(t)$ is a {\it time dependent} function. This time dependence can arise through a dependence on the state variables $q(t)$, $\dot q(t)$, and $S(t)$, but an explicit dependence on time is also allowed. For simplicity, we have assumed that the exterior force $F^{\rm ext}$ depends on time only through the state variables $(q(t),\dot q(t), S(t))$. However, this is not necessarily the case, and our formalism may include the more general case of an explicit dependence of $F^{\rm ext}$ on time.

We shall show that the coupled differential equations describing the mechanical and thermal evolution of a simple closed system can be obtained by a generalization of the Lagrange-d'Alembert principle of nonholonomic mechanics with a \textit{nonlinear constraint} given by the evolution part of the second law of thermodynamics  in Stueckelberg's formulation. In absence of thermal effects, this principle recovers the Hamilton principle of classical mechanics.

\begin{definition}[\textbf{Variational formalism for nonequilibrium thermodynamics of closed simple systems}]\label{LdA_def} Consider a simple closed system. Let $L: TQ \times \mathbb{R}  \rightarrow \mathbb{R}$ be the Lagrangian, $F^{\rm ext}:TQ\times \mathbb{R}  \rightarrow T^*Q$  the external force, $F^{\rm fr}:TQ\times \mathbb{R}  \rightarrow T^*Q$ the friction force, and $P^{\rm ext}_H$ the power due to heat transfer between the system and the exterior. The variational formalism for the thermodynamics of the simple closed system is defined as
\begin{equation}\label{LdA_thermo} 
\delta \int_{ t _1 }^{ t _2 }L(q , \dot q , S)dt +\int_{ t _1 }^{ t _2 }\left\langle F^{\rm ext}(q, \dot q, S), \delta q\right\rangle  =0, \quad\;\;\; \textsc{Variational Condition}
\end{equation}
where the curves $q(t)$ and $S(t)$ satisfy the nonlinear nonholonomic constraint 
\begin{equation}\label{Kinematic_Constraints} 
\frac{\partial L}{\partial S}(q, \dot q, S)\dot S  = \left\langle F^{\rm fr}(q, \dot q, S) , \dot q \right\rangle -  P_H^{\rm ext}, \quad \textsc{Phenomenological Constraint}
\end{equation} 
and with respect to the variations $ \delta q $ and $\delta S$  subject to
\begin{equation}\label{Virtual_Constraints} 
\frac{\partial L}{\partial S}(q, \dot q, S)\delta S= \left\langle F^{\rm fr}(q , \dot q , S),\delta q \right\rangle,\qquad\qquad\, \textsc{Variational Constraint}
\end{equation}
and with $\delta q(t_1 )= \delta q(t _2 )=0$.
\end{definition} 
We first note that the explicit expression of the constraint \eqref{Kinematic_Constraints}  involves phenomenological laws for the friction force $ F^{\rm fr}$, this is why we refer to it as a {\bfi phenomenological constraint}. The associated constraint \eqref{Virtual_Constraints} is called a {\bfi variational constraint}  since it is a condition on the variations to be used in \eqref{LdA_thermo}. 
The constraint \eqref{Virtual_Constraints} follows from \eqref{Kinematic_Constraints} by formally replacing the velocity by the corresponding virtual displacement, and by removing the contribution from the exterior of the system. Such a simple correspondence between the phenomenological and virtual constraints will still hold in the more general thermodynamic systems considered later. 
We also note that since \eqref{Kinematic_Constraints} is a nonlinear constraint in general, our principle does not follow from the standard Lagrange-d'Alembert principle \eqref{LdA_VP}.

From a mathematical point of view our variational formalism (Definition \ref{LdA_def}) falls into the setting studied in \cite{CeIbdLdD2004} for nonholonomic mechanics but does not follows Chetaev's approach, that is, if we derive the variational constraint from the nonlinear nonholonomic constraint \eqref{Kinematic_Constraints} following Chetaev's approach, then we will not obtain \eqref{Virtual_Constraints}. As we shall see below, {\it energy is preserved when the system is isolated}, i.e., when $P^{\rm ext}_W=P^{\rm ext}_H=P^{\rm ext}_M=0$, consistently with the first law of thermodynamics.

\paragraph{Equations of motion.} The derivation of the equations of evolution proceeds as follows. Taking variations of the integral in \eqref{LdA_thermo}, integrating by part and using $ \delta q(t _1 )= \delta q(t _2 )=0$, it follows
\[
\int_{t _1 }^{ t _2 }\left(  \left\langle  \frac{\partial L}{\partial q}- \frac{d}{dt} \frac{\partial L}{\partial \dot q} +F^{\rm ext}, \delta q \right\rangle + \frac{\partial L}{\partial S}\delta S \right) dt=0,
\]
where the variations $ \delta q $ and $ \delta S$ have to satisfy the variational constraint \eqref{Virtual_Constraints}. Now, replacing $ \frac{\partial L}{\partial S}\delta S $ by the virtual work expression $\left\langle F^{\rm fr}(q, \dot q, S), \delta q \right\rangle $ according to \eqref{Virtual_Constraints}, it provides the following differential equation for the simple closed system with {\it coupled  mechanical and thermal processes}:
\begin{equation}\label{thermo_mech_equations}
\left\{ 
\begin{array}{l}
\displaystyle\vspace{0.2cm}\frac{d}{dt} \frac{\partial L}{\partial \dot q}- \frac{\partial L}{\partial q}= F^{\rm ext}(q , \dot q, S) +  F^{\rm fr}(q, \dot q, S),\\
\displaystyle\frac{\partial L}{\partial S}\dot S= \left\langle F^{\rm fr} (q , \dot q, S) , \dot q\right\rangle  -  P_H^{\rm ext}.
\end{array} 
\right.
\end{equation}

\medskip

In order to illustrate this formalism, let us consider the following particular form of the Lagrangian as
\begin{equation}\label{special_form}
L(q  , \dot q  , S):=K_{\rm mech}( q  , \dot q )-U(q , S),
\end{equation} 
where $K_{\rm mech}:TQ \rightarrow \mathbb{R}  $ denotes the kinetic energy of the mechanical part of the system (assumed to be independent of $S$) and $U: Q \times \mathbb{R}  \rightarrow \mathbb{R}$ denotes the {\it potential energy}, which is a function of both the mechanical displacement $q$ and the entropy $S$.
This situation will be illustrated with the example of a mass-spring system with friction in \S\ref{example-mech-thermal}.

By introducing the {\it generalized internal forces}
\begin{equation}\label{F_int} 
F^{\rm int}(q , S):=-\frac{\partial U}{\partial q}(q,S),
\end{equation} 
we can rewrite the system \eqref{thermo_mech_equations} in terms of $K_{\rm mech}$ and $U$ as
\begin{equation}\label{thermo_mech_equations_special}
\left\{ 
\begin{array}{l}
\displaystyle\vspace{0.2cm} 
\frac{d}{dt} \frac{\partial K_{\rm mech}}{\partial \dot q }- \frac{\partial K_{\rm mech}}{\partial q }= F^{\rm int}( q,S)+F^{\rm ext}(q , \dot q , S) +  F^{\rm fr} (q, \dot q, S),\\
\displaystyle\frac{\partial U}{\partial S}\dot S= -\left\langle F^{\rm fr} (q , \dot q, S) , \dot q\right\rangle  +   P_H^{\rm ext}.
\end{array} 
\right.
\end{equation}
Examples of such systems have been derived in \cite{Gr1997}, \cite{GrBr2011} by using exclusively the first and second laws of thermodynamics in \cite{StSc1974}.

\paragraph{Energy balance law.} The energy associated with $L:TQ \times \mathbb{R}  \rightarrow \mathbb{R}  $ is the function $E:TQ \times \mathbb{R}  \rightarrow \mathbb{R}  $ defined by $E(q, \dot q, S):= \left\langle \frac{\partial L}{\partial \dot q}, \dot q \right\rangle -L(q, \dot q, S)$.
Using the system \eqref{thermo_mech_equations},  we obtain the {\bfi energy balance law}:
\begin{equation}\label{ddt_E} 
\frac{d}{dt} E= \left\langle \frac{d}{dt} \frac{\partial L}{\partial \dot q}- \frac{\partial L}{\partial q} , \dot q \right\rangle - \frac{\partial L}{\partial S}\dot S=P_W^{\rm ext}+P_H^{\rm ext},
\end{equation} 
which is consistent with the first law of thermodynamics.

Note that in the particular case of the Lagrangian \eqref{special_form}, the energy decomposes as $E(q,\dot q, S)=E_{\rm mech}(q, \dot q)+U(q,S)$, where $E_{\rm mech}: TQ \to \mathbb{R}$, defined by $E_{\rm mech}(q, \dot q)= \left\langle \frac{\partial K_{\rm mech}}{\partial \dot q}, \dot q \right\rangle -K_{\rm mech}(q, \dot q)$ is the mechanical energy. It is instructive to consider the energy balances for both $E_{\rm mech}$ and $U$. Using \eqref{F_int} and \eqref{thermo_mech_equations_special}, we have
\begin{align*}
\frac{d}{dt} E_{\rm mech}&= \left\langle F^{\rm fr}, \dot q \right\rangle+\left\langle F^{\rm int},\dot q \right\rangle+ \left\langle F^{\rm ext},\dot q \right\rangle= \left\langle F^{\rm fr}, \dot q \right\rangle+\left\langle F^{\rm int},\dot q \right\rangle+P^{\rm ext}_W,\\
\frac{d}{dt} U&=\frac{\partial U}{\partial S} \dot S + \left\langle \frac{\partial U}{\partial q}, \dot q \right\rangle   = - \left\langle F^{\rm fr}, \dot q \right\rangle-\left\langle F^{\rm int},\dot q \right\rangle+ P_H^{\rm ext},
\end{align*}
consistently with \eqref{ddt_E}. Note that the internal and friction forces are thus responsible for exchanges between the mechanical and internal energies.

\paragraph{Entropy production.} The temperature is given by minus the partial derivative of the Lagrangian with respect to the entropy, $T=- \frac{\partial L}{\partial S}$, which is assumed to be positive. So the second equation in \eqref{thermo_mech_equations} reads
\[
T \dot S= P_H^{\rm ext}- \left\langle F^{\rm fr}(q, \dot q, S), \dot q\right\rangle .
\]
According to the second law of thermodynamics, for an adiabatically closed systems, i.e., when $P^{\rm ext}_H=P^{\rm ext}_M=0$, entropy is increasing. So the friction force $F^{\rm fr}$ must be dissipative, that is $\left\langle F^{\rm fr}(q, \dot q, S), \dot q \right\rangle \leq 0$, for all $(q, \dot q, S) \in TQ \times \mathbb{R}  $. For the case in which the force is linear in velocity, i.e., $F^{\rm fr}(q, \dot q, S)=- \lambda  (q,S) ( \dot q,\_\,)$, where $ \lambda  (q,S):T_qQ  \times T_qQ \rightarrow \mathbb{R}  $ is a {\it $2$ covariant tensor field}, this implies that the symmetric part $ \lambda ^{\rm sym}$ of $ \lambda $ has to be positive. For a simple system, an {\bfi internal entropy production} is therefore of the form
\[
I(t)= -\frac{1}{T}\left\langle F^{\rm fr}(q, \dot q, S), \dot q \right\rangle.
\]
We will see later in \S\ref{discrete_systems} that for discrete systems $I(t)$ can have a more general expression.

\begin{remark}[Free energy formalism]{\rm It is possible to write a variational formalism based on the free energy for simple systems. In this case the temperature, rather than the entropy, is seen as a primitive variable. This will be done later in \S\ref{section_free_energy} in the more general case of discrete (not necessarily simple) systems and needs the introduction of the new variables $ \gamma $ and $ \sigma $.}
\end{remark}

\begin{remark}{\rm In the above macroscopic description, it is assumed that the macroscopically ``{\it slow\,}'' or collective motion of the system can be described by $q(t)$, while the time evolution of the entropy $S(t)$ is determined from the microscopically ``{\it fast\,}'' motions of molecules through statistical mechanics under the assumption of  {\bfi local equilibrium conditions}.
At the macroscopic level, the internal energy $U(q,S)$, given as a potential energy, is essentially coming from the total kinetic energy associated with the microscopic motion of molecules, which is directly related to the temperature of the system.}
\end{remark}

We now illustrate our Lagrangian variational formalism (Definition \ref{LdA_def}) with various examples of simple systems. We first consider examples from mechanical systems coupled with thermodynamics. Then we treat the case of the
nonequilibrium thermodynamics of chemical reaction, electric circuits. Finally, we also consider the case of matter diffusion transfer through a membrane as well as its coupling with chemical
reactions.

\subsubsection{Mechanical systems with thermodynamics}\label{example-mech-thermal}

\begin{figure}[h]
\begin{center}
\includegraphics[scale=0.77]{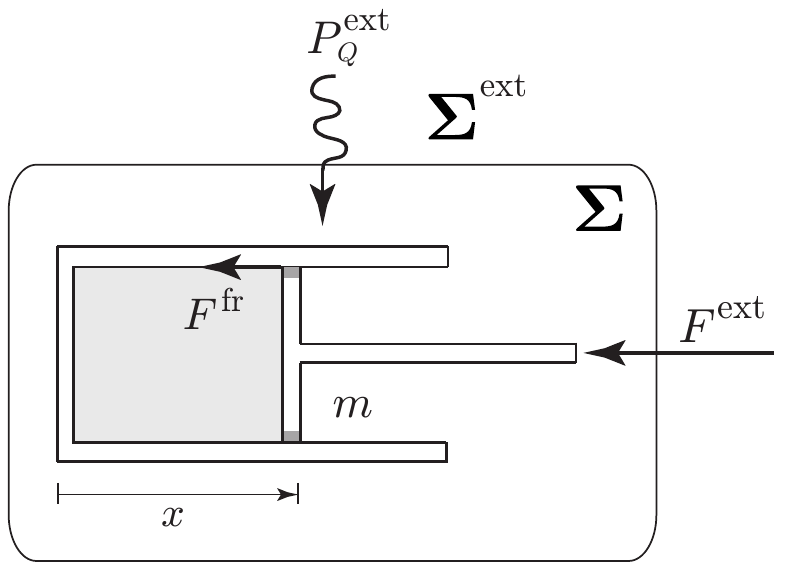}
\caption{One-cylinder}
\label{one_cylinder}
\end{center}
\end{figure}

\paragraph{Example: the one-cylinder problem.} Let us consider a gas confined by a piston in a cylinder as in Fig. \ref{one_cylinder}. This is supposed to be a closed system (i.e., $P^{\rm ext}_M=0$). The state of the system can be characterized by the three variables $(x, \dot x, S)$.
The equations of motion for dynamics of this system were derived in \cite{Gr1999} by using exclusively the two laws of thermodynamics, as formulated by \cite{StSc1974}. Here we shall derive these equations by using the variational formalism of Definition \ref{LdA_def}.

The Lagrangian is $L(x, \dot x, S)=K_{\rm mech}(x, \dot x)-U(x,S)=\frac{1}{2} m \dot x ^2 -U(x,S)$, where $m$ is a mass of the piston, $U(x,S):=U(S,V=Ax,N_0)$ is the internal energy of the gas$^{2}$, $N_0$ is a number of moles, $V=Ax$ is the volume, and $A$ is the area of the cylinder.
\addtocounter{footnote}{1}
\footnotetext{The state functions for a perfect gas are $U=cNRT$ and $pV=NRT$,
where $c$ is a constant depending exclusively on the gas (e.g. $c=\frac{3}{2}$ for monoatomic gas, $c=\frac{5}{2}$ for diatomic gas) and $R$ is the universal gas constant. From this, it is deduced that the internal energy reads
\[
U(S,N,V)= U _0 e^ {\frac{1}{cR}\left(\frac{S}{N}- \frac{S _0 }{N _0 }   \right)  }\left( \frac{N}{N_0}\right) ^{ \frac{1}{c}+1} \left( \frac{V _0 }{V}\right) ^{\frac{1}{c}}.
\]
}

The temperature and the generalized internal force are respectively given by
\[
{T(x,S)= \frac{\partial U}{\partial S}(x,S), \quad \textrm{and} \quad  F^{\rm int}(x, S)=- \frac{\partial U}{\partial x}(x,S)= p(x,S)A,}
\]
where $p=- \frac{\partial U}{\partial V}$ is the pressure.
The friction force reads $F^{\rm fr}(x, \dot x, S)=- \lambda (x, S) \dot x$, where $ \lambda (x, S)\geq 0$ is the phenomenological coefficient,  determined experimentally. Hence, it follows from Definition \ref{LdA_def} that the phenomenological constraint is 
\[
\frac{\partial U}{\partial S}\dot S= \lambda (x, S) \dot x ^2 +P_H^{\rm ext}
\]
and the variational formalism reads
\[
\delta \int_{t _1 }^{t _2 } \left[ \frac{1}{2} m \dot x ^2 -U(x,S) \right] dt +\int_{ t _1 }^{ t _2 }\left\langle F^{\rm ext}(q, \dot q, S), \delta q\right\rangle=0,
\]
subject to the variational constraint
\[
\frac{\partial U}{\partial S}\delta  S= \lambda (x, S) \dot x \delta x.
\]
It yields the time evolution equation of the coupled  mechanical and thermal system for the one-cylinder problem:
\[
m\ddot x =p(x,S)A+F^{\rm ext}- \lambda (x, S)\dot x, \qquad T\dot S= \lambda (x,  S)\dot x ^2 +P^{\rm ext}_H,
\]
consistently with the equations derived by \cite[\S4]{Gr1999}.
In particular, the balance of mechanical and internal energies are
\[
\frac{d}{dt} E_{\rm mech}= p(x,S)A\dot x+F^{\rm ext}\dot x- \lambda (x, S)\dot x^2 , \quad \frac{d}{dt} U= -p(s,x)A\dot x+ \lambda (x, S)\dot x ^2 +P^{\rm ext}_H, 
\]
by which one can easily verify the energy balance law $\frac{d}{dt} E=F^{\rm ext}\dot x+P^{\rm ext}_H$,
where $E=E_{\rm mech}+ U$ is the total energy.

\begin{figure}[h]
\begin{center}
\includegraphics[scale=0.6]{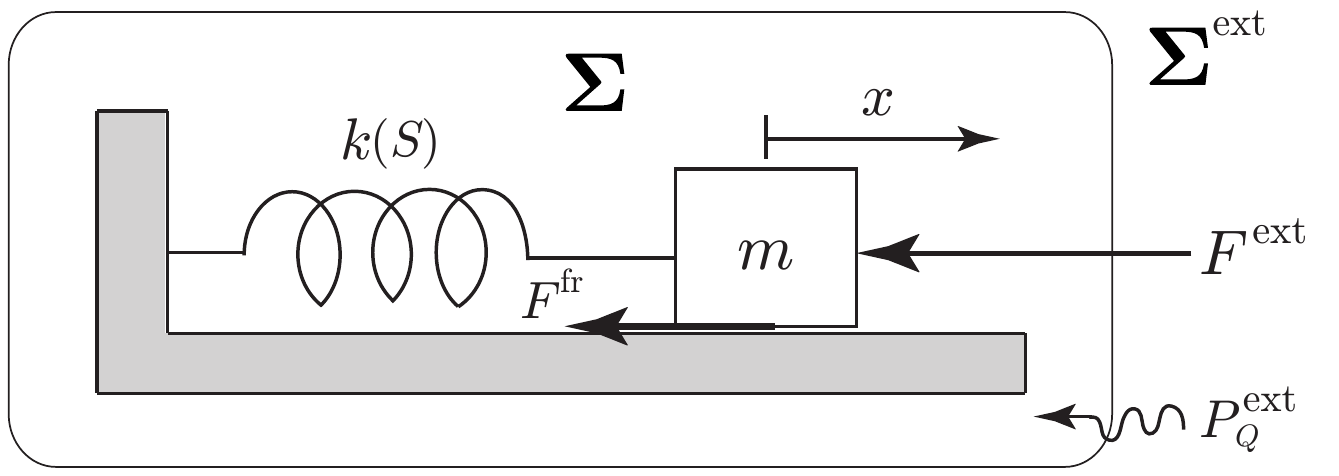}
\caption{Friction of a solid attached to a spring}
\label{FrictionMassSpring}
\end{center}
\end{figure}

\paragraph{Example: a mass-spring system with friction.} The nonequilibrium thermodynamics of a mass-spring  system with friction was considered in \cite{FeGr2010}, see Fig. \ref{FrictionMassSpring}.
In order to apply our approach, we consider the Lagrangian of the thermodynamic system given by $L(x,\dot x, S)=\frac{1}{2} m\dot x ^2 - \frac{1}{2} k(S) x ^2 -U(S)$, where $m$ is a mass and $k(S)$ is a spring constant. In general, $k$ does not depend very strongly on $S$ (or equivalently on temperature), but we
have included this dependence here to illustrate the generality of our approach. The friction force is $F^{\rm fr}(x, \dot x, S)= - \lambda (x, S) \frac{\dot x}{|\dot x|}$ and we assume that the system is subject to an external force $F^{\rm ext}$. As before, the power due to heat transfer between the system and the exterior is denoted by $P^{\rm ext}_H$. From the thermodynamic point of view, the microscopic variables have disappeared at the macroscopic level and have been replaced by the single variable $S$.

The temperature reads
\[
T(x, S)= - \frac{\partial L}{\partial S}=  \frac{1}{2} \frac{\partial k}{\partial S}x ^2 + \frac{\partial U}{\partial S}.
\]
Since the internal energy does not depend on $x$, there is no generalized internal forces.
The system \eqref{thermo_mech_equations} yields the time evolution equation of the coupled mechanical and thermal system:
\begin{equation}\label{equ_FeGr} 
\frac{d}{dt} \left( m\dot x\right)  =- k(S)x+F^{\rm ext}  - \lambda (x,  S) \frac{\dot x}{|\dot x|}, \qquad \dot S= \frac{1}{T} \lambda (x,  S)|\dot x|  +\frac{1}{T}P^{\rm ext}_H.
\end{equation} 
The total energy is $E(x, \dot x, S)= \frac{1}{2} m\dot x ^2 +  \frac{1}{2} k(S) x ^2 + U(S)$ verifies $ \frac{d}{dt} E= P^{\rm ext}_W+P^{\rm ext}_H$.
\medskip

Note that in general, we have $T= -\frac{\partial L}{\partial S}$ and not necessarily $T= \frac{\partial U}{\partial S}$, where $U(S)$ is the internal energy of the system. In the present example this is due to the fact that the mechanical potential energy also depends on $S$, through the coefficient $k(S)$. If the dependence of $k$ and $ \lambda $ on $S$ can be neglected, then the first equation in \eqref{equ_FeGr} can be solved independently of the thermodynamic equation. This justifies the conventional study of friction in mechanics without taking into account the thermodynamic effects (\cite{FeGr2010}).

\subsubsection{Electric circuits with entropy production}\label{subsec_simple_EC}

In any real (i.e., non-ideal) electric circuit, an {\it  irreversible energy conversion from electrical energy into heat} occurs and hence leads to {\it entropy production}. We shall describe the thermodynamics of such electric circuits by using our nonholonomic Lagrangian formalism.

In absence of thermodynamic effects, the Lagrangian formalism for electric circuits has been well established. It is based on the electric and magnetic energies in the circuit and the interconnection constraints expressed in the Kirchhoff laws. We refer to \cite{ChMc1974} for the Lagrangian formalisms of electric circuits. In particular, regarding the variational formalism as degenerate Lagrangian systems with KCL constraints and Dirac structures, see \cite{YoMa2006a,YoMa2006b,YoMa2006c}.
\medskip

\begin{figure}[h]
\vspace{3mm}
\begin{center}
\includegraphics[scale=0.75]{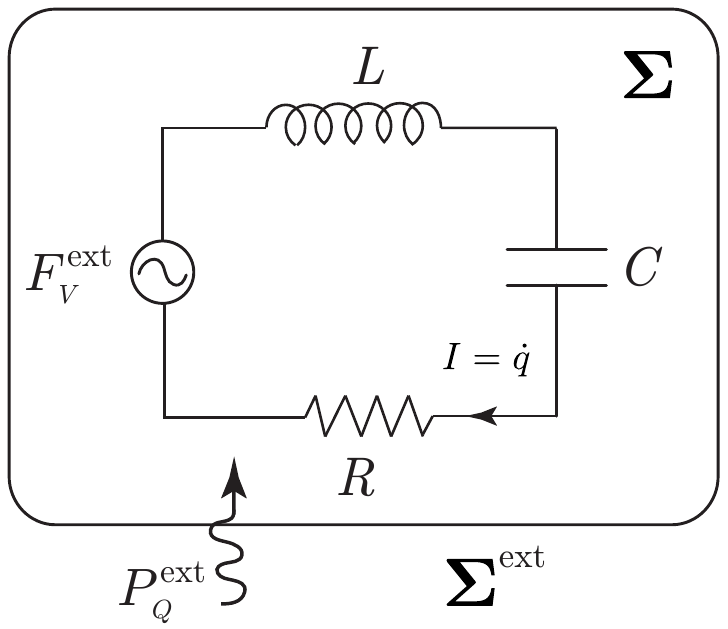}
\caption{A nonlinear RLC circuit with a voltage source}
\label{simple_nonlinear_circuit}
\end{center}
\end{figure}

Consider a serial nonlinear RLC circuit with a voltage source, as illustrated in Fig. \ref{simple_nonlinear_circuit}, with one single entropy variable. The general case will be discussed in \S\ref{discrete_systems}.

Since our formalism is well adapted for both linear and nonlinear circuit elements such as charge-controlled capacitors, current controlled inductors, current controlled resistors, etc., we shall directly consider the nonlinear case. Recall that the constitutive relations of these elementary circuit elements are respectively given by $V=V_C(q)$, and $ \varphi = \varphi _L(I)$, $V=V_R(I)$, where $q$ is the charge, $I=\dot q$ is the current, $ \varphi $ is the magnetic flux, and $V_L=\dot \varphi $ is the inductor voltage.
The total energy stored in this circuit is
\[
E(q, \dot q, S)= K_L(\dot q)+U_C(q)+U(S),
\]
where $K_L(\dot q)$ is the magnetic energy stored by the inductor, i.e., $ \varphi _L (\dot q)=\frac{\partial K_L}{\partial \dot q}$, $U_C(q)$ is the electric energy stored by the capacitor, i.e., $ V_C(q)=\frac{\partial U_C}{\partial q}$, and $U(S)$ is the internal energy, i.e.,  $T= \frac{\partial U}{\partial S}$. 
The Lagrangian of the nonlinear circuit is therefore given by
\[
L(q, \dot q, S)= K_L(\dot q)-U_C(q)-U(S).
\]
For the linear case, the constitutive relations are given by $V= \frac{1}{C}q$ and $ \varphi = LI$, so that we recover the usual expressions $U_C(q)= \frac{1}{2C}q ^2 $ and $K_L(\dot q)= \frac{1}{2} L \dot q ^2 $.
\medskip

In this simple case, the dissipative effect due to friction, which contributes to the entropy production, is modeled by a single resistor in series, so $F^{\rm fr}(q,\dot q, S)=V_R(q, \dot q, S)=-R(q,S)\dot{q}$, where  $R(q,S)$ is a positive coefficient of the resistance. 
The voltage source is given by an external force $F^{\rm ext}(q, \dot q, S)= V^{\rm ext}$ and we assume an external heat power $P^{\rm ext}_H$.

The system \eqref{thermo_mech_equations} yields the equations
\[
\frac{d}{dt}\varphi _L (\dot q)+V_C(q) =V_R(q, \dot q, S)+ V^{\rm ext}, \quad \dot S= -\frac{1}{T}V_R(q, \dot q, S) \dot q+\frac{1}{T}P^{\rm ext}_H,
\]
where $V_R(q, \dot q, S)= - R(q,S) \dot q$ is dissipative, i.e., $V_R(q, \dot q, S)=-R(q,S) \dot{q}^{2} \leq 0$, for all $\dot q$. The entropy production can be rewritten exclusively in terms of the stored energy associated to the capacitor and the inductor as
\begin{equation}\label{S_production_nonlin_circuit}
\begin{aligned} 
\dot S&= -\frac{1}{T}V_R(q, \dot q, S) \dot q+\frac{1}{T}P^{\rm ext}_H\\
&= \frac{1}{T}\left[ \left( -\frac{d}{dt}\varphi _L (\dot q)-V_C(q)+V^{\rm ext}\right) \dot q+P^{\rm ext}_H \right]\\
&=\frac{1}{T}\left[- \frac{d}{dt} \left( K_L(\dot q)+U_C(q) \right) +P^{\rm ext}_W+P^{\rm ext}_H \right],
\end{aligned}
\end{equation}  
where the equality $(\frac{d}{dt}\varphi _L (\dot q)+V_C(q))\dot{q}= \frac{d}{dt} \left(K_L(\dot q)+U_C(q) \right) $ follows from a simple computation using the definition of $ \varphi _L$ and $U_C$. By analogy with mechanics, we denoted $P_W^{\rm ext}:= V^{\rm ext} \cdot \dot q$ the external  "mechanical" power exerted by the voltage source.
\medskip

For the linear case, the rate of entropy production obtained in \eqref{S_production_nonlin_circuit} recovers usual expressions, see for example (17.1.25) and (17.1.26) in \cite{KoPr1998} for the entropy production in real capacitors and inductors.

\subsubsection{Dynamics of simple systems with chemical reactions}\label{DCR}

\paragraph{Setting for  chemical reactions.}

Consider a system of several chemical components undergoing chemical reactions. Let $I=1,...,R$ be the {\it chemical components}  and $a=1,...,r$ the chemical reactions. We denote by $N_I$ the {\it number of moles} of the component $I$. Chemical reactions may be represented by
\begin{equation}\label{chemical_reaction} 
\sum_I {\nu '}_{I}^a\,I\; \stackrel[a _{(2)} ]{ a _{(1)} }{\rightleftarrows} \; \sum_I{\nu ''}^a_{I}\, I, \quad a=1,...,r,
\end{equation} 
where $a_{(1)}$ and $a_{(2)}$ are {\it forward and backward reactions} associated to the reaction $a$, and ${\nu''}^a _{I}$, ${\nu '}^a_I$ are {\it forward and backward stoichiometric coefficients} for the component $I$ in the reaction $a$. From this relation, the number of moles $N_I$ has to satisfy
\begin{equation}\label{NH_constraint_reaction}  
\frac{d}{dt} N_I= \sum_{a=1}^{r} \nu^a _{I} \frac{d}{dt} \psi _a, \quad I=1,...,R,
\end{equation} 
where $\nu^a _{I}:= {\nu ''}^a_{I}- {\nu'}^a _{I}$, $ \psi _a $  is the {\it degree of advancement of reaction} $a$, and $\dot \psi _a $ is the {\it  rate of the chemical reaction} $a$. 

The state of the system is given by the internal energy $U=U(S, N_{1}, ..., N_{R},V)$, which is a function of the entropy $S$, the number of moles $N_{1}, ..., N_{R}$, and the volume $V$. The {\it chemical potential} of the component $I$ is defined by $\mu ^I = \frac{\partial U}{\partial N_I}$
and the temperature is given by $T= \frac{\partial U}{\partial S}$. An {\it affinity} of the reaction $a$ is the state function defined by
\begin{equation}\label{def_affinity} 
\mathcal{A} ^a= - \sum_{I=1}^{R} \nu^a _{I} \mu ^I, \quad a=1,...,r.
\end{equation}

\paragraph{I) Variational formalism for chemical reactions.} We shall now develop two variational formalisms for the nonequilibrium thermodynamics of chemical reactions. We assume that the reaction is \textit{isochoric} $V=V_0$ (constant) and \textit{closed} $P_M^{\rm ext}=0$. The first formalism follows from Definition \ref{LdA_def} and uses the degree of advancement of reactions as a generalized coordinate. The second formalism, that we will present in Definition \ref{Al_LdA_def-chem} below, needs the introduction of new variables, but it has the advantage to admit a corresponding version in the continuum case, that will be of crucial use in the case of a multicomponent fluid with chemical reactions (\S\ref{sec_multi_comp}).

\paragraph{First approach.} We shall now show that the coupled reaction and thermal evolution of the system can be obtained via the variational formalism of Definition \ref{LdA_def}. In order to do this, we shall first use the degree of advancement of reactions $ \psi _a, \, a=1,...,r $ as well as the entropy $S$ as the thermodynamic variables, which characterize the nonequilibrium thermodynamics of the chemical reactions.

It follows from the time integral of \eqref{NH_constraint_reaction} that we have
\begin{equation}\label{N_psi} 
N_I(t)=N_I(t_1)+ \nu^a _{I} \psi  _a (t), \quad \psi_a  (t _1 )=0.
\end{equation} 
Replacing this expression in the internal energy, we can define the Lagrangian by
\[
L( \psi_{a}, S):=-U(N_1,...,N_R, V_0, S),
\]
where we note that there is no dependence on $\dot \psi_{a}$.
\medskip

The variational formalism in Definition \ref{LdA_def} reads
\[
-\delta \int_{t_1}^{t_2}U(S, N_1,...,N_R, V_0)dt= 0,
\]
with phenomenological and variational constraints
\[
-\frac{\partial U}{\partial S}\dot S= \sum_{a=1}^{r} \mathcal{F}^{{\rm fr}\,a}( \psi_{a} , \dot \psi_{a}, S)  \dot \psi_{a} -P^{\rm ext}_H \quad\text{and}\quad - \frac{\partial U}{\partial S}\delta S= \sum_{a=1}^{r}  \mathcal{F}^{{\rm fr}\, a}(\psi_{a} , \dot \psi_{a}, S) \delta  \psi_{a}.
\]
This variational formalism provides the evolution equations \eqref{thermo_mech_equations} given here by
\begin{equation*}
\left\{ 
\begin{array}{l}
\displaystyle\vspace{0.2cm} \frac{\partial U}{\partial \psi^{a} }= \mathcal{F}^{{\rm fr}\, a}(\psi_{a} , \dot \psi_{a}, S),\\
\displaystyle-\frac{\partial U}{\partial S}\dot S= \sum_{a=1}^{r} \mathcal{F}^{{\rm fr}\, a}(\psi_{a} , \dot \psi_{a}, S)  \dot \psi_{a} -P^{\rm ext}_H.
\end{array} 
\right.
\end{equation*} 
In the above, the variables $\mathcal{F}^{\rm fr\, a} (\psi_{a} , \dot \psi_{a}, S)$ are responsible for entropy increase, which may have the form 
\[
\mathcal{F}^{\rm fr\, a} (\psi_{a} , \dot \psi_{a}, S)=- \lambda ^{ab}(\psi_{a}, S)\dot \psi _b, 
\]
where the symmetric part of the matrix $ \lambda^{ab}$ is positive from the second principle.
It follows from \eqref{def_affinity} that we have 
\[
\frac{\partial L}{\partial \psi _a }=-\frac{\partial U}{\partial \psi _a }=-\frac{\partial N_{I}}{\partial \psi _a }\frac{\partial U}{\partial N_{I}}=-\sum_{I=1}^{R}\nu^a_{I}\mu^{I}=\mathcal{A} ^a, 
\]
and hence the evolution equations read
\[
\mathcal{A} ^a = \lambda ^{ab}\dot \psi _b \;\;\text{and}\;\;  T\dot S= \lambda ^{ab}\dot \psi _a \dot \psi _b +P_H^{\rm ext}.
\]
Assuming that the matrix $ \lambda _{ab}$ is invertible, with its inverse denoted $ \mathcal{L} ^{ab}$, we can rewrite these equations as
\begin{equation}\label{chemical_reaction_1} 
\dot \psi _a = \mathcal{L} _{ab} \mathcal{A} ^b \;\;\text{and}\;\; T\dot S= \mathcal{L} _{ab} \mathcal{A} ^a \mathcal{A} ^b +P_H^{\rm ext},
\end{equation}
which are the well-known coupled equations for chemical reactions and thermal evolution, see for example (3.45) in \cite{Gr1997}.


\paragraph{Second approach.} The previous approach, while simpler, is not well adapted for a generalization to the continuum case (see \S\ref{sec_multi_comp}).
We shall now present an alternative variational formalism for chemical reactions, inspired from Definition \ref{LdA_def}, that 
admits such a generalization. This approach does not describe the evolution of the variables $ \psi_a (t)$, but focus directly on the number of moles $N_I(t)$. To do this, we shall introduce the new variables $ W^I(t)$ and $ \nu ^a (t)$ defined such that 
\begin{equation}\label{def_W-nu}
\dot{W}^{I}=\mu^{I} \quad\text{and}\quad 
\dot{\nu}^{a}:=-\mathcal{A}^{a}
\end{equation}
and whose interpretation will be given in the context of continuum systems in \S\ref{sec_multi_comp}. The alternative variational formalism for chemical reactions is stated in the following definition.

\begin{definition}[\textbf{Alternative variational formalism for chemical reaction dynamics}]\label{Al_LdA_def-chem} Consider chemical reactions \eqref{chemical_reaction} and define the Lagrangian
\begin{equation*}\label{al_Lag_chem}
L(N_1,...,N_R ,S):=-U(N_1,...,N_R ,S, V_0),
\end{equation*}
where $U$ is the internal energy. 
Let  $J^{\rm fr}_{a}= J^{\rm fr}(N_1,...,N_R , S)_a :  \mathbb{R}^{R}  \times \mathbb{R} \rightarrow \mathbb{R}$ be the friction rate of the chemical reactions $a=1,...,r$ and let $P_H^{\rm ext}$ be the heat power exchange between the system and the exterior. The alternative variational formalism is given by
\begin{equation*}\label{al_GLdA_thermo_chem} 
\delta \int_{ t _1 }^{ t _2 }    \left(L(N_1,...,N_R , S)+ \sum_{I=1}^{R}\dot{W}^{I}N_{I} \right)   dt  =0, \quad \textsc{Variational Condition}
\end{equation*}
where the curves  $S(t)$, $N_{I}(t)$, and $W^{I}(t)$ satisfy the nonholonomic constraints
\begin{equation*}\label{al_GNonholonomic_Constraints1_chem} 
\frac{\partial L}{\partial S}\dot S   =  \sum_{a=1}^{r}J^{\rm fr}_{a} \dot{\nu}^{a}-  P_H^{\rm ext},\qquad\qquad\quad\;\;\, \textsc{Phenomenological Constraint}
\end{equation*}
\begin{equation*}\label{al_GNonholonomic_Constraints2_chem}
\dot{\nu}^a=\sum_{I=1}^{R} \nu^a_{I} \dot{W}^{I},\quad\hspace{4.5cm}\;\; \textsc{Chemical Constraints}
\end{equation*} 
and with respect to the variations $\delta S$ and $\delta {\nu}^{a}$ subject to
\begin{equation*}\label{al_GVariational_Constraints_chem} 
\frac{\partial L}{\partial S}\delta S=  \sum_{a=1}^{r}J^{\rm fr}_{a} \delta{\nu}^{a}  \;\;\text{and}\;\; \delta{\nu}^{a}=\sum_{I=1}^{R} \nu^a_{I} \delta{W}^{I}, \;\;\; \textsc{Variational constraints}
\end{equation*}
with $ \delta W ^I ( t _i)=0$ for $i=1,2$. 
\end{definition}

\medskip

By this principle, one deduces that the equations associated to the variations $ \delta N_I$ and $ \delta W^I $ are, respectively
\begin{equation}\label{conditions_second_version} 
-\frac{\partial U}{\partial N_I}+\dot W ^I =0\quad\text{and}\quad \dot N _I = J^{\rm fr}_a \nu ^a_{I}.
\end{equation} 
Taking into account of the two nonholonomic constraints and the definition $ \mathcal{A} ^a=- \nu ^a_{I} \mu ^I $, we obtain
\begin{equation}\label{chemical_reaction_2} 
\dot N _I= J^{\rm fr}_{a} \nu^a _{I} \quad\text{and}\quad T\dot S=J^{\rm fr}_{a}  \mathcal{A} ^a+P_H^{\rm ext}.
\end{equation}

If we make the choice $J^{\rm fr}_a=\mathcal{L} _{ab} \mathcal{A} ^b$, then the equations \eqref{chemical_reaction_2} are obtained from \eqref{chemical_reaction_1} by multiplying the first equation by the stoichiometric coefficients $\nu^a_{I}$. This eventually induces $J^{\rm fr}_a=\dot{\psi}_{a}$.

\begin{figure}[h]
\begin{center}
\includegraphics[scale=0.79]{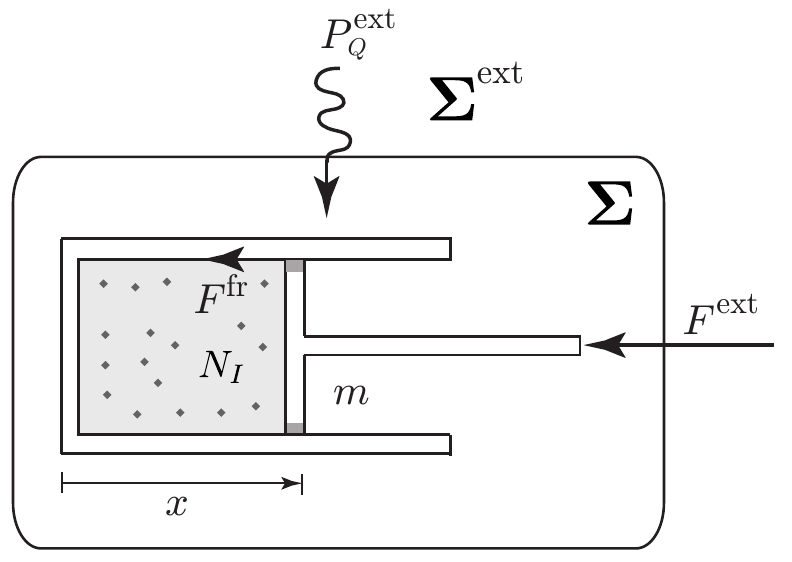}
\caption{Piston-cylinder with chemical reactions}
\label{piston_cylinder_chem}
\end{center}
\end{figure}

\paragraph{II) Variational formalism for simple systems with chemical reactions.}
Let us consider the more general case of a mechanical system involving chemical reactions. As above, we assume that the system is closed $P^{\rm ext}_M=0$. This setting is well illustrated with the example of chemical reactions occurring in a piston-cylinder system, as illustrated in Fig. \ref{piston_cylinder_chem}. We will see that our variational approach allows a very efficient treatment for the derivation of the evolution equations for the nonequilibrium thermodynamics of this system. 

As before, we will present two variational formalisms. The first is based on Definition \ref{LdA_def}. The second one generalizes Definition \ref{Al_LdA_def-chem} and has a corresponding version in the continuum setting.

\paragraph{First approach.} Assume that there are $R$ chemical components and $r$ chemical reactions. Let $\psi_{a}$ be the degree of advancement of reactions $a=1,...,r$ and $x$ the displacement of the piston from the bottom of the cylinder. We now apply the setting of Definition \ref{LdA_def} to the system of the piston-cylinder with chemical reactions, where the state of the system is characterized by $(x, \dot x, S, \psi_{a})$.  The Lagrangian of this system is
\begin{equation}\label{chemical_reaction_piston} 
L( x,\dot x, S, \psi_{a})=\frac{1}{2} m \dot x ^2 - U(S, N_1,...,N_R,V=Ax),
\end{equation} 
where as before $ N_I(t)=N_I(t_0)+ \nu ^a_{I} \psi  _a (t)$, $\psi_a  (t _0 )=0$ and the cross sectional area of the cylinder $A={\rm const}$. It follows from \eqref{thermo_mech_equations} that we can immediately obtain the equations of motion
\[
\frac{d}{dt} \frac{\partial L}{\partial \dot x}-\frac{\partial L}{\partial x}=F^{\rm ext}+F^{\rm fr}, \quad -\frac{\partial L}{\partial \psi _a }=\mathcal{F}^{{\rm fr}\, a}, \quad \frac{\partial L}{\partial S}\dot S= F^{\rm fr} \dot x+ \mathcal{F}^{{\rm fr}\, a} \dot \psi _a -P^{\rm ext}_H,
\]
for friction forces given by $ F^{\rm fr} =-\lambda \dot x$ and $\mathcal{F}^{{\rm fr}\, a} =-\lambda^{ab} \dot \psi_{b}$. From the expression \eqref{chemical_reaction_piston} of the Lagrangian, the evolution equations are 
\[
m\ddot x=pA- \lambda \dot x+F^{\rm ext}, \quad\dot \psi_a =\mathcal{L} _{ab} \mathcal{A} ^b \quad\text{and}\quad T\dot S= \lambda \dot x ^2 + \mathcal{L} _{ab} \mathcal{A} ^a \mathcal{A} ^b +P_H^{\rm ext},
\]
where $p=p(x, \psi _a, S)=- \frac{\partial U}{\partial V}$ and we recall that $\mathcal{A} ^a=\frac{\partial L}{\partial \psi _a }$.

\paragraph{Second approach.} This approach generalizes the variational formalism of Definition \ref{Al_LdA_def-chem} to the case when chemical reactions are coupled to a mechanical system.

\begin{definition}[\textbf{Alternative variational formalism for simple systems with chemical reactions}]\label{Alt_LdA_def} Let $N_{I},\,I=1,...,R$ be the number of moles of  chemical components associated with chemical reactions $a=1,...,r$, given by \eqref{N_psi}. Let $L:TQ \times \mathbb{R} \times \mathbb{R}^R  \rightarrow \mathbb{R}$ be the Lagrangian,  $J^{\rm fr}: TQ\times \mathbb{R}    \rightarrow T^*Q$ the friction flow, and $P^{\rm ext}_H$ the heat power exchange between the system and the exterior. Let $F^{\rm ext}$ and $F^{\rm fr}:TQ\times \mathbb{R}    \rightarrow T^*Q$ be external and friction forces. The variational formalism for simple closed systems with chemical reactions is defined as
\begin{equation}\label{AltGLdA_thermo} 
\delta \int_{ t _1 }^{ t _2 }\left( L(q , \dot q , S, N_{I}) +\sum_{I=1}^{R} \dot{W}^{I} N_{I} \right)dt +\int_{ t _1 }^{ t _2 }\left\langle F^{\rm ext}(q, \dot q, S), \delta q\right\rangle  =0,
\end{equation}
where the curves $q(t)$, $S(t)$, $W^{I}(t)$, and $N_{I}(t)$, $I=1,...,R$, satisfy the nonholonomic constraints
\begin{equation*}\label{AltGNonholonomic_Constraints1} 
\frac{\partial L}{\partial S}\dot S   = \left\langle F^{\rm fr}(q, \dot q, S) , \dot q \right\rangle + \sum_{a=1}^rJ^{\rm fr}_a \dot \nu ^a -  P_H^{\rm ext}\quad\text{and}\quad \dot \nu ^a =\sum_{I=1}^R\nu ^a_{I}\dot W ^I
\end{equation*} 
and with respect to the variations $ \delta q $, $\delta S$, and $\delta{W}^{I}$ subject to the constraints
\begin{equation}\label{AltGVariational_Constraints} 
\frac{\partial L}{\partial S}\delta S= \left\langle F^{\rm fr}(q , \dot q , S),\delta q \right\rangle+\sum_{a=1}^rJ^{\rm fr}_a \delta  \nu ^a\quad\text{and}\quad\delta  \nu ^a =\sum_{I=1}^R\nu ^a_{I}\delta  W ^I,
\end{equation}
with $ \delta q(t _i )=0$ and $\delta W ^I ( t _i)=0$ for $i=1,2$. 
\end{definition} 


\paragraph{Equations of motion.} 
By taking variations of the integral in \eqref{AltGLdA_thermo}, integrating by parts and using $ \delta q(t _1 )= \delta q(t _2 )=0$, the variational formalism of Definition \ref{Alt_LdA_def} yields
\begin{equation*}
\begin{split}
\int_{t _1}^{ t _2 }& \left[  \left\langle  \frac{\partial L}{\partial q}- \frac{d}{dt} \frac{\partial L}{\partial \dot q} +F^{\rm ext}+F^{\rm fr}, \delta q \right\rangle + \left(\frac{\partial L}{\partial N_{I}}+ \dot{W}^{I}\right)\delta N_{I} +\left(\sum_{a=1}^rJ^{\rm fr}_a \nu ^a_{I}-\dot{N}_{I}\right)\delta{W}^{I}
 \right] dt=0,
\end{split}
\end{equation*}
for all the variations $ \delta q $, $ \delta N_{I}$,  and $\delta W^{I}$, where we used both variational constraints in \eqref{AltGVariational_Constraints}. This yields the {\it coupled  mechanical and thermochemical system}:
\begin{equation*}\label{Alt_Gthermo_mech_chem_equations-1}
\left\{ 
\begin{array}{l}
\displaystyle\vspace{0.2cm}\frac{d}{dt} \frac{\partial L}{\partial \dot q}- \frac{\partial L}{\partial q}= F^{\rm ext}(q , \dot q, S) +  F^{\rm fr}(q , \dot q, S),\\[2mm]
\displaystyle \dot{W}^{I}=-\frac{\partial L}{\partial N_{I}},\quad \dot{N}_{I}=\sum_{a=1}^rJ^{\rm fr}_a \nu ^a_{I},\\[1mm]
\displaystyle \frac{\partial L}{\partial S}\dot S   = \left\langle F^{\rm fr}(q, \dot q, S) , \dot q \right\rangle + \sum_{a=1}^rJ^{\rm fr}_a \dot \nu ^a -  P_H^{\rm ext}.
\end{array} 
\right.
\end{equation*} 
The application of this approach to the system of Fig. \ref{piston_cylinder_chem} is left to the reader.

\subsubsection{Dynamics of a simple system with diffusion due to matter transfer}\label{DMT_CR} 

\begin{figure}[h]
\begin{center}
\includegraphics[scale=0.55]{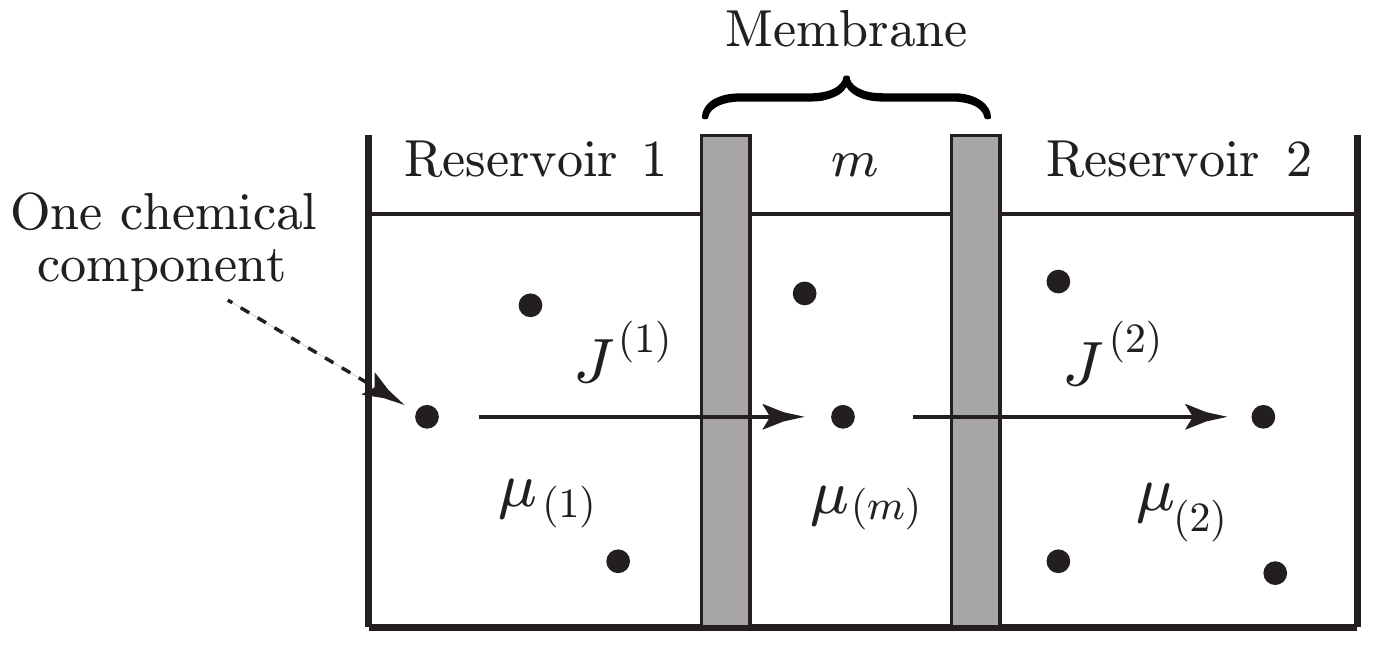}
\caption{Nonelectrolyte diffusion through a homogeneous membrane}
\label{MatterTransport}
\end{center}
\end{figure}

\paragraph{1) Nonelectrolyte diffusion through a homogeneous membrane.}
We consider a {\it system with diffusion due to (internal) matter transfer} through a homogeneous membrane separating two reservoirs. We suppose that the system is simple (so it is described by a single entropy variable) and involves a single chemical component.  We assume that the membrane consists of three regions; namely, the central layer denotes the membrane capacitance in which energy is stored without dissipation, while the outer layers indicate transition region in which dissipation occurs with no energy storage. We denote by $N^{(m)}$ the number of mole of this chemical component in the membrane and also by $N^{(1)}$ and $N^{(2)}$ the numbers of mole in the reservoirs $1$ and $2$, as shown in Fig. \ref{MatterTransport}. Define the Lagrangian by $L(S, N^{(1)}, N^{(2)}, N^{(m)})=-U(S, N^{(1)}, N^{(2)}, N^{(m)})$, where $U(S, N^{(1)}, N^{(2)}, N^{(m)})$ denotes the internal energy of the system and we suppose that the volumes are constant and the system is  isolated. We denote by $\mu _{(k)}= \frac{\partial U}{\partial N ^{(k)} }$ the chemical potential of the chemical components in the reservoirs $(k=1,2)$ and in the membrane $(k=m)$. We denote by $J^{(1)}$ the flux from the reservoir $1$ into the membrane and $J^{(2)}$ the flux from the membrane into the reservoir $2$. 

The {\it variational formalism for the diffusion process} is provided by
\begin{equation}\label{var_cond_matter} 
\delta \int_{ t _1 }^{ t _2 }\left( L(S, N^{(1)}, N^{(2)}, N^{(m)})+\dot{W}_{(1)} N^{(1)}+\dot{W}_{(2)} N^{(2)}+\dot{W}_{(m)} N^{(m)} \right)dt =0,
\end{equation}
subject to the phenomenological and mass conservation constraints
\begin{equation}\label{phenom_constraint_matter} 
\frac{\partial L}{\partial S}\dot S   = J^{(1)}(\dot W_{(1)} - \dot W_{(m)})  +J^{(2)}(\dot W_{(m)} - \dot W_{(2)})\quad\text{and}\quad \dot N^{(1)}+\dot N^{(2)}+\dot N^{(m)}=0,
\end{equation} 
together with the variational constraints
\begin{equation}\label{var_constraint_matter} 
\frac{\partial L}{\partial S}\delta  S   = J^{(1)}( \delta  W_{(1)} - \delta  W_{(m)})  +J^{(2)}(\delta  W_{(m)} - \delta  W_{(2)})\quad\text{and}\quad \delta  N^{(1)}+\delta  N^{(2)}+\delta  N^{(m)}=0,
\end{equation}
with $\delta W_{(k)}( t _i )=0$, for $k=1,2,m$ and $i=1,2$.
  
By applying this variational formalism and using the first variational constraint, it follows
\begin{align*} 
&\int_{t_1}^{t_2} \left( (J^{(1)}-\dot N^{(1)}) \delta W_{(1)}+(-J^{(1)}+J^{(2)}-\dot N^{(m)}) \delta W_{(m)}+(-J^{(2)}-\dot N^{(2)}) \delta W_{(2)}\phantom{\frac{\partial U}{\partial N^2}} \right .\\
&\qquad \qquad  \left . +\left( \dot W _{(1)} - \frac{\partial U}{\partial N^{(1)}}\right) \delta N ^{(1)} + \left( \dot W _{(2)} - \frac{\partial U}{\partial N^{(2)}}\right) \delta N ^{(2)}+\left( \dot W _{(m)} - \frac{\partial U}{\partial N^{(m)}}\right) \delta N ^{(m)}\right) dt=0.
\end{align*}
Since the variations $ \delta W _{(k)} $ are free, we have
\begin{equation}\label{N_dot_equ_diff}
\dot N^{(1)} = J^{(1)},\quad \dot N ^{(m)} = -J^{(1)}+J^{(2)}, \quad \dot N ^{(2)} =-J^{(2)}.
\end{equation} 
Using the second variational constraint as to $ \delta N^{(k)}$, it follows $\dot W _{(1)} - \mu _{(1)}=\dot W _{(2)} - \mu _{(2)}= \dot W _{(m)} - \mu _{(m)}$
and so the first constraint in \eqref{phenom_constraint_matter}, in view of $T=-\frac{\partial L}{\partial S}$,  reads
\begin{equation}\label{S_dot_equ_diff}
-T\dot S= J^{(1)}(\mu _{(1)}- \mu _{(m)})+ J^{(2)}( \mu _{(m)}- \mu _{(2)}).
\end{equation}
Equations \eqref{N_dot_equ_diff} and \eqref{S_dot_equ_diff} are consistent with those derived in \cite[\S2.2]{OsPeKa1973}. From the equations \eqref{N_dot_equ_diff} and \eqref{S_dot_equ_diff}, we have $\frac{d}{dt}U= 0$, in agreement with the fact that the system is isolated.

\paragraph{2) Coupling of chemical reactions and diffusion.} We now assume  that the above system contains several chemical components $I=1,...,R$ that can diffuse through the membranes and can undergo chemical reactions. We denote by $N_I^{(1)}$, $N_I^{(2)}$, $N_I^{(m)}$, the numbers of moles of the component $I$ in the reservoir $1$, the reservoir $2$, and the membrane $m$, respectively. Define the Lagrangian by $L(S,\{N^{(1)}_I, N^{(2)}_I, N^{(m)}_I\}):=-U(S,\{N^{(1)}_I, N^{(2)}_I, N^{(m)}_I\})$, where $U(S,\{N^{(1)}_I, N^{(2)}_I, N^{(m)}_I\}):=U(S, N^{(1)}_1, N^{(2)}_1, N^{(m)}_1, ...,N^{(1)}_R, N^{(2)}_R, N^{(m)}_R)$ the internal energy of the system. The chemical potentials are denoted $ \mu _{(k)}^I= \frac{\partial U}{\partial N^{(k)}_I}$. 

\begin{figure}[h]
\begin{center}
\hspace{-2cm}
\includegraphics[scale=0.55]{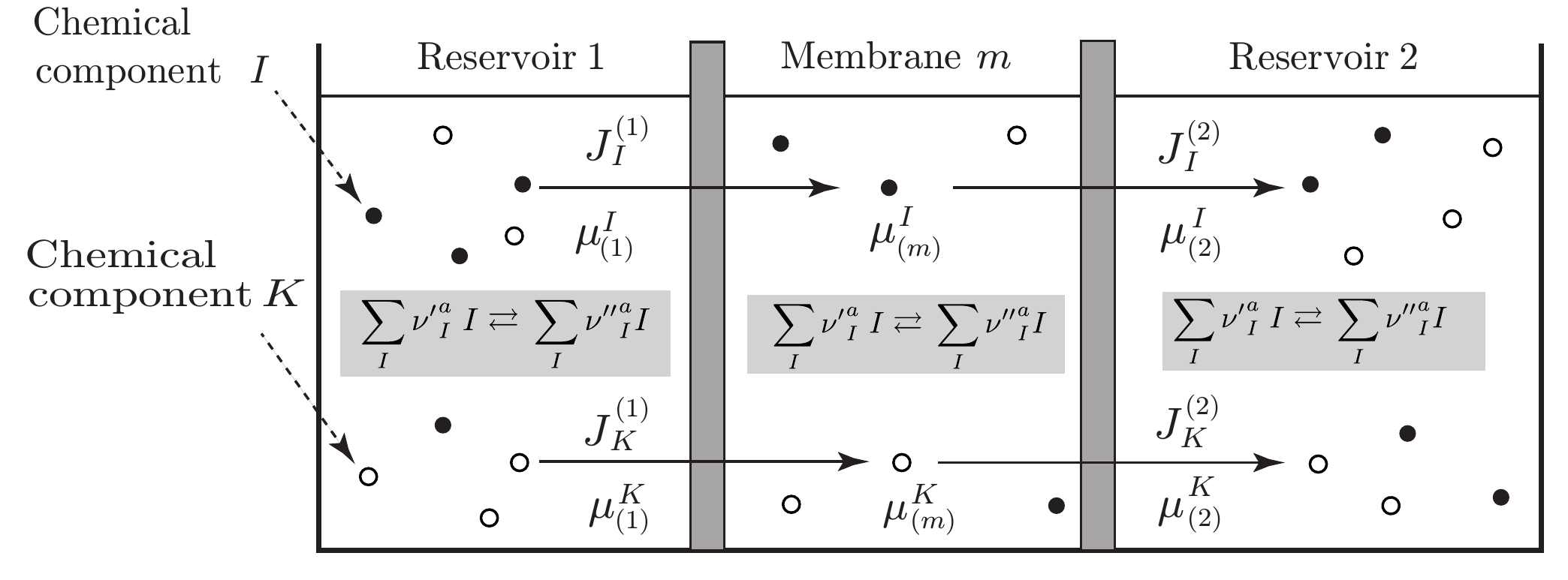}
\caption{Chemical reactions and diffusion through membranes}
\label{MatterTransportReaction}
\end{center}
\end{figure}

The {\it variational formalism for chemical reactions and diffusion} is
\begin{equation}\label{var_cond_matter_chem} 
\delta \int_{ t _1 }^{ t _2 }\Big( L(S, \{N^{(1)}_I, N^{(2)}_I, N^{(m)}_I\})+\sum_{I,k}\dot{W}^I_{(k)} N^{(k)}_I \Big)dt =0
\end{equation}
subject to the phenomenological constraints
\begin{equation}\label{phenom_constraint_matter_chem}
\begin{aligned}
\frac{\partial L}{\partial S}\dot S  & = \sum_I \left( J^{(1)}_I(\dot W_{(1)} ^I  - \dot W_{(m)}^I )  +J^{(2)}_I(\dot W_{(m)} ^I - \dot W_{(2)}^I )\right) + \sum_{k,a}J^{{\rm fr}\,(k)}_a \dot \nu ^a_{(k)},\\
\dot \nu ^a_{(k)} &= \sum_{I}\nu^a _{I} \dot W _{(k)} ^I, \quad k=1,2, m, \;\; a=1,...,r,
\end{aligned}
\end{equation} 
and with respect to the variational constraints
\begin{equation}\label{variat_constraint_matter_chem}
\begin{aligned} 
 \frac{\partial L}{\partial S}\delta  S   &= \sum_{I=1}^R \left( J^{(1)}_I( \delta  W_{(1)}^I - \delta  W^I_{(m)})  +J^{(2)}_I(\delta  W_{(m)}^I - \delta  W_{(2)}^I)\right) + \sum_{k,a}J^{{\rm fr}\,(k)}_a \delta  \nu ^a_{(k)},\\
 \delta  \nu ^a_{(k)} &= \sum_{I}\nu^a _{I} \delta  W _{(k)} ^I, \quad k=1,2, m, \;\; a=1,...,r,
\end{aligned}
\end{equation} 
with $\delta W_{(k)}( t _i )=0$, for $k=1,2,m$ and $i=1,2$.
\medskip

The variations $ \delta W _{(k)}^I  $ yield
\begin{equation}\label{N_dot_equations}
\begin{aligned} 
&\dot N^{(1)} _I= J^{(1)}_I+\sum_{a=1}^r J^{{\rm fr}\, (1)}_ a \nu _{aI},\qquad \dot N ^{(m)}_I = -J^{(1)}_I+J^{(2)} _I+\sum_{a=1}^r J^{{\rm fr}\,(m)}_a \nu ^a_{I},\\
&\dot N ^{(2)}_I =-J^{(2)}_I+\sum_{a=1}^r J^{{\rm fr}\, (2)}_ a \nu _{aI}, \qquad I=1,...,R. 
\end{aligned}
\end{equation}  
Using the variations $ \delta N^{(k)}_I$, we obtain $\dot W^I  _{(1)} =\mu^I _{(1)}$, $\dot W^I _{(2)} = \mu^I _{(2)}$, $\dot W^I _{(m)}= \mu^I _{(m)}$,
and therefore the phenomenological constraint \eqref{phenom_constraint_matter_chem} becomes
\begin{equation}\label{S_dot_equation} 
-T\dot S= \sum_I \left( J^{(1)}_I(\mu ^I_{(1)}- \mu ^I_{(m)})+ J^{(2)}_I( \mu^I _{(m)}- \mu ^I_{(2)}) \right) -\sum_{k,a} J^{{\rm fr}\,(k)}_a\mathcal{A} ^a_{(k)},
\end{equation} 
where the affinity is defined by $\mathcal{A} ^a_{(k)}:= - \sum_{I=1}^{R} \nu ^a_{I} \mu ^I_{(k)}, \quad a=1,...,r$.
Equations \eqref{N_dot_equations} and \eqref{S_dot_equation} recover those derived in \cite[\S VI]{OsPeKa1973}. Again, internal energy conservation $\frac{dU}{dt}=0$ is obtained from \eqref{N_dot_equations} and \eqref{S_dot_equation} since the system is isolated.
 
\begin{remark}{\rm
In presence of chemical reactions, the mass conservation during each reaction arises from the condition $\sum_I \nu^a _{I}=0$ for $a=1,...,r$ (Lavoisier law). The variational formalism \eqref{var_cond_matter_chem}--\eqref{variat_constraint_matter_chem} for chemical reactions and diffusion is therefore consistent with the variational formalism \eqref{var_cond_matter}--\eqref{var_constraint_matter} in absence of chemical reaction, as mass conservation is already contained in the chemical constraint. Note also that the variational formalism \eqref{var_cond_matter_chem}--\eqref{variat_constraint_matter_chem} recovers  the variational formalism of Definition \ref{Al_LdA_def-chem} in absence of diffusion processes. Of course, one can easily develop a similar generalization of the variational formalism of Definition \ref{Alt_LdA_def} in order to include diffusion processes. This yields the dynamical equations for simple systems involving the coupling of mechanical variables with chemical reactions and diffusion processes. A continuum version of such a variational formalism will be also developed in \S\ref{sec_multi_comp} for a multicomponent fluid subject to the irreversible processes associated to viscosity, heat transport, (internal) matter transport, and chemical reactions.}
\end{remark}

\subsection{Variational formalism for the nonequilibrium thermodynamics of discrete systems}\label{discrete_systems} 

We now consider the case of a closed discrete system $ \boldsymbol{\Sigma}  = \cup_{A=1}^N \boldsymbol{\Sigma}  _A$, composed of interconnecting simple systems $ \boldsymbol{\Sigma}  _A$ that can exchange heat and mechanical power, and interact with external heat sources $ \boldsymbol{\Sigma}  _R$. As will be shown, we need to extend the formalism of Definition \ref{LdA_def} in order to take account of {\it internal heat exchanges}. Before presenting the variational formalism, we first review from \cite{StSc1974} and \cite{Gr1997} the description of discrete systems.

By definition, a {\it heat source} of a system is uniquely defined by a single variable $S_R$. Therefore, its energy is given by $U_R=U_R(S_R)$, the temperature is $T^R:= \frac{\partial U_R}{\partial S_R}$, so that $ \frac{d}{dt} U_R=T^R\dot S_R=P^{R \rightarrow \boldsymbol{\Sigma}  }_H$, where $P_H^{R \rightarrow \boldsymbol{\Sigma}  }$ is the heat power flow due to heat exchange with $ \boldsymbol{\Sigma}  $.

\paragraph{Discrete systems.} The state of a discrete system is described by geometric variables $q \in Q_{ \boldsymbol{\Sigma} } $ and entropy variables $S_A$, $A=1,...,N$. Note that the entropy $S_A$ has the index $A$ since it is associated to the system $ \boldsymbol{\Sigma}  _A$. The geometric variables, however, are not indexed by $A$ since in general they are associated to several systems $ \boldsymbol{\Sigma}  _A$ that can interact. Note that in practice, it can be a difficult task to identify such independent geometric variables $q$ for a given complex (interconnected) discrete system. In this case, it is often useful to first consider the (in general) non independent geometric variables $q^A$ associated to each of the simple systems $ \boldsymbol{\Sigma}  _A $, and subject to an \textit{interconnection constraint}, see Remark \ref{Remark_interconnection}. Such an approach also allows the treatment of nonholonomic interconnections, which is a more general situation that we do not consider in the present paper, see also Remark \ref{Remark_interconnection}.

For discrete systems the variables $V_A$ and $N_A$ (volume and number of moles) are in general defined from (a subset of) the geometric variables, as we have seen in the examples of the piston problem ($V$ in terms of $x$) and chemical reactions ($N_A$ in terms of $ \psi ^a $, \eqref{N_psi}).

\begin{figure}[h]
\begin{center}
\hspace{2cm}
\includegraphics[scale=0.67]{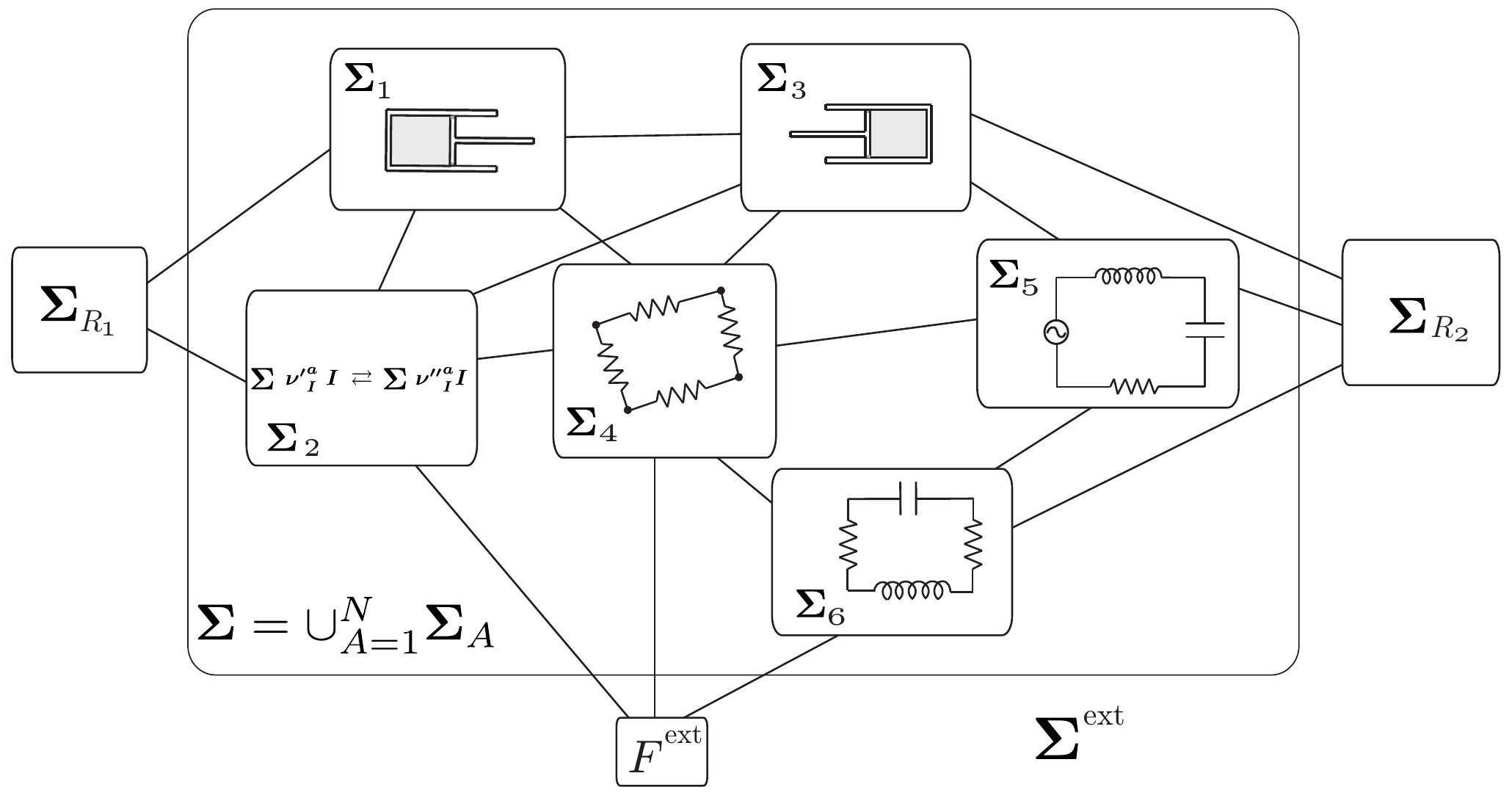}
\caption{An illustration of a discrete system}
\label{discrete_system}
\end{center}
\end{figure}

As before, the action from the exterior is given by external forces and by transfer of heat. For simplicity, we ignore matter exchange in this section so, in particular, the system is closed. The external force reads $F^{\rm ext}:=\sum_{A=1}^N F^{{\rm ext} \rightarrow A}$, where $F^{{\rm ext} \rightarrow A}$ is the external force acting on the system $\boldsymbol{\Sigma}  _A$. The associated external mechanical power is $P^{\rm ext}_W=\langle F^{\rm ext}, \dot q \rangle = \sum_{I=1}^N \langle F^{{\rm ext} \rightarrow A} , \dot q \rangle = \sum_{A=1}^N P^{{\rm ext}\rightarrow A}$.
The external heat power associated to heat transfer reads
\[
P^{\rm ext}_H =\sum_{R}P^{R \rightarrow \boldsymbol{\Sigma}  }_H=\sum_R \left(\sum_{A=1}^NP_H^{R \rightarrow A}\right)=\sum_{A=1}^NP_H^{{\rm ext} \rightarrow A},
\]
where $P_H^{R \rightarrow A}$ denotes the power of heat transfer  between the external heat source $ \Sigma _R$ and the system $ \boldsymbol{\Sigma}  _A$, and $P^{R \rightarrow \boldsymbol{\Sigma}  }_H=\sum_{A=1}^NP_H^{R \rightarrow A}$ denotes the power of heat transfer between $\boldsymbol{\Sigma} _{R}$ and the system $ \boldsymbol{\Sigma} $.
In addition to the above external actions,  there are also {\it internal actions} on every $ \boldsymbol{\Sigma}  _A$ due to the mechanical and heat power from $ \boldsymbol{\Sigma}  _B$. First, there are {\it internal forces} $F^{B \rightarrow A}=- F^{ A \rightarrow B}: TQ_\Sigma \times \mathbb{R}  ^N \rightarrow T^*Q_ {\boldsymbol{\Sigma}} $ exerted by $ \boldsymbol{\Sigma}  _B $ on $ \boldsymbol{\Sigma}  _A$, with associated power denoted by $P_W^{B \rightarrow A}=\langle F^{B \rightarrow A}( q , \dot q , S _1 , ..., S _N ),\dot q\rangle $. Secondly, there are friction forces
\[
F^{{\rm fr}(A)}:TQ_{ \boldsymbol{\Sigma}  } \times \mathbb{R}  ^N \rightarrow T^* Q_ {\boldsymbol{\Sigma} }\quad\text{with}\quad F^{\rm fr}:=\sum_{A=1}^N F^{{\rm fr}(A)}
\]
associated to $ \boldsymbol{\Sigma}  _A$, i.e., involved in the entropy production $\dot S _A$. Finally there is an internal heat power exchange between $ {\boldsymbol{\Sigma} } _A$ and $ {\boldsymbol{\Sigma}} _B$, denoted by $P_H ^{B \rightarrow A}$. We have
\begin{equation*}\label{action_reaction} 
P^{B \rightarrow A}_W =- P^{A \rightarrow B}_W\quad\text{and}\quad P^{B \rightarrow A}_H=-P^{A \rightarrow B}_H.
\end{equation*}
 An illustration of a discrete system is provided in Fig. \ref{discrete_system}.

\paragraph{Irreversible internal heat transfer as a generalized friction.} Let us recall that from the first law and part $(a)$ of the second law, it is shown (see e.g. \cite{StSc1974}) that the internal heat power exchange can be described by
\begin{equation*}\label{heat_exchange_A_B} 
P_H^{B \rightarrow A}= \kappa _{AB}(q , S ^A , S ^B)(T^B-T^A),
\end{equation*} 
where $ \kappa _{AB}= \kappa _{BA}\geq 0$ is the heat transfer phenomenological coefficient. In order to include the irreversible effect of internal heat transfer in the variational formalism, it is useful to rewrite it in a form similar to the expression of the power associated to friction. Indeed, we can rewrite the total heat power supplied to $A$ as

\begin{equation}\label{friction_like} 
\sum_{B=1}^NP_H^{B \rightarrow A}=-\sum_{B=1}^N  J^{\rm fr(A)}_B(q , \dot q ,S ^A , S ^B)T^B,
\end{equation} 
where one may regard  $J^{\rm fr (A)}_B$ as a ``friction force" associated with $\boldsymbol{\Sigma} _{A}$ by setting
$$
J^{\rm fr (A)}_B:=-\left( \kappa _{AB} - \delta_{AB}\sum_{C=1}^N\kappa _{AC}\right). 
$$
This eventually verifies
\[
J^{\rm fr}_B:= \sum_{A=1}^NJ^{\rm fr (A)}_B=0.
\]
In view of the formula \eqref{friction_like}, one may interpret the temperatures $T^B$ as velocities. This may be achieved by introducing new variables $ \Gamma ^B$ such that $ \dot \Gamma ^B = T ^B$. \cite{GrNa1991} called such a variable $\Gamma$ a {\bfi thermal displacement}, which was originally developed by \cite{He1884}. As to its historical details, see the Appendix of \cite{Po2009}.
\medskip

Further, in order to include this variable into the variational formalism, we also need to introduce the additional variables $ \boldsymbol{\Sigma}  _A $, 
whose rate of change are associated to {\it total entropy production}. The complete physical meaning and roles of $ \boldsymbol{\Sigma}  _A $ will be clarified in the more general context of continuum systems later.

\subsubsection{Nonequilibrium thermodynamics of general discrete systems}

With this discussion in mind, the generalization of Definition \ref{LdA_def} for arbitrary discrete closed systems reads as follows.
\begin{definition}[\textbf{Variational formalism for nonequilibrium thermodynamics of discrete systems}]\label{LdA_def_discrete_Systems} Given the Lagrangian $L: TQ_ { \boldsymbol{\Sigma} }  \times \mathbb{R} ^N  \rightarrow \mathbb{R}$, the external, friction, and internal forces $F^{{\rm ext} \rightarrow A}, F^{{\rm fr} (A)}, F^{B \rightarrow A}:TQ_ { \boldsymbol{\Sigma} } \times \mathbb{R} ^N   \rightarrow T^*Q_ { \boldsymbol{\Sigma} } $, the external heat power $ P_H^{R \rightarrow A}$, and the generalized friction $J^{\rm fr (A)}_B$, the variational formalism for thermodynamics of discrete systems read
\begin{equation}\label{LdA_thermo_discrete_systems} 
\begin{split}
\delta \int_{ t _1 }^{ t _2 }\Big[ L(q  , \dot q , S _1 , ... S _N )+ \sum_{A=1}^N(S_A - \Sigma_A)\dot{\Gamma}^A \Big]   dt +&\int_{ t _1 }^{ t _2 } \left\langle F^{\rm ext}, \delta q \right\rangle dt =0,\quad \\& \textsc{Variational Condition} 
\end{split}
\end{equation}
where the curves $q(t), S _A (t), \Gamma ^A (t) , \Sigma _A (t) $ satisfy the constraint 
\begin{equation}\label{Kinematic_Constraints_Systems} 
\begin{split}
\frac{\partial L}{\partial S _A }  \dot{\Sigma}_A= \left\langle F^{{\rm fr}(A)}(...), \dot q  \right\rangle +\sum_{B=1}^N& J^{\rm fr (A)}_B(...) \dot \Gamma ^B -  P_H^{\rm ext}, \\\qquad &\qquad\textsc{Phenomenological Constraints} 
\end{split}
\end{equation} 
$($no sum on $A)$, and with respect to the variations $ \delta q , \delta S_A , \delta \Gamma^A  , \delta \Sigma_A$ subject to
\begin{equation}\label{Virtual_Constraints_Systems} 
\begin{split}
\frac{\partial L}{\partial S_A } \delta \Sigma _A
=  \left\langle F^{{\rm fr}(A)}(...), \delta q  \right\rangle+\sum_{B=1}^NJ&^{\rm fr (A)}_B(...) \delta \Gamma  ^B, \\& \;\;\textsc{Variational Constraints}
\end{split}
\end{equation}
with $\delta q( t _i )=0$ and $ \delta \Gamma ( t _i )=0$, for $i=1,2$.
\end{definition}

\paragraph{Equations of motion of a discrete system.} Taking the variations of the action integral and integrating by part, we have
\begin{equation}\label{vari_act_integ_part}
\begin{array}{l}
\displaystyle\int_{t _1 }^{t _2 } \left[ \left\langle \frac{\partial L}{\partial q } - \frac{d}{dt} \frac{\partial L}{\partial \dot q }+F^{\rm ext}, \delta q  \right\rangle 
+ \sum_{A=1}^N\left(\frac{\partial L}{\partial S_A }+\dot{\Gamma}^{A} \right)\delta S_A 
 \right.\\[5mm]
\left. \hspace{4cm}\displaystyle+
\sum_{A=1}^N (- \dot{\Gamma}^{A}\delta{\Sigma}_{A}) - \sum_{A=1}^N\delta \Gamma^{A}(\dot  S_A- \dot \Sigma _A) \right] dt=0.
\end{array} 
\end{equation} 
Using the variational constraint \eqref{Virtual_Constraints_Systems} and the fact that $ \frac{\partial L}{\partial S_A}\neq 0$, and collecting the terms associated to the variations $ \delta q $ and $ \delta \Gamma ^A $, $ \delta S_A$, one obtains
\begin{align} 
\delta q :&\;\;\frac{d}{dt} \frac{\partial L}{\partial \dot q }- \frac{\partial L}{\partial q }= F^{\rm ext}-\sum_{A=1}^N  \dot{\Gamma}^{A}\left(\frac{\partial L}{\partial S _A }\right)^{-1} F^{{\rm fr}(A)},\label{delta_zeta} \\
\delta \Gamma ^A :&\;\; \dot S_A=\dot \Sigma _A-\sum_{B=1}^N \dot{\Gamma}^{B}\left(\frac{\partial L}{\partial S_B } \right)^{-1} J_A^{{\rm fr}(B)}, \quad A=1,...,N\label{delta_gamma}\\
\delta S _A :&\;\; \dot\Gamma ^A=- \frac{\partial L}{\partial S_A}. \label{delta_SA} 
\end{align} 
Since $\sum_{B=1}^NJ_A^{{\rm fr}(B)}=0$, we can eventually deduce that $\dot S_A=\dot \Sigma _A$. 

\begin{remark}\rm
We will see later that the equality $\dot S_A=\dot \Sigma _A$ will no longer be true in the case of continuum systems. 
\end{remark}

The condition \eqref{delta_SA} gives the desired velocity interpretation of the temperature:
\[
T^A= -\frac{\partial L}{\partial S_A}=\dot \Gamma ^A.
\]
Using \eqref{delta_zeta}--\eqref{delta_SA}, and \eqref{Kinematic_Constraints_Systems}, we obtain the equations of evolution for the thermodynamics of a discrete system
\begin{equation}\label{thermo_mech_equations_discrete_systems}
\left\{ 
\begin{array}{l}
\displaystyle\vspace{0.2cm}\frac{d}{dt} \frac{\partial L}{\partial \dot q }- \frac{\partial L}{\partial q }=  
\sum_{A=1}^N  F^{{\rm fr}(A)}
+ F^{\rm ext},\\[2mm]
\displaystyle\frac{\partial L}{\partial S_A }\dot S_A = \left\langle F^{{\rm fr}(A)}(...)  , \dot q \right\rangle + \sum_{B=1}^NJ^{\rm fr (A)}_B(...) T ^B -P_H^{{\rm ext} \rightarrow A}, \quad A=1,...,N.\end{array} 
\right.
\end{equation} 
We recall that
\[
\sum_{B=1}^NJ^{\rm fr (A)}_B(...) T^B= -\sum_{B=1}^NP_H^{B \rightarrow A}= -\sum_{B=1}^N\kappa _{AB}(...)(T^B-T^A) \quad\text{and}\quad P_H^{{\rm ext} \rightarrow A}=\sum_{R}P_H^{R \rightarrow A}.
\]
These are the general evolution equations for the nonequilibrium thermodynamics of a discrete system with no matter transfer. This system generalizes the one obtained in \cite{Gr1997}, see equation (3.129), following \cite{StSc1974}. It is recovered by choosing $L( q , \dot q , S _1 , ..., S _N )=-\sum_{A=1}^N U^A ( q , S_A)$, $A=1,...,N$. 

\begin{remark}[Interconnection of discrete systems]\label{Remark_interconnection}{\rm As we already mentioned earlier, we have assumed that the given mechanical constraints 
due to the interconnections between the simple systems $ \boldsymbol{\Sigma} _A$ are holonomic and are already solved to produce the independent geometric variables $(q,\dot{q})$ in the current variational formulation. There are more general cases of discrete systems in which nonholonomic constraints due to the mechanical interconnection have to be considered. For purely mechanical systems this issue has been considered from a variational perspective in \cite{JaYo2014}. The extension of this approach to the setting of nonequilibrium thermodynamics will be explored in a future work.
}
\end{remark}  
\paragraph{Energy balance and entropy production.} From \eqref{thermo_mech_equations_discrete_systems}, we obtain the energy balance
\[
\frac{d}{dt} E= P_W^{\rm ext}+P_H^{\rm ext},
\]
consistently with the first law, where $E( q, \dot q, S_{1}, ..., S_{N})$ is the energy associated to the Lagrangian $L( q, \dot q, S_{1}, ..., S_{N})$.

Being an {\it extensive} quantity, the {\bfi total entropy} of the system is given by $S=\sum_{A=1}^NS_A$. We thus obtain
\begin{equation}\label{S_dot} 
\frac{d}{dt} S=-\sum_{A=1}^N\frac{1}{T^A}\left\langle F^{{\rm fr}(A)} (...) , \dot q \right\rangle +\sum_{A=1}^N\sum_{B\ne A}^N\frac{1}{T^A}P_H^{B \rightarrow A} +\sum_{A=1}^N\frac{1}{T^A}P_H^{{\rm ext} \rightarrow A}.
\end{equation} 
It follows from part $(a)$ of the second law that the first two terms on the right hand side must be positive.

\begin{remark}[Internal entropy production]{\rm Note that the {\it internal entropy production} for system $ \Sigma ^A$ is
\[
I^A=-\frac{1}{T^A}\left\langle F^{{\rm fr}(A)} (...) , \dot q \right\rangle ,
\]
and the internal entropy production for the {\it interconnected system} $ \Sigma $ is given by
\[
I=\sum_{A=1}^N I^A+ \frac{1}{2} \sum_{A,B=1}^NI^{AB}=\sum_{A=1}^N I^A+ \frac{1}{2} \sum_{A=1}^N\sum_{B\ne A}^N \kappa_{AB} (q , S _A , S _B)\frac{(T^B-T^A) ^2 }{T^AT^B} \neq \sum_{A=1}^NI ^A,
\]
where $I^{AB}=P_H^{B \rightarrow A} \left(T ^B - T ^A \right)=\frac{1}{2}\kappa_{AB} (q , S _A , S _B)\frac{(T^B-T^A) ^2 }{T^AT^B}$.
In this case the entropy production is due to both friction forces $F^{\rm fr(A)}$ on each subsystem as well as to heat transfer between the simple subsystems.
}
\end{remark} 

\begin{remark}[External heat power]{\rm Concerning the heat power due to external heat source, we also have $P_H^{R \rightarrow A}= \kappa _{AR}(q , S _A , S _R) \left(T ^R - T ^A \right)$, where we recall that $T^R= \frac{\partial U^R}{\partial S_R}$ is the temperature of the heat source $ \Sigma ^R $. From this we deduce that $\frac{1}{T^A} P_H^{R \rightarrow A}\geq \frac{1}{T^R} P_H^{R \rightarrow A}$ and, therefore, it follows from \eqref{S_dot} that
\begin{equation}\label{important_inequality} 
\frac{d}{dt} S \geq  \sum_{A=1}^N\frac{1}{T^A}P_H^{{\rm ext} \rightarrow A}=\sum_{A=1}^N\sum_R\frac{1}{T^A}P_H^{R \rightarrow A} \geq  \sum_{A=1}^N\sum_R \frac{1}{T^R} P_H^{R \rightarrow A}=\sum_R\frac{1}{T^R}P_H^{R \rightarrow \Sigma }.
\end{equation} 
This property relates the entropy variation of the whole system to a property of the exterior of the system, namely, the heat sources $ \boldsymbol{\Sigma}  ^R $ (\cite{Gr1997}).}
\end{remark} 

\begin{remark}[Reversibility]\normalfont
Note that we have $\sum_R\dot S_R=-\sum_R\frac{1}{T^R}P_H^{R \rightarrow  \boldsymbol{\Sigma}  }$. So, by the definition of reversibility and assuming that there is no entropy associated to the external forces, the evolution of this system is reversible if and only if we have equalities in \eqref{important_inequality}. Therefore, by \eqref{important_inequality} and \eqref{S_dot}, the evolution is reversible if and only if
\[
T^1=...=T^N=T^R, \;\text{for all $R$} \qquad\text{and}\qquad F^{{\rm fr} (A)} =0, \; \text{for all $A$.} 
\]
\end{remark} 

\paragraph{Structures of the variational formalism.} For both discrete and continuum systems (the continuum case will be shown later in \S\ref{VHCF}), our variational formalism has the following structure:
\begin{itemize}
\item
Our variational formalism of nonequilibrium thermodynamics is a generalization of the Hamilton variational principle of classical mechanics to the case in which irreversible effects associated with thermodynamics are included.
\item
The nonholonomic constraint is given by the entropy production of the system. In general, it consists of a sum of terms, each of them being the product of a thermodynamic affinity $X^{\alpha}$ and a thermodynamic flux $J_{\alpha}$ characterizing an irreversible process, the relation between them being given by phenomenological laws. It happens that in our formalism, we are able to attribute to each of the irreversible process a rate $\dot {\Lambda}  ^ \alpha  $ such that $\dot \Lambda  ^ \alpha  =X^{\alpha}$ (e.g. here $\dot \Gamma ^A $ identified with the temperature $T ^A$). It follows that the entropy production appears as a sum of internal (mechanical and thermodynamic) power due to generalized frictions (e.g. $\langle F^{\rm fr (A)},\dot q \rangle $ and $ \langle J^{\rm fr (A)}_B, \dot \Gamma ^B \rangle $). The same idea will be developed in the case of continuum systems in \S\ref{continuum_section}.
\item
The variational constraint is obtained by formally replacing all the rate variables $\dot {\Lambda } ^ \alpha $, or, in thermodynamic language, the  thermodynamic affinities $X^{\alpha}=\dot {\Lambda }^ \alpha $, by the corresponding virtual displacements $\delta { \Lambda } ^ \alpha $. In addition, in passing from the phenomenological constraint to the variational constraint, the effects from the exterior are removed.
\end{itemize}

\begin{remark}\rm
The reader will observe that the same equations \eqref{thermo_mech_equations_discrete_systems} can be also obtained with a more simpler variational formalism, namely, one of the type of Definition \ref{LdA_def} given as follows:
\begin{equation}\label{LdA_thermo_not_physical} 
\delta \int_{ t _1 }^{ t _2 } L(q  , \dot q , S _1 , ... S _N )  dt +\int_{ t _1 }^{ t _2 }\left\langle F^{\rm ext}, \delta q \right\rangle dt =0,
\end{equation}
where the curve $q(t), S _A (t)$ satisfy the constraint
\begin{equation}\label{Kinematic_Constraints_Systems_not_physical} 
\frac{\partial L}{\partial S _A}  \dot S_A = \left\langle F^{{\rm fr}(A)}(...), \dot q  \right\rangle +\sum_{B=1}^NJ^{\rm fr (A)}_B(...) \dot \Gamma ^B - P_H^{{\rm ext} \rightarrow A},
\end{equation} 
$($no sum on $A)$ and with respect to variations $ \delta q , \delta S_A $ subject to
\begin{equation}\label{Virtual_Constraints_Systems_not_physical} 
\frac{\partial L}{\partial S _A}  \delta  S_A  = \left\langle F^{{\rm fr}(A)}, \delta q  \right\rangle.
\end{equation}
This variational formalism is however less consistent from the physical point of view, since it interprets the whole expression $\sum_{B=1}^NJ^{\rm fr (A)}_B(...) \dot \Gamma ^B - P_H^{\rm ext}$ as an external power, appearing only in the phenomenological constraint, and not present in the variational constraint. This is not consistent with the fact that $\sum_{B=1}^NJ^{\rm fr (A)}_B(...) \dot \Gamma ^B$ is associated to an {\it internal} irreversible process in the same way as $\left\langle F^{{\rm fr}(A)}(...), \dot q  \right\rangle$ and therefore it has to appear in both the phenomenological and variational constraints. In the continuum situation, it will be clear that the variational formalism given in Definition \ref{LdA_def_discrete_Systems} (as opposed to one of the type \eqref{LdA_thermo_not_physical}--\eqref{Virtual_Constraints_Systems_not_physical}) is the one that is appropriate when internal heat transfer occurs.
\end{remark}

\subsubsection{Formalism based on the free energy}\label{section_free_energy}

Let us consider a variational formalism in terms of the free energy, i.e., when the temperature, as opposed to the entropy, is chosen as a primitive variable. Recall that the {\it Helmholtz free energy} $\mathcal{F}$ is defined in terms of the internal energy by the Legendre transform
\[
\mathcal{F}(q,T):= U(q,S(q,T))-TS(q,T),
\]
where $S(q,T)$ is such that $ \frac{\partial U}{\partial S}(q,S)=T$, for all $q,S$, and where, for simplicity, we do not explicitly write the dependence on the variables $V$ and $N$.  

In order to treat the nonequilibrium case, we now employ this partial Legendre transform to obtain a free energy  Lagrangian. Namely, given a Lagrangian $L(q, \dot q, S)$, we define the {\bfi free energy Lagrangian} $ \mathcal{L}$ by
\[
\mathcal{L}(q,\dot q,T):= L(q,\dot q, S(q, T))+TS(q, T),
\]
where we assumed that $S$ does not depend on $\dot q$, as it is the case in most physical examples.
For example, in the particular case of a Lagrangian of the form  \eqref{special_form}, i.e., $L(q  , \dot q  , S):=L_{\rm mec}( q  , \dot q )-U(q , S)$, we obtain
\begin{equation}\label{special_form_free} 
\mathcal{L}(q,\dot q,T)= L_{\rm mec}( q  , \dot q )-\mathcal{F}(q , T).
\end{equation}

\begin{definition}[\textbf{Variational formalism for discrete systems based on the free energy}]\label{LdA_def_discrete_Systems_free} Let $\mathcal{L}: TQ_ { \boldsymbol{\Sigma} }  \times \mathbb{R} ^N  \rightarrow \mathbb{R}$ be a free energy Lagrangian. Let $F^{{\rm ext} \rightarrow A}, F^{{\rm fr} (A)}, F^{B \rightarrow A}:TQ_ {\boldsymbol{\Sigma} } \times \mathbb{R} ^N   \rightarrow T^*Q_ {\boldsymbol{\Sigma} }$ be external, friction, and internal forces respectively. Let $J^{\rm fr (A)}:TQ_ { \boldsymbol{\Sigma} } \times \mathbb{R} ^N   \rightarrow (\mathbb{R} ^N)^{\ast}$ be generalized friction forces. Given an external heat power $ P_H^{R \rightarrow A}$, the variational formalism reads
\begin{equation}\label{LdA_thermo_discrete_systems-new}
\begin{aligned} 
\delta \int_{ t _1 }^{ t _2 }\Big( \mathcal{L}(q  , \dot q , T^{1},...,T^{N} )&- \sum_{A=1}^N S_AT^A+\sum_{A=1}^N(S_{A}-\Sigma_A)\dot{\Gamma}^A  \Big)   dt \\
&+\int_{ t _1 }^{ t _2 }\left\langle F^{\rm ext}, \delta q \right\rangle dt =0,\quad \textsc{Variational Condition} 
\end{aligned}
\end{equation}
where the curves $q(t), S _A (t), T^{A}(t), \Gamma ^A (t) , \Sigma _A (t) $ satisfy the constraint
\begin{equation}\label{Kinematic_Constraints_Systems-new} 
\begin{split}
T^A \dot{\Sigma}_A = -\left\langle F^{{\rm fr}(A)}(...), \dot q  \right\rangle -\sum_{B=1}^NJ^{\rm fr (A)}_B(...) \dot \Gamma ^B + P_H^{{\rm ext} \rightarrow A}, \hspace{2.2cm} \\\textsc{\hspace{6.3cm} Phenomenological Constraints}
\end{split}
\end{equation} 
$($no sum on $A)$, and with respect to the variations $ \delta q , \delta S_A , \delta \Gamma^A  , \delta \Sigma_A$ subject to
\begin{equation}\label{Virtual_Constraints_Systems-new} 
\begin{split}
T ^A  \delta \Sigma _A  = - \left\langle F^{{\rm fr}(A)}(...), \delta q  \right\rangle - \sum_{B=1}^NJ^{\rm fr (A)}_B(...) \delta \Gamma  ^B,\qquad 
\hspace{2cm}  \\\textsc{\hspace{7.5cm} Variational Constraints}
\end{split}
\end{equation}
with $ \delta q( t _i )=0$ and $ \delta \Gamma^A ( t _i )=0$ for $i=1,2$. 
\end{definition}

\paragraph{Equations of motion of a discrete system.} Taking the variations of the action integral and integrating by part, we have
\begin{equation*}\label{alt-vari-discrete}
\begin{array}{l}
\displaystyle\int_{t _1 }^{t _2 } \left[ \left\langle \frac{\partial \mathcal{L}}{\partial q } - \frac{d}{dt} \frac{\partial \mathcal{L}}{\partial \dot q }+F^{\rm ext}, \delta q  \right\rangle +
\sum_{A=1}^N \delta T^A \left( \frac{\partial \mathcal{L}}{\partial T^{A}}-S_{A} \right) 
 \right.\\[4mm]
\left. \hspace{2cm}\displaystyle+
\sum_{A=1}^N \delta S _A (-T^{A} + \dot{\Gamma}^{A} )- \sum_{A=1}^N\delta \Gamma^{A}(\dot  S_A- \dot \Sigma _A)- \dot{\Gamma}^{A}\delta \Sigma_{A} \right] dt=0.
\end{array} 
\end{equation*} 
Using the variational constraint, written in terms of $ \delta \Sigma _A $ and using $ T^A\neq 0$, we obtain
\begin{align}
\delta T^{A}:&\;\;S_A=\frac{\partial \mathcal{L}}{\partial T^{A}}, \quad A=1,...,N, \label{al_delta_TA} \\[1mm]
\delta S_A:&\;\; \dot \Gamma^{A}=T^{A}, \quad A=1,...,N. \label{al_delta_SA}\\
\delta q :&\;\;\frac{d}{dt} \frac{\partial \mathcal{L}}{\partial \dot q }- \frac{\partial \mathcal{L}}{\partial q }=\sum_{A=1}^N  \dot{\Gamma}^{A}(T^A)^{-1}  F^{{\rm fr}(A)}+F^{\rm ext},\label{al_delta_zeta} \\
\delta \Gamma ^A :&\;\; \dot S_A= \dot \Sigma_A+ \sum_{B=1}^N\dot \Gamma ^B (T^B)^{-1} J_A^{{\rm fr}(B)}, \quad A=1,...,N\label{al_delta_gamma}.
\end{align}
Since $\sum_{B=1}^NJ_A^{{\rm fr}(B)}=0$, we obtain $\dot S_A=\dot \Sigma _A$.

Note that the condition \eqref{al_delta_TA} consistently defines the entropy as
\[
S_A=\frac{\partial \mathcal{L}}{\partial T^{A}} \qquad \text{i.e.,}\quad S_A= - \frac{\partial \mathcal{F} }{\partial T ^A }\quad \text{if $\mathcal{L} $ is of the special form \eqref{special_form_free}}.
\]
Note also that the condition \eqref{al_delta_SA} provides the desired velocity interpretation of the temperature $T^A=\dot \Gamma ^A$.

Using \eqref{al_delta_SA}--\eqref{al_delta_gamma}, and \eqref{Kinematic_Constraints_Systems-new} we obtain the equations of evolution for the thermodynamics of a discrete system
\begin{equation*}\label{al_thermo_mech_equations_discrete_systems}
\left\{ 
\begin{array}{l}
\displaystyle\vspace{0.1cm}\frac{d}{dt} \frac{\partial \mathcal{L}}{\partial \dot q }- \frac{\partial \mathcal{L}}{\partial q }= \sum_{A=1}^N F^{{\rm fr}(A)}+F^{\rm ext},\\
\displaystyle T^{A}\dot S_A =- \left\langle F^{{\rm fr}(A)}(...)  , \dot q \right\rangle - \sum_{B=1}^N J^{\rm fr (A)}_B(...)  T^{B}+P_H^{{\rm ext} \rightarrow A}, \quad A=1,...,N,\\
\displaystyle S_A=\frac{\partial \mathcal{L}}{\partial T^{A}},
\end{array} 
\right.
\end{equation*} 
which are equivalent to \eqref{thermo_mech_equations_discrete_systems}.

\begin{remark}{\rm
In this paper, we assume that the internal energy, the Helmholtz free energy and hence the associated Lagrangians are {\it not explicitly} dependent on the thermal displacement $\Gamma$. In general, the Lagrangian for nonequilibrium thermodynamics can be understood as a function of variables including explicitly {\it thermodynamic displacements} such as $ W^I$, $ \nu ^a$ as well as $\Gamma$. This general setting will be interesting from the geometric viewpoint of nonequilibrium thermodynamics. We will explore in details this direction in a future work.}
\end{remark}
 
\subsubsection{The connected piston-cylinder problem}\label{section-two-piston}
\begin{figure}[h]
\begin{center}
\hspace{2cm}
\includegraphics[scale=0.75]{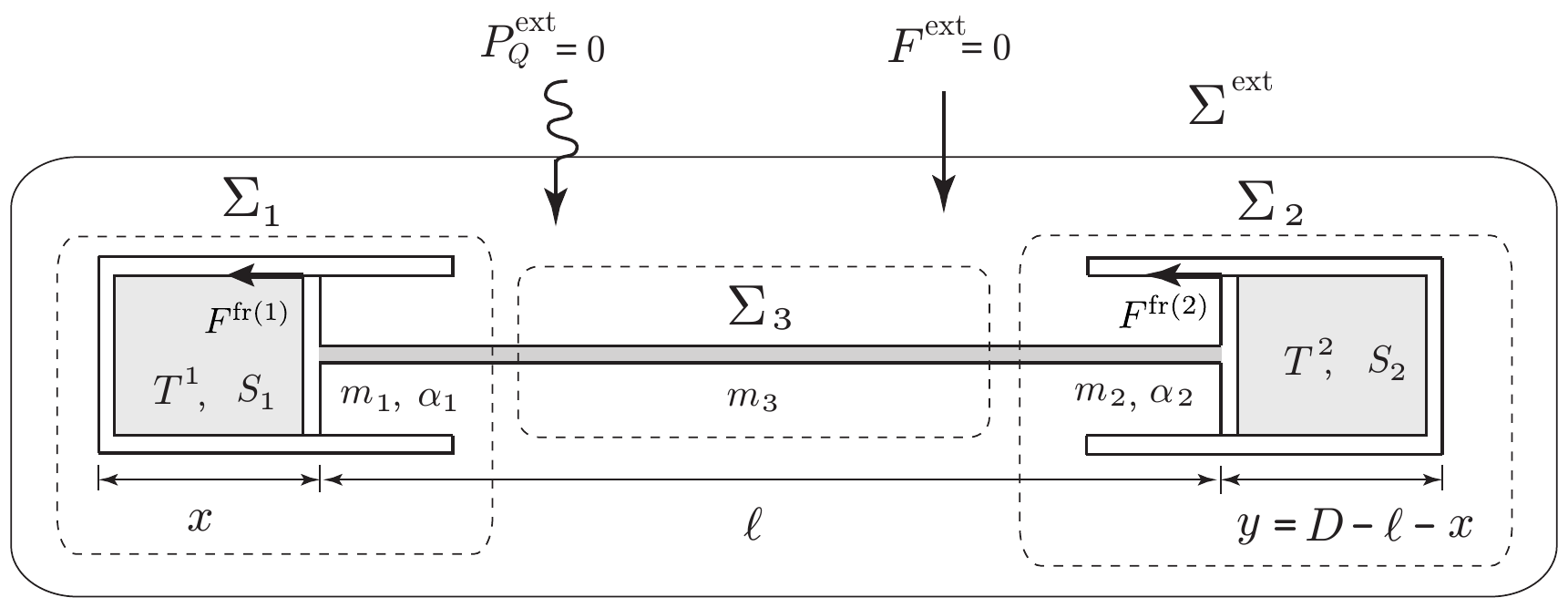}
\caption{The two-cylinder problem}
\label{two_pistons}
\end{center}
\end{figure}

We consider a piston-cylinder system composed of two connected cylinders which contain two fluids separated by an adiabatic or (internally) diathermic piston, as shown in Fig. \ref{two_pistons}. We assume that the system is isolated. Despite its apparent simplicity, this system has attracted a lot of attention in the literature because there has been some controversy about the final equilibrium state of this system. We refer to \cite{Gr1999} for a review of this challenging problem and for the derivation of the time evolution of this system, based on the approach of \cite{StSc1974}. 
The system $ \boldsymbol{\Sigma} $ may be regarded as an interconnected system consisting of three simple systems; namely, the two pistons $ \boldsymbol{\Sigma} _1 , \boldsymbol{\Sigma} _2 $ of mass $ m_1 , m _2 $ and the connecting rod $ \boldsymbol{\Sigma} _3 $ of mass $ m _3 $. As illustrated in Fig.\ref{two_pistons}, $x$ and $y=D-l-x$ denote respectively the distance between the bottom of each piston to the top, where $D$ is a constant.
In this setting, we choose the variables $(x, \dot x, S_1, S_2)$ (there is no entropy associated to $ \boldsymbol{\Sigma} _3 $) to describe the dynamics of the interconnected system and the Lagrangian is given by
\begin{equation}\label{Lagrangian_2_pistons}
L(x, \dot x, S _1, S _2  )= \frac{1}{2} M\dot x ^2 -U_1 (x, S _1 )-U_2(x, S _2 ),
\end{equation} 
where $M:= m_1  + m _2 + m _3$. The equations of evolution can be obtained either by the formalism of Definition \ref{LdA_def_discrete_Systems} based on the entropy or by the formalism of Definition \ref{LdA_def_discrete_Systems_free} based on the temperature. We shall choose the latter, and hence we consider the free energy Lagrangian associated to \eqref{Lagrangian_2_pistons}, given by
\[
\mathcal{L}(x, \dot x, T^{1}, T^{2})= \frac{1}{2} M\dot x ^2 -\mathcal{F}_1 (x, T^{1})-\mathcal{F}_2(x, T^{2}),
\] 
where the Helmholtz free energies are given by $\mathcal{F}_1 (x, T^{1} )=U_1 (x, S _1 )-T^{1}S_{1}$ and $\mathcal{F}_2(x, T^{2})=U_2(x, S _2 )-T^{2}S_{2}$. Since the system is isolated, $F^{\rm ext \rightarrow A}=0$ and $P^{\rm ext \rightarrow A}_H =0$, for all $A=1,2,3$. We also have $P_H^{A \rightarrow 3}=0$, $F^{{\rm fr} (A)}(x, \dot x,  T^A )= - \lambda _A (x,T^{A} )\dot x$, for all $A=1,2$, $F^{1 \rightarrow 2}=0$, and $P^{2 \rightarrow 1}_H = \kappa (x, T^{1}, T^{2} )(T ^2 - T ^1 )$, where $ \kappa $ is the heat conductivity. Using $\frac{\partial \mathcal{F}_1 }{\partial x}=-p _1 (x, T^{1})\alpha_1 $ and $ \frac{\partial \mathcal{F}_2}{\partial x}=p _2(x, T^{2}) \alpha_2 $, where $ p _1 , p _2 $ are the pressures of the fluids and $\alpha_1, \alpha_2$ are the areas of the cylinders, equations \eqref{thermo_mech_equations_discrete_systems} yield the time evolution equations of the system as
\begin{equation*}\label{2_pistons_equ}
\left\{ 
\begin{array}{l}
\displaystyle\vspace{0.2cm}M\ddot x= p _1(x, T^{1} ) \alpha_1 - p _2(x, T^{2}) \alpha_2 - (\lambda _1 + \lambda _2 )\dot x,\\[2mm]
\displaystyle T^1\dot S_1= \lambda _1 \dot x^2  +\kappa(x, T^{1},T^{2}) \left( T ^2-T^1\right),\\[4mm]
\displaystyle  T^2\dot S_2= \lambda _2 \dot x^2  +\kappa(x,  T^{1},T^{2}) \left(T^1-T^2\right),\\[2mm]
\displaystyle  S_{1}=-\frac{\partial \mathcal{F}_1}{\partial T^{1}}, \quad S_{2}=-\frac{\partial \mathcal{F}_2}{\partial T^{2}},
\end{array} 
\right.
\end{equation*}
where $M:= m_1  + m _2 + m _3 $. 
These time evolution equations recover the equations of motion for the diathermic piston derived in \cite{Gr1999}, (51)--(53). We have $ \frac{d}{dt} E=0$, where $E= \frac{1}{2} M\dot x^2 +U_1(x,S_1)+U(x, S_2 )$, consistently with the fact that the system is isolated. The rate of entropy production is
\[
\frac{d}{dt} S= \left( \frac{\lambda _1}{T ^1} + \frac{\lambda _2}{T ^2}  \right) \dot x ^2 + \kappa (x, T^{1},T^{2})\frac{(T ^2 - T ^1 ) ^2 }{T ^1 T ^2 } .
\]
The equations of motion for the adiabatic piston are obtained by setting $ \kappa =0$.

\subsubsection{Thermodynamics of interconnected electric circuits}

We explored a simple electric circuit with thermodynamics in \S\ref{subsec_simple_EC}. In this section, we shall apply our Lagrangian formalism to a case of interconnected electric circuits. Namely, we will show a generalization of the variational formalism in \S\ref{subsec_simple_EC} to the case of several entropy variables and internal heat transfer.

\begin{figure}[h]
\begin{center}
\includegraphics[scale=0.6]{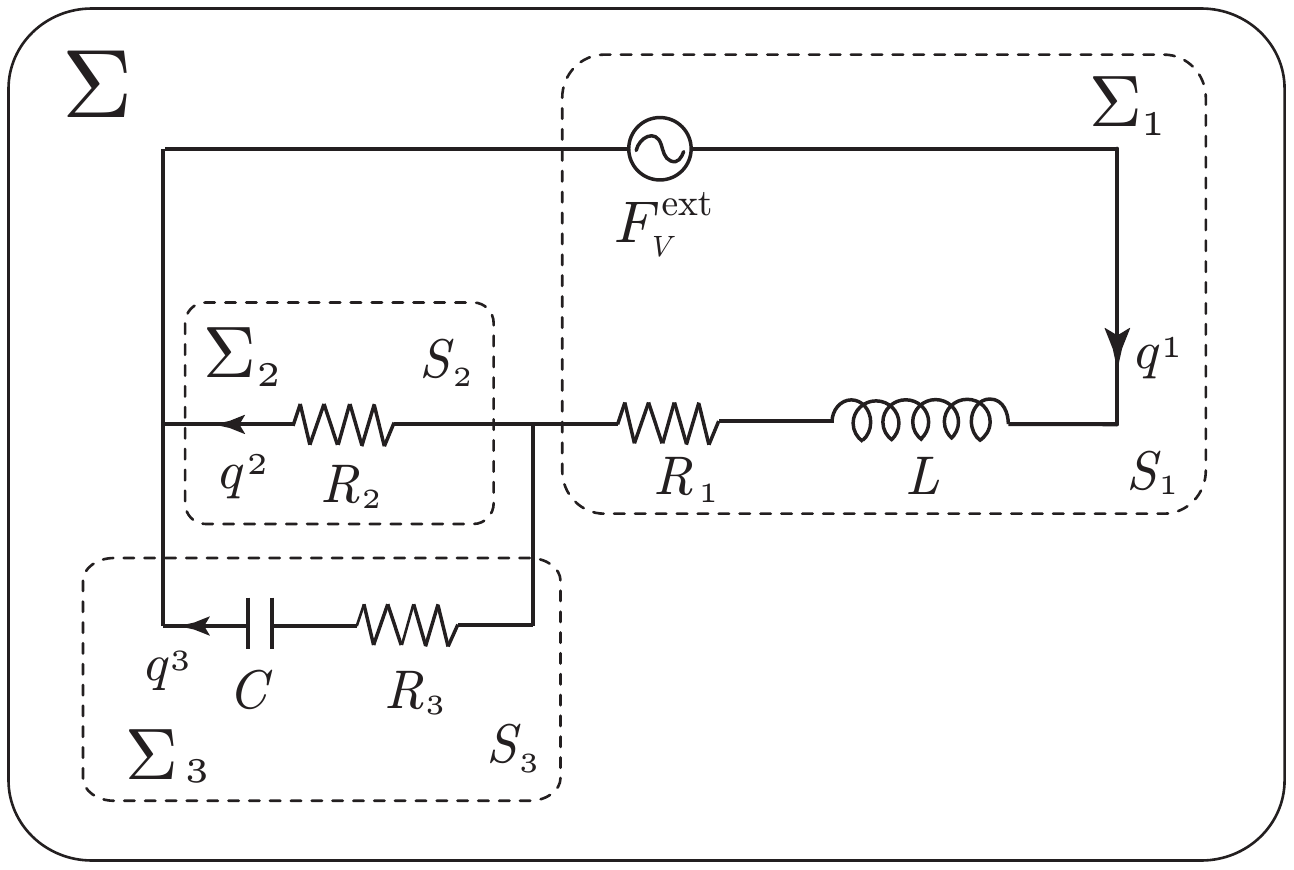}
\caption{An interconnected electric circuit}
\label{EC_RCLS}
\end{center}
\end{figure}

Consider the interconnected electric circuit $\boldsymbol{\Sigma}=\cup_{A=1}^{3}  \boldsymbol{\Sigma} _A$ illustrated in Fig. \ref{EC_RCLS}. We assume that there is no heat transfer to the exterior, namely, $P_{Q}^{\rm ext}=0$, but there is heat transfer between the simple systems $\boldsymbol{\Sigma}_{A}$, $A=1,2,3$. Further, we assume that $P^{\rm ext}_{M}=P^{\rm ext}_{W}=0$, i.e., the system is isolated. Following the general approach to discrete systems described above, since  
the system $\boldsymbol{\Sigma}$ is decomposed into the simple systems $ \boldsymbol{\Sigma} _A $, we can appropriately choose charge and current variables $ q , \dot q $ through Kirchhoff's current to describe $ \Sigma $ and the entropy variables $ S _A $ associated to each of the simple systems $ \boldsymbol{\Sigma} _A $. As before, we shall assume (possibly) nonlinear constitutive equations for circuit elements, namely, $ \varphi = \varphi _L (I)$, $V=V_C(q)$, $V=V_{R_i}(I)$, $i=1,2,3$. 
Let us denote by $\dot  q ^1 , \dot q^2 , \dot q ^3$ the currents as indicated on the figure. Since we have the KCL constraint $\dot q^1 = \dot q^2 + \dot q^3 $, we can choose the charge variables $ q =(q^2 , q ^3 )$ and write $q^1(t)=q^2(t)+q^3(t)+q^0$, where $q^{0}$ denotes a constant charge. We denote by $ \kappa _{AB}(...)$ the heat transfer coefficients between each of the systems. We thus have the following expressions
\begin{align*} 
\Sigma ^1 : &\;\; L_1 ( q^2 , q ^3 ,\dot q^2, \dot q ^3 , S _1 )= K_L(\dot q ^2 + \dot q ^3 )-U_1 (S _1 ), && F^{{\rm fr}(1)}(...)= V_{R _1}(\dot q ^2 + \dot q^3)(dq ^2 + d q^3 ),\\
&\; &&F^{{\rm ext} \rightarrow 1}(...)=V^{\rm ext} ( dq ^2 + d q ^3 ),\\
\Sigma ^2 : &\;\; L_2 ( q ^2 , q ^3 ,\dot q ^2, \dot q ^3 , S _1 )= -U_2 (S _2), &&F^{{\rm fr}(2)}(...)= V_{R _2}(\dot q ^2)dq ^2,\\
\Sigma ^3 : &\;\; L_3 ( q ^2 , q ^3 ,\dot q ^2, \dot q ^3 , S _1 )= -U_C( q ^3 )-U_3 (S _3), &&F^{{\rm fr}(3)}(...)= V_{R _3}(\dot q ^3)dq ^3.
\end{align*} 
Applying the variational formalism in Definition \ref{LdA_def_discrete_Systems}, together with $L=L _1 + L _2 + L _3 $, yields the evolution equations
\begin{equation*}\label{real_circuit}
\begin{aligned}
&\frac{d}{dt} \varphi _L (\dot q ^2 +\dot q ^3 )=V^{\rm ext}+V_{R_1}(\dot q ^2 +\dot q ^3 )+V_{R_2}(\dot q ^2 ),\\
&\frac{d}{dt} \varphi _L (\dot q ^2 +\dot q ^3 )+V_C(q ^3 )=V^{\rm ext}+V_{R_1}(\dot q ^2 +\dot q ^3 )+V_{R_3}(\dot q ^3),\\
&\dot S_1 = - \frac{1}{T ^1 }V_{R_1}(\dot q ^2 +\dot q ^3 )(\dot q _2 +\dot q ^3 )+ \frac{1}{T ^1 } \kappa _{12} (T ^2 - T ^1 )+\frac{1}{T ^1 } \kappa _{13} (T ^3 - T ^1 ),\\
&\dot S_2 = - \frac{1}{T ^2}V_{R_2}(\dot q ^2 )\dot q ^2+ \frac{1}{T ^2 } \kappa _{21} (T ^1 - T ^2 )+\frac{1}{T ^2 } \kappa _{23} (T ^3 - T ^2 ),\\
&\dot S_3 = - \frac{1}{T ^3}V_{R_3}(\dot q ^3 )\dot q ^3+ \frac{1}{T ^3 } \kappa _{31} (T ^1 - T ^3 )+\frac{1}{T ^3 } \kappa _{32} (T ^2 - T ^3 ).
\end{aligned}
\end{equation*} 
One easily checks that $ \frac{d}{dt} E=0$, consistently with the fact that the system is isolated.
The total entropy production reads
\[
\dot S= - \frac{1}{T ^1 }V_{R_1}(\dot q ^2 +\dot q ^3 )(\dot q ^2 +\dot q ^3 )- \frac{1}{T ^2}V_{R_2}(\dot q ^2 )\dot q ^2- \frac{1}{T ^3}V_{R_3}(\dot q ^3 )\dot q ^3+\sum_{A<B} \kappa _{AB} \frac{(T ^B - T ^A ) ^2 }{T ^A T ^B } ,
\]
where $V_{R_A}( \dot q^{i}) \dot q^{i}\leq 0$, for all $\dot q^{i}$, $A=1,2,3$ and $i=1,2,3$. 

\begin{remark}{\rm Note that in our approach we first imposed the KCL constraint $\dot q ^1 = \dot q ^2 + \dot q ^3 $ (which is holonomic) to eliminate the variable $ q ^1 $ and then we applied the variational formalism \eqref{LdA_thermo_discrete_systems} with variables $(q ^2, q ^3 , S_1 ,S_2 , S_3 )$.

It is also possible to derive the same equations by keeping the dependent variables $q ^1 , q ^2 , q^3 $ and treating the holonomic constraint $\dot q ^1 = \dot q ^2 + \dot q ^3 $ as an additional constraint besides the nonholonomic constraints \eqref{Kinematic_Constraints_Systems} associated to entropy production. The equations are obtained by using the formalism \eqref{LdA_thermo_discrete_systems} with the variational constraints $ \delta q ^1 = \delta q^2 + \delta q ^3 $ and \eqref{Virtual_Constraints_Systems}. This approach will be pursued in a more general setting in a future work on the geometric description for the interconnection of thermodynamic systems.
}
\end{remark}

\section{Nonequilibrium thermodynamics of continuum systems}\label{continuum_section} 

In this section, we adapt the variational formalism of Definition \ref{LdA_def_discrete_Systems} to the case of continuum systems. By working first with the material representation of continuum mechanics, we show that the same variational formalism, now formulated in an infinite dimensional geometric setting (manifolds of embeddings or diffeomorphism groups), yields the evolution equation for the nonequilibrium thermodynamics of continuum mechanics. When written in the spatial representation, this variational formalism becomes more involved, but is naturally explained by a reduction process that mimics the Lagrangian reduction processes used in reversible continuum mechanics, and takes into account of the nonholonomic constraint.

\subsection{Preliminaries on variational formalisms for continuum mechanics}

In this section we recall some required ingredients concerning the geometry and kinematics of continuum systems such as fluids and elastic materials. We follow the notations and conventions of \cite{MaHu1983}, to which we refer for a detailed treatment.

\subsubsection{Geometric preliminaries}

Let $ \mathcal{B} \subset \mathbb{R}^3 $ be the reference configuration of the system, assumed to be the  closure of an open subset of $ \mathbb{R}  ^3 $ with piecewise smooth boundary.
A configuration is a smooth embedding $ \varphi : \mathcal{B} \rightarrow \mathcal{S} $ of $ \mathcal{B} $ into a three dimensional manifold $ \mathcal{S} $. We will write $x= \varphi (X)$, where $x=(x ^1 , x ^2 , x ^3 )$ and are called spatial points and $X=(X ^1 , X ^2 , X ^3 )$ are material points. The manifold of all {\it smooth embeddings} is denoted by $ \operatorname{Emb}( \mathcal{B} , \mathcal{S} )$.

For problems with fixed boundaries, we can choose $ \mathcal{S}= \mathcal{B} $, so that the configuration is a smooth diffeomorphism of $ \mathcal{B} $, whereas for free boundary problems, we have $ \mathcal{S} = \mathbb{R}  ^3$ and in this case the configuration $\varphi $ is a diffeomorphism onto the current configuration $ \varphi ( \mathcal{B} ) \subset \mathbb{R}  ^3 $.
We will denote by $ \operatorname{Diff}( \mathcal{B} )= \operatorname{Emb}( \mathcal{B} , \mathcal{B} )$ the group of all smooth diffeomorphisms of $ \mathcal{B} $.

Even if in this paper, we will restrict to fixed boundary problems, we still consider the general situation $ \varphi : \mathcal{B} \rightarrow \mathcal{S} $ in this preliminary section, as a notational way to make a clear distinction between material and spatial quantities. This is a crucial point for the rest of the paper.

Let us denote by $A,B,...$ local coordinates on $ \mathcal{B} $ and by $ a, b,...$ local coordinates on $ \mathcal{S} $. The deformation gradient $F$ is the tangent map (derivative) of the configuration $ \varphi $, that is $F(X):=T_X \varphi :T_X \mathcal{B} \rightarrow T_x \mathcal{S} $. In coordinates, we denote it by $F^a_A$. The deformation gradient can be interpreted as a $(1,1)$ two-point tensor over $ \varphi $, since we can write $F(X):T_X \mathcal{B} \times T_x ^\ast \mathcal{S} \rightarrow \mathbb{R}  $, where $x = \varphi (X)$. By a $(p,q)$ tensor field we mean a $p$ times contravariant and $q$ times covariant tensor field.

We endow $\mathcal{B} $ with a Riemannian metric $G$ and $ \mathcal{S} $ with a Riemannian metric $g$. We will denote by $\mu _G$ and $\mu _g $ the associated Riemannian volume forms. Recall that the Jacobian of $ \varphi $ relative to $ \mu _G$ and $ \mu _g $ is the function $J _\varphi $ on $ \mathcal{B} $ defined by $ \varphi ^\ast \mu _g = J_\varphi\mu _G $.

The Levi-Civita covariant derivative relative to $G$ and $g$ will both be denoted by $ \nabla  $. If $\mathbf{T} $ and $\mathbf{t} $ are tensor fields on $ \mathcal{B} $ and $ \mathcal{S} $, respectively, with local coordinates $T^{A...C}_{E...G}$ and $t^{a...c}_{e...g}$, then the covariant derivatives $ \nabla \mathbf{T} $ and $ \nabla \mathbf{t} $ will be denoted locally by $T_{E...G|K}^{A...D}$ and $t^{a...c}_{e...g|k}$. The divergence operator on $ \mathcal{B} $ and $ \mathcal{S} $, will be denoted by $ \operatorname{DIV}$ and $ \operatorname{div}$, respectively. They are obtained by contraction of the last covariant and contravariant indices of the covariant derivative. For example, if $\mathbf{T}$ and $\mathbf{t}$ are $(p,q)$ tensor fields, then $ \operatorname{DIV}\mathbf{T} $ and $ \operatorname{div}\mathbf{t} $ are the $(p-1,q)$ tensor fields given, in local coordinates, by $ (\operatorname{DIV}T)^{A...C}_{E...G}=T^{A...C}_{E...G|C}$ and $(\operatorname{div}t)^{a...c}_{e...g}=t^{a...c}_{e...g|c}$. We will occasionally use the following instance of Stokes' theorem. Let $S$ be a $(2,0)$ tensor field and $ \alpha$ a one-form on $ \mathcal{B}$. Then we have the following {\it integration by parts formula}:
\begin{equation}\label{generalized_Stokes_theorem} 
\int_ \mathcal{B} (\alpha \cdot \operatorname{DIV}\mathbf{S} ) \mu _G=  -\int_ \mathcal{B} (\nabla \alpha : \mathbf{S} )\mu _G+\int_{ \partial \mathcal{B} } \mathbf{S} (\alpha, \mathbf{N}^{\flat _G}) \mu _G^ \partial, 
\end{equation} 
where $ \mu _G^ \partial $ is the Riemannian volume form induced on the boundary and $\mathbf{N}$ is the outward pointing unit vector field on $ \partial \mathcal{B} $ relative to $G$. Note that, in coordinates, we have $\alpha \cdot \operatorname{DIV}\mathbf{S}  = \alpha _A S^{AB}{}_{|B}$, $ \nabla \alpha : \mathbf{S}=  \alpha _{A|B}S^{AB}$, and $\mathbf{S} ( \alpha , \mathbf{N} ^{\flat_g})= S^{AB} \alpha _A N ^C G_{BC}$. See, Problem 7.6, Chap. 1 in \cite{MaHu1983}.

Given a vector field $\mathbf{u}$ on $ \mathcal{S} $, the {\it Piola transformation} of $\mathbf{u}$ is the vector field $\mathbf{U}$ on $ \mathcal{B} $ defined by $\mathbf{U}:= J_\varphi\,\varphi ^\ast \mathbf{u}$. In local coordinates, we have $U^a = J_\varphi \, (F ^{-1} )^A_a u ^a $. The Piola identity relates the divergences $ \operatorname{DIV}\mathbf{U}$ and $ \operatorname{div} \mathbf{u} $ as follows
\[
\operatorname{DIV}\mathbf{U}= J_\varphi (\operatorname{div} \mathbf{u})\circ \varphi.
\]
We can also make a Piola transformation on any index of a tensor field. For
example, let $\boldsymbol{\sigma} $ be a given $(2,0)$ tensor field on $ \mathcal{S} $. If we make a Piola transformation on the
last index, we obtain the two-point tensor field $\mathbf{P} (X):T_X ^\ast \mathcal{B} \times T_x ^\ast \mathcal{S} \rightarrow \mathbb{R}$ over $ \varphi $ defined by $\mathbf{P} (X)(\alpha _x , \beta _X):=J_\varphi\, \boldsymbol{\sigma}  (x)( \alpha _x , T ^\ast _x\varphi^{-1} (\beta_X) )$. In local coordinates, we have $P^{aA}=J_ \varphi \sigma ^{ab} (F ^{-1} )^B_b $ and the Piola identity is
\[
\operatorname{DIV}\mathbf{P} = J_ \varphi (\operatorname{div} \boldsymbol{\sigma} )\circ \varphi,  
\]
or, in local coordinates $P^{aA}{}_{|A}= J_ \varphi \sigma ^{ab}{}_{|b}$. The Piola identity will be crucial later, when passing from the material to the spatial representation.

A motion of $ \mathcal{B} $ is a time-dependent family of configurations, written $x= \varphi _t (X)= \varphi (X,t)$. It can therefore be thought of as a curve $\varphi _t \in \operatorname{Emb}( \mathcal{B} , \mathcal{S} )$ in the infinite dimensional configuration manifold. The material velocity is defined by $\mathbf{V} _t (X):= \mathbf{V} (X,t):= \frac{d}{dt}\varphi _{X}(t)\in T_{ \varphi (X)} \mathcal{S} $ and the spatial or Eulerian velocity is $\mathbf{v} _t (x):= \mathbf{v}(x,t)= \mathbf{V}_t  ( \varphi _t ^{-1} (x))$. We have
\[
\mathbf{V} _t \in T_{ \varphi _t } \operatorname{Emb}( \mathcal{B} , \mathcal{S} )\quad\text{and}\quad \mathbf{v}_t  = \mathbf{V} _t \circ \varphi_t  ^{-1} \in \mathfrak{X}  ( \varphi _t( \mathcal{B} ) ), 
\]
where $T_ \varphi \operatorname{Emb}( \mathcal{B} , \mathcal{S} )$ is the tangent space to $ \operatorname{Emb}( \mathcal{B} , \mathcal{S} )$ at $ \varphi $ and $ \mathfrak{X}  (\varphi _t( \mathcal{B} ))$ is the space of vector fields on $\varphi _t( \mathcal{B} ) \subset \mathcal{S}$.

\subsubsection{Equations of motion}

Let $ \varphi _t : \mathcal{B} \rightarrow \mathcal{S} $ be a motion of $\mathcal{B} $, let $ \mathbf{v} (x,t)$ and $ \mathbf{V} (X,t)$ be its spatial and material velocities, and let $ \rho (x,t)$ be the mass density function. The body undergoing the motion $ x=\varphi (X,t)$ is acted on by several kind of forces such as \textit{body forces} $ \mathbf{b} (x,t)$ per unit mass, \textit{surface forces} $ \boldsymbol{\tau}(x,t)$ per unit area, and \textit{stress forces} $ \mathbf{t} (x,t, \mathbf{n} )$ per unit area across any surface element with unit normal $ \mathbf{n} $. For simplicity, from now on we shall assume $ \boldsymbol{\tau}=0$.

As a consequence of the mass conservation one obtains the continuity equation $ \partial _t \rho + \operatorname{div}(\rho \mathbf{v} )=0$. This implies the relation $(\rho\circ \varphi )J_ \varphi= \rho _{\rm ref}$, where $ \rho _{\rm ref}$ is the mass density in the reference configuration, i.e., written in terms of material coordinates. As a consequence of the integral form of momentum balance we obtain the existence of a $(2,0)$ tensor field $ \boldsymbol{\sigma} $, called the Cauchy stress tensor, such that $ \mathbf{t}(x,t, \mathbf{n} )  = \boldsymbol{\sigma}  (x,t)(\_\,, \mathbf{n} ^{\flat_g} )$ or, in coordinates, $t^a= \sigma ^{ab} g_{bc}n^c$. Then using again momentum balance and mass conservation we obtain the following motion equation in spatial coordinates
\[
\rho (\partial _t \mathbf{u} +  \mathbf{u} \cdot \nabla \mathbf{u} )= \operatorname{div} \boldsymbol{\sigma} + \rho \mathbf{b},
\]
where we recall that $ \nabla $ and $\operatorname{div}$ are, respectively, the covariant derivative and the divergence associated to the Riemannian metric $g$ on $ \mathcal{S} $. 
Using balance of moment of momentum, one obtains that the Cauchy stress tensor $\boldsymbol{\sigma}$ is symmetric, i.e. $ \sigma ^{ab}= \sigma ^{ba}$.

In order to write the equations of motion in material coordinates, one has to consider the {\it Piola transformation} of the Cauchy stress tensor $ \boldsymbol{\sigma} (x,t)$, called the {\bfi first Piola-Kirchhoff tensor} $ \mathbf{P}(X,t)$. We also define the material body force $ \mathbf{B} (X,t):= \mathbf{b} ( \varphi (X,t),t)$. By using the Piola identity together with the relation $(\rho\circ \varphi )J_ \varphi= \rho _{\rm ref}$, one arrives at the equations 
\begin{equation}\label{equation_material} 
\rho _{\rm ref}\frac{D \mathbf{V}}{D t}=  \operatorname{DIV}\mathbf{P} + \rho _{\rm ref} \mathbf{B},
\end{equation}
where $D/Dt$ denotes the covariant derivative relative to the Riemannian metric $g$.
As stressed in \S2.2 of \cite{MaHu1983}, when $ \mathcal{S} $ is not the Euclidean space, one cannot derive the equations of motion from the integral form of momentum balance. One can however derive the equations from an energy principle that does not require
$ \mathcal{S} $ to be linear.

\subsubsection{Reversible continuum mechanics - material representation}\label{VP_no_therm_mat}

In material representation, the variational formalism follows from the standard Hamilton principle. One considers the Lagrangian defined on the tangent bundle $T \operatorname{Emb}( \mathcal{B} , \mathcal{S} )$ and  given by the kinetic minus potential energy, whose most general form is
\begin{equation}\label{general_Lagrangian}
\begin{aligned}
L_{g,G_{\rm ref}, R _{\rm ref}, Z_{\rm ref}}( \varphi , \dot \varphi )=&\int_ \mathcal{B} \mathfrak{L}(X, \varphi (X), \dot \varphi (X), T_X \varphi ) \mu _{G_{\rm ref}}(X)\\
=&\int_ \mathcal{B}\frac{1}{2}\rho _{\rm ref}(X)|\dot \varphi(X) | ^2 _g \mu _{G_{\rm ref}}(X)\\
&-\int_ \mathcal{B} \mathfrak{E} (g( \varphi (X)), G_{\rm ref}^\sharp(X), T_X\varphi,R_{\rm ref}(X),Z_{\rm ref} (X))\mu _{G_{\rm ref}} (X)\\
& -\int_ \mathcal{B} \rho _{\rm ref} \mathcal{V} ( \varphi (X)) \mu _{G_{\rm ref}}(X),
\end{aligned}
\end{equation} 
where $\mathfrak{L}$ denotes the Lagrangian density, $R_{\rm ref}:= \rho _{\rm ref}\mu _{G_{\rm ref}}$ is the mass form, and $Z_{\rm ref} $ represents other tensor fields on the reference configuration, on which the internal energy density $\mathfrak{E}$ depends, such as the entropy density $S_{\rm ref}$ or the magnetic field $ \mathbf{B} _{\rm ref}$. The $2$-contravariant symmetric tensor field $G^{\sharp}_{\rm ref}$ denotes the inverse of the Riemannian metric $G_{\rm ref}$. The notation $L_{g,G_{\rm ref}, R_{\rm ref}, Z_{\rm ref}}$ for the Lagrangian is used to recall that it depends parametrically on the tensor fields $g, G_{\rm ref}, R_{\rm ref}, Z _{\rm ref}$. These fields are time independent. The function $\mathcal{V} $ is a potential energy density such as the gravitation.

It is essential to explicitly write the dependence of the internal energy density $ \mathfrak{E}$ on the Riemannian metric $g$ in order to introduce the notion of covariance. Covariance is the statement on left invariance of $ \mathfrak{E} $ relative to the group $ \operatorname{Diff}( \mathcal{S} )$ of spatial diffeomorphisms and is a fundamental physical requirement. We refer to \cite{MaHu1983} for a detailed account and to \cite{SiMaKr1988} and \cite{GBMaRa2012} for the corresponding consequences in the reduced Hamiltonian and Lagrangian formulations.

Hamilton's principle reads
\begin{equation}\label{HP_principle} 
\delta \int_{ t _1 }^{ t _2 }L_{g,G_{\rm ref},R_{\rm ref}, Z _{\rm ref}}( \varphi(t)  , \dot \varphi (t) )dt=0,
\end{equation} 
for arbitrary variations $ \delta \varphi $ of the configuration $ \varphi $ such that $ \delta \varphi (t_1 )= \delta \varphi ( t _2 )=0$. It yields the Euler-Lagrange equations. In order to derive the equations of motion, one has to impose appropriate boundary conditions.

Let us first assume {\it prescribed boundary values}, that is,  $ \varphi (t,X)= \varphi_0(X)$, for all $X \in\partial \mathcal{B} $. In this case, Hamilton variational principle \eqref{HP_principle} for the Lagrangian \eqref{general_Lagrangian} yields the equations \eqref{equation_material}, with
\[
\mathbf{P} =\mathbf{P} ^{\rm cons}= - \left( \frac{\partial \mathfrak{L} }{\partial T_{X} \varphi } \right) ^{\sharp_g}= \left( \frac{\partial\mathfrak{E}}{\partial T_{X}\varphi } \right) ^{\sharp_g} \quad\text{and}\quad  \mathbf{B} = \mathbf{B} ^{\rm cons}=-( \mathbf{d}  \mathcal{V}\circ \varphi )^{\sharp_g}=- \operatorname{grad} \mathcal{V} \circ \varphi,
\]
where $ \sharp_g$ denotes the sharp operator associated to the Riemannian metric $g$.
We use the notations $\mathbf{P} ^{\rm cons}$ and $\mathbf{B} ^{\rm cons}$ to emphasize the fact that the stress forces and body forces arising from the Euler-Lagrange equations are {\it conservative}. Note that the case of prescribed boundary values includes for example {\it no-slip boundary conditions} when $ \mathcal{S} = \mathcal{B} $ and $ \varphi _0= id$, so that the configuration space is the group $ \operatorname{Diff}_0( \mathcal{B} )$ of diffeomorphisms of $ \mathcal{B} $ that fix each point of the boundary.

In the case of a {\it free boundary}, then Hamilton variational principle \eqref{HP_principle} for the Lagrangian \eqref{general_Lagrangian} yields, in addition to the equations \eqref{equation_material}, the zero-traction boundary condition
\begin{equation*}\label{BC_1} 
\mathbf{P} ^{\rm cons}( \mathbf{N} ^{\flat_G},\_\,)=0\quad \text{on} \quad \partial \mathcal{B}. 
\end{equation*}

In the case with a {\it fixed boundary} and {\it tangential boundary condition}, that is $ \mathcal{S} = \mathcal{B} $, so that $ \varphi \in \operatorname{Diff}( \mathcal{B} )$, then the zero-traction boundary condition reads
\begin{equation}\label{BC_2} 
\mathbf{P}^{\rm cons} ( \mathbf{N} ^{\flat_G},\_\,)|_{ T \partial \mathcal{B} }=0\quad \text{on} \quad \partial \mathcal{B},
\end{equation} 
since the variation $ \delta \varphi $ is parallel to the boundary $ \partial \mathcal{B} $.

\paragraph{Including external forces.}Nonconservative stress forces and body forces can be easily included in the variational principle by considering the Lagrange-d'Alembert principle obtained by adding the virtual work done by the forces along a displacement $ \delta \varphi $. It reads
\begin{equation}\label{Forced_HP_elasticity} 
\begin{split}
\delta \int_{ t _1}^{ t _2 }&L_{g,G_{\rm ref},R_{\rm ref}, Z _{\rm ref}}( \varphi , \dot \varphi )dt\\
&\hspace{2cm}+ \int_{ t _1 }^{ t _2 }\left[ \int_ \mathcal{B}\left(  \rho _{\rm ref} \mathbf{B} ^{\rm ext}\cdot  \delta \varphi  - (\mathbf{P} ^{\rm ext}) ^{\flat_g} : \nabla \delta \varphi \right) \mu _{G_{\rm ref}} \right] dt =0,
\end{split}
\end{equation} 
for arbitrary variations $ \delta \varphi $ of the configuration $ \varphi $ such that $ \delta \varphi (t_1 )= \delta \varphi ( t _2 )=0$. 
In the above, the flat operator is computed relative to the metric $g$ and we dropped the argument $X$ for simplicity. The notation $\nabla ^g \delta \varphi $ stands for the Levi-Civita covariant derivative (with respect to $g$) of the two-point tensor $ \delta \varphi $ (see, e.g, \cite{MaHu1983} for a complete treatment). The principle \eqref{Forced_HP_elasticity} yields the equations \eqref{equation_material} with $ \mathbf{P} = \mathbf{P} ^{\rm cons}+ \mathbf{P} ^{\rm ext}$ and $ \mathbf{B} = \mathbf{B} ^{\rm cons}+ \mathbf{B} ^{\rm ext}$. In the case of free, respectively, fixed boundary, we obtain the zero-traction boundary condition $\mathbf{P}( \mathbf{N} ^{\flat_G},\_\,)=0$, respectively, $\mathbf{P}( \mathbf{N} ^{\flat_G},\_\,)|_{T \partial \mathcal{B} }=0$ on $\partial \mathcal{B}$.
However, in presence of external forces, one usually imposes no-slip boundary conditions in which case no additional boundary conditions arise from the variational formalism.

\paragraph{Example 1: ideal compressible fluids.} For this particular case, the tensor field $Z_{\rm ref}$ is taken as the entropy density $S_{\rm ref}$, and hence the internal energy density is of the form
\begin{equation}\label{fluid_energy_pot}
{\footnotesize\begin{aligned} 
&\mathfrak{E}\left( g( \varphi (X)), G_{\rm ref}^\sharp(X),T_X \varphi, R_{\rm ref}(X), S_{\rm ref}(X) \right) \mu _{G_{\rm ref}}(X)  = \varepsilon  \left( \frac{\rho _{\rm ref}(X)}{J_ \varphi(X)  }, \frac{S _{\rm ref}(X)}{J_ \varphi(X)  }\right)J_ \varphi(X) \mu _{G_{\rm ref}}(X)\\
&\qquad \quad = \varepsilon ( \rho ( \varphi (X)), s( \varphi (X)))\varphi ^\ast \mu _g(X) = \varphi ^\ast \left(\varepsilon ( \rho , s) \mu _g \right)(X)  ,
\end{aligned}}
\end{equation}
where $ \varepsilon ( \rho , s) \mu _g$ is the energy density in spatial representation, taken here to be an arbitrary function of $  \rho (x)$ and $ s(x)$. In \eqref{fluid_energy_pot} we used the relations between the spatial and material quantities as $(\rho\circ \varphi )J_ \varphi= \rho _{\rm ref}$, $(s\circ \varphi )J_ \varphi= S _{\rm ref}$, and $ J_\varphi \mu_g=\varphi ^\ast \mu _{G_{\rm ref}}$. Note that for ideal compressible fluids the energy $\mathfrak{E}$ depends on the deformation gradient $T_X \varphi $ only through its Jacobian $J_ \varphi $. It is important to note from \eqref{fluid_energy_pot} that whereas both  $ \mathfrak{E}$ and $ \mu _{G_{\rm ref}}$ depend on $G_{\rm ref}$, the whole expression $\mathfrak{E}(...) \mu _{G_{\rm ref}}$ does {\it not} depend on $G_{\rm ref}$ for ideal compressible fluid. For this reason, there is no spatial variable associated to $G_{\rm ref}$ and we can write $L_{g, G_{\rm ref}, R_{\rm ref}, S_{\rm ref}}=L_{g,R_{\rm ref}, S_{\rm ref}}$. In this case, the introduction of a Riemannian metric $G_{\rm ref}$ on $ \mathcal{B} $ is only needed to facilitate the computations. We will see below that for viscoelastic fluids, $\mathfrak{E}$ depends explicitly on $G_{\rm ref}$.

The Cauchy stress tensor is computed to be
\begin{equation}\label{sigma_fluid} 
\boldsymbol{\sigma} ^{\rm cons} = - p g^\sharp =- \left( \frac{\partial \varepsilon}{\partial \rho }\rho + \frac{\partial \varepsilon }{\partial s}s - \varepsilon \right)   g^\sharp,
\end{equation} 
where $p$ is the pressure. Using the Piola identity (see, Theorem 7.20 of Chapter 1 in  \cite{MaHu1983}), we obtain $ \operatorname{DIV} \mathbf{P}^{\rm cons} =- (\operatorname{grad}p) \circ \varphi J_ \varphi$ and hence the equation of motion reads
\begin{equation*}\label{fluid_equation_material} 
\rho _{\rm ref}\frac{D \mathbf{V}}{D t}= -( \operatorname{grad}p) \circ \varphi \, J_ \varphi  + \rho _{\rm ref} \mathbf{B}^{\rm cons}
\end{equation*}
for ideal compressible fluids in material representation.

For fluids with fixed boundary,  i.e., $ \mathcal{B} = \mathcal{S} $, the boundary condition \eqref{BC_2} vanishes because of the expression \eqref{sigma_fluid}.  Indeed, by definition of the Piola transformation, it follows
\begin{align*}
\mathbf{P}^{\rm cons} \left( \delta \varphi (X)^{\flat _g}, \mathbf{N} ^{\flat _G} (X)\right)& = J_ \varphi (X) \boldsymbol{\sigma} ^{\rm cons}(x) \left(  \delta \varphi (X)^{\flat _g},T^* _x \varphi ^{-1} (\mathbf{N} ^{\flat_G } (X) )\right) \\
&=- p(x)J_ \varphi (X) g(x)\left(  \delta \varphi (X)^{\flat _g}, T^* _x \varphi ^{-1} (\mathbf{N} ^{\flat_G } (X) ) \right) \\
&=- p(x)J_ \varphi (X) \left\langle T_x \varphi ^{-1}  \delta \varphi (X), \mathbf{N} ^{\flat_G } (X)\right\rangle \\
&=- p(x)J_ \varphi (X) G_{\rm ref}(X) \left( T_x \varphi ^{-1}  \delta \varphi (X), \mathbf{N}  (X)\right) =0,
\end{align*} 
since $T_x \varphi ^{-1}  \delta \varphi (X)$ is parallel to the boundary.

In absence of thermodynamic effects, viscosity and other nonconservative stresses can be included in the variational formalism by using the Lagrange-d'Alembert principle \eqref{Forced_HP_elasticity}. In this case one usually imposes no-slip boundary conditions.

\paragraph{Example 2: multicomponent fluids.} In the ideal case, to describe a multicomponent fluid with $K$ chemical components $A=1,...,K$, one has to introduce the number density $n_A(t,x)$ of the substance $A$ in spatial representation. The material Lagrangian is now indexed as $L_{g, \{N_{A\,\rm ref}\}, S_{\rm ref}}$, where $N_{A\,\rm ref}(X)$, $A=1,...,2$ is the number density of the substance $A$ evaluated at $X \in \mathcal{B}$ in material representation, hence $N_{A\,\rm ref}$ is constant in time. The Lagrangian is of the form \eqref{general_Lagrangian}, where $ \rho _{\rm ref}(X) :=\sum_{A=1}^K M^A N_{A\,\rm ref}(X) $ in the first term denotes the total mass density, with $M^A$ the molar mass of the substance $A$, and the internal energy density has the form
\begin{equation}\label{int_energy_multi_comp} 
 \mathfrak{E}\left( g( \varphi (X)), G_{\rm ref}^\sharp(X),T_X \varphi,\{N_{A\,\rm ref}(X)\}, S_{\rm ref}(X) \right)\mu _{G_{\rm ref}} (X) =\varphi ^\ast \left(\varepsilon ( \{n_A\} , s) \mu _g \right)(X).
\end{equation} 
The Cauchy stress tensor is computed to be
\begin{equation*}\label{sigma_Multi_fluid} 
\boldsymbol{\sigma} ^{\rm cons} = - p g^\sharp =- \left( \sum_{A=1}^K\frac{\partial \varepsilon}{\partial n_A }n_A + \frac{\partial \varepsilon }{\partial s}s - \varepsilon \right)   g^\sharp
\end{equation*} 
and, as before, the boundary condition \eqref{BC_2} is automatically satisfied.

\paragraph{Example 3: viscoelastic fluids.} For completeness, we briefly explain how this geometric setting applies to viscoelastic fluids, even though we shall not pursue the study of the thermodynamics of viscoelastic fluids in this paper.  This will be the object of a future work.

For viscoelastic fluids, 
the material energy density is of the form
\begin{equation}\label{fluid_energy_pot_viscoel}
{\footnotesize\begin{aligned} 
&\mathfrak{E}\left( g( \varphi (X)), G_{\rm ref}^\sharp(X),T_X \varphi, R_{\rm ref}(X), S_{\rm ref}(X) \right) \mu _{G_{\rm ref}}(X)\\
&\qquad  \quad  = \varepsilon  \left( \frac{\rho _{\rm ref}(X)}{J_ \varphi(X)  }, \frac{S _{\rm ref}(X)}{J_ \varphi(X)  },\left( \varphi _\ast G_{\rm ref}^\sharp\right)  ( \varphi(X))\right)J_ \varphi(X) \mu _{G_{\rm ref}}(X)\\
&\qquad \quad = \varepsilon ( \rho ( \varphi (X)), s( \varphi (X)),b( \varphi (X)))\varphi ^\ast \mu _g(X) = \varphi ^\ast \left(\varepsilon ( \rho , s,b) \mu _g \right)(X)  ,
\end{aligned}}
\end{equation}
where $b:=\varphi _\ast G_{\rm ref}^\sharp$ is the {\it Finger deformation tensor} (also called the {\it left Cauchy-Green tensor}). It reads $b ^{ab}(x)=G^{AB}_{\rm ref}(X)F^a_A(X)F^b_B(X)$ in local coordinates.
So, for viscoelastic fluids, the material internal energy density $\mathfrak{E}$ depends on the deformation gradient through its Jacobian and through the tensor $b$ that describes the elastic deformation of the viscoelastic fluid.
The Cauchy stress tensor is computed to be
\begin{equation*}\label{sigma_fluid_elast} 
\boldsymbol{\sigma} ^{\rm cons} = - p g^\sharp + \boldsymbol{\sigma} _{\rm el}^{\rm cons}, \quad p=- \left( \frac{\partial \varepsilon}{\partial \rho }\rho +\frac{\partial \varepsilon }{\partial s}s - \varepsilon \right),\quad \boldsymbol{\sigma} _{\rm el}^{\rm cons}= 2 \left( \frac{\partial \varepsilon }{\partial b}\cdot b \right) ^\sharp,
\end{equation*}
where the extra conservative stress $ \boldsymbol{\sigma} _{\rm el}^{\rm cons}$ is caused by the elastic deformation of the fluid. In coordinates, we have $(\boldsymbol{\sigma} _{\rm el}^{\rm cons}) ^{ab}=2g^{ac} \frac{\partial \varepsilon}{\partial b^{cd}}b^{db}$, which is necessarily symmetric by the covariance assumption.

In material representation, the equations of motion are thus given by
\begin{equation*}\label{viscoelasticfluid_equation_material} 
\rho _{\rm ref}\frac{D \mathbf{V}}{D t}= -( \operatorname{grad}p) \circ \varphi \, J_ \varphi  + \operatorname{DIV}\, \mathbf{P} ^{\rm cons}_{\rm el}+ \rho _{\rm ref} \mathbf{B}^{\rm cons},
\end{equation*}
where the two-point tensor $ \mathbf{P} ^{\rm cons}_{\rm el}$ is the Piola transformation of $ \boldsymbol{\sigma} ^{\rm cons}_{\rm el}$, namely, $\operatorname{DIV}\mathbf{P}^{\rm cons}_{\rm el} = J_ \varphi (\operatorname{div} \boldsymbol{\sigma}^{\rm cons}_{\rm el} )\circ \varphi.$ They result from Hamilton's principle \eqref{HP_principle}. 
\medskip

Contrary to the case of the adiabatic compressible fluid above, the presence of the extra term $\mathbf{P} ^{\rm cons}_{\rm el}$ in the total conservative material stress tensor $ \mathbf{P} ^{\rm cons}$ implies that the boundary condition \eqref{BC_2} is not automatically satisfied. It implies the boundary condition
\[
\mathbf{P} ^{\rm cons}_{\rm el}( \mathbf{N} ^{\flat_G},\_\,)|_{ T \partial \mathcal{B} }=0 \quad\text{on}\quad  \partial \mathcal{B}.
\]
As shown above, viscosity can be included in the equations by using the Lagrange-d'Alembert principle \eqref{Forced_HP_elasticity}. In this case one usually imposes no-slip boundary conditions.

\subsubsection{Reversible continuum mechanics - spatial representation}\label{VP_no_therm_spa}

In spatial representation, the variational formalisms are much more involved than in the material representation. Indeed, the variations of spatial quantities are subject to constraints, such as the well-known {\it Lin constraints}. In this section, we shall restrict to the case $ \mathcal{S} = \mathcal{B} $ so that the configuration is a Lie group, the group $ \operatorname{Diff}( \mathcal{B} )$ of all smooth diffeomorphisms. In this case, the variational formalisms in spatial representation have been developed in \cite{HMR1998} by using the theory of {\it Euler-Poincar\'e reduction on Lie groups with advected quantities}. The general case $ \mathcal{B} \neq \mathcal{S} $ needed for free boundary continuum mechanics, has been developed in \cite{GBMaRa2012}. In conjunction with the Hamilton-Pontryagin principle with advected parameters, it has been shown in \cite{GBYo2015} that second order Rivlin-Ericksen fluids can be formulated in the context of the infinite dimensional nonholonomic Lie-Dirac reduction.

In order to obtain the spatial form of the equations from a reduced Lagrangian point of view, one has to assume that the Lagrangian \eqref{general_Lagrangian} is right-invariant under the action of the group $ \operatorname{Diff}( \mathcal{B} )$. This happens if the material internal energy density $ \mathfrak{E}$ is $ \operatorname{Diff}( \mathcal{B} )$ equivariant, that is,  
\[
\mathfrak{E}\left( g \circ \varphi\circ \psi  , \psi ^\ast G^\sharp_{\rm ref}, T( \varphi \circ \psi ), \psi ^\ast R _{\rm ref}, \psi ^\ast Z_{\rm ref}\right) = \mathfrak{E}\left(g \circ \varphi , G^\sharp_{\rm ref}, T \varphi , R_{\rm ref},Z _{\rm ref}\right)  \circ \psi,
\]
for all $ \psi \in \operatorname{Diff}( \mathcal{B} )$. This is an hypothesis on the material energy density $ \mathfrak{E}$; see \cite{MaHu1983}, \cite{GBMaRa2012}. It is clear that both \eqref{fluid_energy_pot} and \eqref{fluid_energy_pot_viscoel} satisfy this hypothesis.

\paragraph{Example 1: ideal compressible fluids.}  Using the change of variables $x= \varphi (X)$, one obtains, from the material Lagrangian \eqref{general_Lagrangian} and the internal energy \eqref{fluid_energy_pot}, the (reduced) Lagrangian in spatial representation as
\[
\ell( \mathbf{v} , \rho ,s)=\int_ \mathcal{S} \frac{1}{2}\rho  | \mathbf{v} | _g ^2 \mu _g -\int_ \mathcal{S} \varepsilon ( \rho , s) \mu _g-\int_ \mathcal{S} \rho \mathcal{V} \mu _g,
\]
where $ \mathbf{v} = \dot \varphi \circ \varphi ^{-1} \in \mathfrak{X} _{\|}( \mathcal{S} )$ is the spatial or Eulerian velocity. Here, $\mathfrak{X} _{\|}( \mathcal{S} )$ denotes the space of vector fields on $ \mathcal{S}= \mathcal{B}  $ parallel to the boundary, i.e., $\mathfrak{X} _{\|}( \mathcal{S} )=\{\mathbf{v} \in \mathfrak{X}( \mathcal{S}) \mid \mathbf{v} \cdot \mathbf{n}=0 \;\;\textrm{on}\;\; \partial\mathcal{B} \}$, where $ \mathbf{n} $ is the unit normal vector field to $ \partial \mathcal{S}$, relative to the metric $g$. Hamilton's principle \eqref{HP_principle} yields the variational principle
\begin{equation}\label{VP_spat} 
\delta \int_{ t _1 }^{ t _2 }\ell( \mathbf{v} , \rho , s)dt=0,
\end{equation} 
for \textit{constrained} variations of the form
\begin{equation*}\label{constrained_V} 
\delta \mathbf{v} = \partial _t \boldsymbol{\zeta}+[ \mathbf{v} , \boldsymbol{\zeta} ], \quad \delta \rho =- \operatorname{div}( \rho \boldsymbol{\zeta} ), \quad \delta s=- \operatorname{div}( s \boldsymbol{\zeta} ),
\end{equation*} 
where $ \boldsymbol{\zeta} = \delta \varphi \circ \varphi ^{-1} $ is an arbitrary curve in $ \mathfrak{X}_{\|} ( \mathcal{S} )$ vanishing at the endpoints. These variations can be obtained by direct computations of the variations of $ \mathbf{v} = \dot \varphi \circ \varphi ^{-1} $, $ \rho = (\rho _{\rm ref} \circ \varphi ^{-1} )J_ { \varphi ^{-1} }$, and $ s = (S _{\rm ref} \circ \varphi ^{-1} )J_ { \varphi ^{-1} }$, induced from the variations $ \delta \varphi $ of the Lagrangian trajectory $ \varphi $. From the reduced variational principle \eqref{VP_spat}, the equations in spatial variables are obtained as
\begin{equation}\label{viscoelastic_spat}
\left\{
\begin{array}{l}
\vspace{0.2cm}\rho ( \partial _t \mathbf{v} + \mathbf{v}  \cdot \nabla  \mathbf{v} )=- \operatorname{grad}  p+ \rho \mathbf{b} ^{\rm cons},\\[1mm]
\partial _t \rho + \operatorname{div} ( \rho \mathbf{v} )=0, \quad \partial _t s+ \operatorname{div}(s \mathbf{v} )=0,
\end{array}
\right.
\end{equation} 
where $p=  \frac{\partial \varepsilon}{\partial \rho }\rho + \frac{\partial \varepsilon }{\partial s}s - \varepsilon$ and $\mathbf{b}^{\rm cons} =- \operatorname{grad}  \mathcal{V} $. We refer to \cite{HMR1998} for the complete description of the Lagrangian reduction theory underlying this example, and its abstract description in terms of the {\it Euler-Poincar\'e reduction with advected parameters}.

\paragraph{Example 2: multicomponent fluids.} 

For the multicomponent fluid, using the change of variables $x= \varphi (X)$, one obtains, from \eqref{general_Lagrangian} with internal energy \eqref{int_energy_multi_comp}, the Lagrangian in spatial representation as
\[
\ell( \mathbf{v} , \{n_A \},s)=\int_ \mathcal{S} \frac{1}{2}\rho  | \mathbf{v} | _g ^2 \mu _g -\int_ \mathcal{S} \varepsilon ( \{n_A\} , s) \mu _g-\int_ \mathcal{S} \rho \mathcal{V} \mu _g.
\]
The variational principle has the same form as \eqref{VP_spat} in which $ \rho $ replaced by the collection of number densities $\{n_A\}$, $A=1,...,K$, with constrained variations $ \delta n _A =- \operatorname{div}(n_A \boldsymbol{\zeta} )$. This results in the equations
\begin{equation}\label{multicomp_spat}
\left\{
\begin{array}{l}
\vspace{0.2cm}\rho ( \partial _t \mathbf{v} + \mathbf{v} \cdot \nabla \mathbf{v} )=- \operatorname{grad}  p+ \rho \mathbf{b} ^{\rm cons},\\[1mm]
\partial _t n_A + \operatorname{div} ( n_A \mathbf{v} )=0,\;A=1,...,K, \quad \partial _t s+ \operatorname{div}(s \mathbf{v} )=0,
\end{array}
\right.
\end{equation} 
where $ \rho =\sum_{A=1}^KM^An_A$ and $p=\sum_{i=1}^K\frac{\partial \varepsilon }{\partial n_A}n_A+ \frac{\partial \varepsilon }{\partial s}s  - \varepsilon $. 
\medskip

Both systems \eqref{viscoelastic_spat} and \eqref{multicomp_spat} will be extended to include irreversible thermodynamics effects in the next sections.

\paragraph{Including external forces.} Nonconservative stress and body forces can be included in the Lagrange-d'Alembert principle by writing \eqref{Forced_HP_elasticity} in spatial representation.  That is, we have
\begin{equation*}\label{Forced_HP_elasticity_spatial} 
\delta \int_{ t _1 }^{ t _2 }\ell( \mathbf{v} , \rho , s,b )dt+ \int_{ t _1 }^{ t _2 }\left[ \int_ \mathcal{S}\left(  \rho \mathbf{b} ^{\rm ext} \cdot \boldsymbol{\zeta} -(\boldsymbol{\sigma}   ^{\rm ext})^{\flat_g}: \nabla  \boldsymbol{\zeta} \right) \mu _g \right] dt =0,
\end{equation*}
where $(\boldsymbol{\sigma}   ^{\rm ext})^{\flat_g}: \nabla  \boldsymbol{\zeta}=(\boldsymbol{\sigma} ^{\rm ext}) ^{\flat _g }: \operatorname{Def} \boldsymbol{\zeta}$ with $ \operatorname{Def} \mathbf{v} := \frac{1}{2} ( \nabla \mathbf{v} + \nabla \mathbf{v} ^\mathsf{T})$ the rate of deformation tensor. In this case, one usually imposes no-slip boundary conditions, i.e., $ \mathbf{v} \in \mathfrak{X}_0( \mathcal{S} )=\{ \mathbf{v} \in \mathfrak{X}  ( \mathcal{S} )\mid \mathbf{v} |_{ \partial \mathcal{S} }=0\}$, and hence it follows
\begin{equation}\label{exterior_forces} 
\rho ( \partial _t \mathbf{v} + \mathbf{v}  \cdot \nabla \mathbf{v} )= \operatorname{div} \boldsymbol{\sigma} + \rho \mathbf{b}, \quad  \boldsymbol{\sigma} =\boldsymbol{\sigma} ^{\rm cons}+ \boldsymbol{\sigma} ^{\rm ext}, \quad \mathbf{b} = \mathbf{b} ^{\rm cons}+\mathbf{b} ^{\rm ext},
\end{equation} 
together with the advection equations for $ \rho , s, n_A$ or $b$.
The energy balance is given by
\begin{equation}\label{energy_density_behavior} 
\partial _t e+ \operatorname{div}( e \mathbf{v} )=  \operatorname{div} (\boldsymbol{\sigma} ^{\flat_g}\cdot \mathbf{v} )-\boldsymbol{\sigma} ^{\rm ext}: \nabla \mathbf{v} + \rho (\mathbf{b} ^{\rm ext})^{\flat_g}\cdot \mathbf{v},
\end{equation} 
where $e:= \frac{1}{2} \rho | \mathbf{v} | ^2_g  + \varepsilon ( \rho , s,b)+ \rho \mathcal{V} $ is the total energy density.

\paragraph{Navier-Stokes equations.} For the case of the Navier-Stokes equations, the external stress tensor $ \boldsymbol{\sigma} ^{\rm ext}$ is given by the viscosity, i.e., 
\begin{equation*}\label{viscous_Cauchy_stress} 
\boldsymbol{\sigma} ^{\rm ext}= 2 \mu  (\operatorname{Def} \mathbf{v})^{\sharp _g }+ \lambda   (\operatorname{div} \mathbf{v} )  g^\sharp= 2 \mu \left[ (\operatorname{Def} \mathbf{v})^{\sharp_g }- \frac{1}{3} ( \operatorname{div} \mathbf{v} ) g^\sharp) \right]  + \zeta( \operatorname{div} \mathbf{v}) g^\sharp,
\end{equation*} 
where $ \mu $ is the first coefficient of viscosity (shear viscosity) and $ \zeta = \lambda + \frac{2}{3} \mu $ is the second coefficient of viscosity (bulk viscosity), while  the Cauchy stress tensor $\boldsymbol{\sigma} ^{\rm cons} $ is given by
\begin{equation*}
\boldsymbol{\sigma} ^{\rm cons} = - p g^\sharp =- \left( \frac{\partial \varepsilon}{\partial \rho }\rho +\frac{\partial \varepsilon }{\partial s}s - \varepsilon \right)g^\sharp.
\end{equation*}
Equations \eqref{exterior_forces} become the {\it compressible Navier-Stokes equations}
\[
\rho ( \partial _t \mathbf{v} + \mathbf{v}\cdot \nabla  \mathbf{v} )=-\mathrm{grad}\,p+ \mu \Delta \mathbf{v} +( \mu + \lambda ) \nabla \operatorname{div} \mathbf{v}+\rho( \mathbf{b}^{\rm cons}+\mathbf{b} ^{\rm ext}),
\]
which are accompanied by the advection equations $\partial _t \rho + \operatorname{div} ( \rho \mathbf{v} )=0$ and $\partial _t s+ \operatorname{div}(s \mathbf{v} )=0$.

The balance of energy \eqref{energy_density_behavior} is computed by using
\[
( \boldsymbol{\sigma} ^{\rm ext}) ^{\flat_g } \!:\! \operatorname{Def} \mathbf{v}= 2 \mu  \operatorname{Def} \mathbf{v}\!: \!\operatorname{Def} \mathbf{v} + \lambda   (\operatorname{div} \mathbf{v} )  \delta \!: \!\operatorname{Def} \mathbf{v}=2 \mu |\operatorname{Def} \mathbf{v}|_g ^2  + \lambda   (\operatorname{div} \mathbf{v} ) ^2,
\]
in \eqref{energy_density_behavior}, where $ \delta $ denotes the $(1,1)$ Kronecker tensor.

In this approach, the viscosity is interpreted as an external force, hence the notation $ \boldsymbol{\sigma} ^{\rm ext}$, leading to a dissipation of energy. This case is an approximation of the Navier-Stokes-Fourier system treated below, where the same stress tensor $ \boldsymbol{\sigma} ^{\rm ext}$ will be considered as a friction $ \boldsymbol{\sigma} ^{\rm fr}$ associated to an irreversible transfer of mechanical energy into heat, for which $\partial _t s+ \operatorname{div}(s \mathbf{v} )=0$ will no longer be true (see \eqref{NSF_spatial}). In this case, $ \boldsymbol{\sigma} ^{\rm fr}$ will no longer be treated a dissipative force in the energy balance.

This ends our preliminaries concerning the variational formalisms in material and spatial representation.

\subsection{Variational formalism for viscous and heat conducting fluids}\label{VHCF} 

In this section, we will present an extension of the variational formalism for nonequilibrium thermodynamics developed for discrete systems  in \S\ref{discrete_systems} to the case of continuum mechanics. We will first consider a compressible fluid with a single (chemical) component subject to the irreversible processes associated to heat transport and viscosity.

\subsubsection{Material representation}\label{Mat_NSF} 

Since the internal heat transfer is now taken into account, the appropriate variational formalism in material representation is a continuum version of the variational formalism that we introduced in Definition \ref{LdA_def_discrete_Systems}. By analogy with the variables $\gamma ^A, \sigma _A$ developed in \S\ref{discrete_systems}, we introduce the variables $ \Gamma (t,X)$ and $ \Sigma (t,X)$ in material coordinates, where the rate $\dot \Gamma (t,X)$ will ultimately be identified with the {\it temperature} in material representation, denoted $ \mathfrak{T}(t,X)$, while the rate $\dot{\Sigma}(t,X)$ will be clarified as the {\it total (internal and external) entropy production}. The introduction of these variables allows us to write the nonholonomic constraint as a sum of thermodynamic affinity densities multiplied by thermodynamic fluxes  (written as time derivatives).

\medskip

The Lagrangian has the same expression as the one for the ideal compressible fluid (i.e. it is given by \eqref{general_Lagrangian}, where $ \mathfrak{E}$ is given by the internal energy density \eqref{fluid_energy_pot}), with a change in the interpretation of the material entropy. Namely, 
the entropy was understood, before, as a fixed material quantity $S_{\rm ref}(X)$ on which the Lagrangian depends parametrically. Now we will interpret it as a {\it dynamic variable} $S(t,X)$ involved in the variational formalism. This crucial difference is illustrated by the change of notation
\[
L_{g, R_{\rm ref},S_{\rm ref}}( \varphi , \dot \varphi ) \;\leadsto \;L_{g,R_{\rm ref}}( \varphi , \dot \varphi , S),\quad S_{\rm ref}(X)\;\leadsto\; S(t,X)
\]
in the material Lagrangian \eqref{general_Lagrangian} with material internal energy density given by \eqref{fluid_energy_pot}.

\paragraph{Variational formalism for nonequilibrium thermodynamics of viscous heat conducting fluids.} The continuum version of the variational formalism of Definition \ref{LdA_def_discrete_Systems} is given by
\begin{equation}\label{VP_NSF} 
\delta \int_{t _1 }^{ t _2 }\left(L_{g, R_{\rm ref}} ( \varphi , \dot \varphi , S)+\int_ \mathcal{B} ( S- \Sigma ) \dot \Gamma \mu _{G_{\rm ref}}\right) dt=0, \qquad \textsc{Variational Condition} 
\end{equation} 
with phenomenological and variational constraints 
\begin{align} 
\frac{\partial \mathfrak{L} }{\partial S}  \dot \Sigma &= -( \mathbf{P} ^{\rm fr})^{\flat_g}: \nabla ^g \dot \varphi +\mathbf{J} _S \cdot\mathbf{d} \dot \Gamma  - \rho _{\rm ref}R,&\;\;\textsc{Phenomenological Constraint} \label{KC_NSF}\\[2mm]
\frac{\partial \mathfrak{L} }{\partial S}  \delta  \Sigma &= -( \mathbf{P} ^{\rm fr})^{\flat_g}: \nabla ^g \delta  \varphi + \mathbf{J} _S\cdot\mathbf{d}  \delta  \Gamma, & \textsc{Variational Constraint}\label{VC_NSF} 
\end{align}
where $\mathbf{P} ^{\rm fr}(t,X)$ is a friction Piola-Kirchhoff tensor, $ \mathbf{J} _S(t,X)$ is an entropy flux density in material representation, and $\rho _{\rm ref}(X)R(t,X)$ is a heat power supply density in material representation.

Let us recall that this formalism means that when we compute the variation of the action functional in \eqref{VP_NSF}, we take any variations $ \delta \varphi , \delta S, \delta \Gamma , \delta \Sigma $ that satisfy the variational constraint \eqref{VC_NSF} and such that $ \delta \varphi $ and $\delta \Gamma$ vanish at the endpoints $(t=t_1,t_2)$. The curve $\varphi(t), S(t), \Gamma (t), \Sigma (t)$ around which the variations are taken has to satisfy the nonholonomic phenomenological constraint \eqref{KC_NSF}.  

\medskip

In the same way as the discrete case, making use of the variables $ \Gamma $ and $ \Sigma $ in the variational formalism provides a very clear and physically meaningful structure of nonequilibrium thermodynamics of continuum systems:

\begin{itemize}
\item 
The variational condition \eqref{VP_NSF} is an extension of Hamilton's principle \eqref{HP_principle} for fluid dynamics in material representation. It is essentially of geometric nature and only involves the knowledge of the configuration space and of the Lagrangian of the system.
\item 
The nonholonomic constraint \eqref{KC_NSF} is the expression of the thermodynamic power density associated to all the irreversible processes (heat transport and viscosity in our case) involved in the entropy production. This constraint is of phenomenological nature, in which each of the "thermodynamic affinities" is related to the thermodynamic fluxes characterizing an irreversible process via phenomenological laws; see Remark \ref{TP_NSF} and Remark \ref{forces_VS_velocity} below. The introduction of the variable $ \Gamma $ allows us to write this constraint as if a sum of force densities were acting on velocity fields, by analogy with classical mechanics; namely, $ \mathbf{P}^{\rm fr}$ "acting" on $ \frac{d}{dt}  \varphi $ and $ \mathbf{J} _S $ "acting" on $ \frac{d}{dt} \Gamma $, resulting in a \textit{power} or \textit{rate of work} density.
\item 
 Concerning the variational constraint \eqref{VC_NSF}, the occurrence of the time derivatives in \eqref{KC_NSF}, also allows us to systematically replace all velocities by "$ \delta $-derivatives", i.e., variational displacements and to formulate the variational constraint as a sum of virtual thermodynamic work densities associated to each of the irreversible processes. It is important to note that this interpretation is possible thanks to the introduction of the variable $ \Gamma(t,X)$ whose time derivative will be identified with the temperature $ \mathfrak{T}(t,X)$:
\begin{equation}\label{continuum_temp}
\frac{d}{dt} \Gamma = - \frac{\partial\mathfrak{L}}{\partial S}=:\mathfrak{T},
\end{equation}
from the stationarity condition associated with the variation $ \delta S$ in the variational formalism, as we will see in \eqref{delta_S}  below. Here we recall that $ \mathfrak{L}$ denotes the Lagrangian density as in \eqref{general_Lagrangian}. 
\end{itemize}

\paragraph{The equations of motion in material coordinates.}
Since we assumed no-slip boundary conditions, we have $ \delta \varphi |_{ \partial \mathcal{B} }=0$ and by computing the variations in \eqref{VP_NSF}, and using the instance of Stokes' theorem recalled in \eqref{generalized_Stokes_theorem}, we obtain
\[
\int_{t _1 }^{ t _2 } \int_ \mathcal{B} \left[ \left( \frac{\partial \mathfrak{L} }{\partial \varphi }- \operatorname{DIV} \frac{\partial \mathfrak{L} }{\partial T_X \varphi }- \frac{D}{Dt}\frac{\partial \mathfrak{L} }{\partial \dot \varphi }\right) \delta \varphi  + \frac{\partial \mathfrak{L}}{\partial S}\delta S-  (\dot S- \dot \Sigma ) \delta \Gamma + ( \delta S- \delta \Sigma )\dot \Gamma \right] \mu _{ G_{\rm ref}}dt=0.
\]
Using the variational constraint \eqref{VC_NSF} and the fact that $ \frac{\partial \mathfrak{L} }{\partial S}\neq 0$, and collecting the terms associated to the variations $ \delta  \varphi  $ and $ \delta \Gamma$, $ \delta S$, one obtains
\begin{align} 
\delta \varphi :&\;\;\rho _{\rm ref} \frac{D \mathbf{V} }{Dt}= \operatorname{DIV} \left(  \mathbf{P} ^{\rm cons}- \Gamma\left( \frac{\partial \mathfrak{L} }{\partial S}\right) ^{-1} \mathbf{P} ^{\rm fr} \right) + \rho _{\rm ref} \mathbf{B} ^{\rm cons}, \label{delta_phi} \\
\delta \Gamma  :&\;\; \dot S =\operatorname{DIV}\left( \Gamma\left( \frac{\partial \mathfrak{L} }{\partial S}\right) ^{-1} \mathbf{J} _S \right)  +\dot \Sigma, \label{delta_Gamma}\\
\delta S:&\;\;\dot \Gamma=-\frac{\partial \mathfrak{L} }{\partial S},\label{delta_S}
\end{align} \color{black} 
where we recall that $\mathbf{V} =\dot \varphi$,
\[
\mathbf{P} ^{\rm cons}:= - \left( \frac{\partial \mathfrak{L} }{\partial (T_X \varphi) }\right) ^{\sharp_g}= \left( \frac{\partial \mathfrak{E} }{\partial (T_X \varphi) }\right) ^{\sharp_g} \quad\text{and}\quad 
\rho _{\rm ref} \mathbf{B}^{\rm cons} =\frac{\partial \mathfrak{L} }{\partial \varphi }= -\rho _{\rm ref}( \mathbf{d}  \mathcal{V}\circ \varphi )^{\sharp_g}.
\]
Using the relations \eqref{delta_S} and \eqref{delta_Gamma} in the phenomenological constraint \eqref{KC_NSF} yields 
\[
\mathfrak{T}  (\dot S+  \operatorname{DIV} \mathbf{J} _S)  = ( \mathbf{P} ^{\rm fr})^{\flat_g}: \nabla ^g \dot \varphi - \mathbf{J} _S \cdot \mathbf{d}  \mathfrak{T}  +\rho _{\rm ref}R,
\]
where we recall that $
 \mathfrak{T}  := - \frac{\partial \mathfrak{L} }{\partial S}$ is the temperature in material representation as in \eqref{continuum_temp}. It is instructive to compare \eqref{delta_S}--\eqref{delta_Gamma}, with the corresponding relations \eqref{delta_SA}--\eqref{delta_gamma} for discrete systems.

If we impose the variation of $ \Gamma $ to vanish at the boundary, i.e., $ \delta \Gamma |_{ \partial \mathcal{B} }=0$, then there are no supplementary boundary conditions arising from the variational formalism. If, however, $ \delta \Gamma $ has no constraints on the boundary, then it implies the condition
\[
\mathbf{J} _S \cdot \mathbf{N}^{\flat_G} =0\quad\text{on}\quad \partial \mathcal{B} ,
\]
where $\mathbf{N}$ is the outward pointing unit normal vector field on $ \partial \mathcal{B} $ relative to $G_{\rm ref}$; namely, it implies that there is {\it no transfer of heat to the exterior}. In this case, it is also consistent with the assumption that {\it no heat sources coming from the exterior}, that is, $\rho _{\rm ref}R=0$ and therefore the fluid is \textit{adiabatically closed}. 
\medskip

Our results are summarized in the box below. 

\begin{framed}
\noindent\underline{\textsf{Variational formalism for the Navier-Stokes-Fourier system -- material representation:}}\\
In material representation, the Navier-Stokes-Fourier equations are given by\phantom{$\int_A^B$}
\begin{equation}\label{summary_NSF} 
\left\{
\begin{array}{l}
\vspace{0.2cm}\displaystyle\rho _{\rm ref} \frac{D \mathbf{V} }{Dt}= \operatorname{DIV}( \mathbf{P} ^{\rm cons}+ \mathbf{P} ^{\rm fr})+ \rho _{\rm ref} \mathbf{B} ^{\rm cons} ,\\[2mm]
\mathfrak{T}  (\dot S+  \operatorname{DIV} \mathbf{J} _S)  = ( \mathbf{P} ^{\rm fr})^{\flat_g}: \nabla ^g \mathbf{V}  - \mathbf{J} _S \cdot \mathbf{d} \mathfrak{T}  +\rho _{\rm ref}R,
\end{array}
\right.
\end{equation}
with no-slip boundary conditions $\mathbf{V} |_{ \partial \mathcal{B} }=0$. They follow from the variational condition \eqref{VP_NSF} with phenomenological and variational constraints \eqref{KC_NSF}, \eqref{VC_NSF}, where we impose $ \delta \Gamma |_{ \partial \mathcal{B} }=0$. If the constraint $ \delta \Gamma |_{ \partial \mathcal{B} }=0$ is not imposed, then the variational formalism yields the condition $ \mathbf{J} _S \cdot \mathbf{N}^{\flat_G}=0$ and if in addition $\rho _{\rm ref}R=0$, then the fluid is adiabatically closed. 
\end{framed}

These equations are not easy to handle because, being in material variables, they are written along the configuration $\varphi _t : \mathcal{B} \rightarrow \mathcal{S} $. We shall below write them in the usual spatial representation. However, we choose to consider the material representation first, since the variational formalism is simpler in material coordinates, which is natural in parallel with the one of the discrete systems. Of course, the variational formalism in spatial variables follows from reduction by symmetry of the variational formalism in material variables.

Exterior stress and body forces can be easily included in the variational picture, by augmenting \eqref{VP_NSF}  with the appropriate virtual force term (see the second term in \eqref{Forced_HP_elasticity}).

\begin{remark}[Interpretation of $ \Sigma $]\label{Interpretation_Sigma}{\rm 
Notice that  \eqref{delta_Gamma} is the {\it entropy balance equation}, namely,
\[
\dot S=- \operatorname{DIV} \mathbf{J} _S+\dot \Sigma,
\]
where $\dot \Sigma$ corresponds to the {\it total (internal and external) entropy production of the system} as
$$
\dot \Sigma= \frac{1}{\mathfrak{T}}\left[  ( \mathbf{P} ^{\rm fr})^{\flat_g}: \nabla  ^g \dot \varphi - \mathbf{J} _S \cdot \mathbf{d} \mathfrak{T}  +\rho _{\rm ref}R\right].
$$
In order to complete the system \eqref{summary_NSF} it is necessary to specify the phenomenological expressions for $ \mathbf{P} ^{\rm fr}$ and $\mathbf{J} _S$ in accordance with the second law of thermodynamics. Recall that the heat power supply density $\rho _{\rm ref}R$ is external, and then the second law, applied locally, requires the inequality that the {\it internal} entropy production density $I$ is positive: 
\begin{equation}\label{Clausius_Duhem_I} 
I=\frac{1}{\mathfrak{T}}\left[  ( \mathbf{P} ^{\rm fr})^{\flat_g}: \nabla^g \dot \varphi - \mathbf{J} _S \cdot \mathbf{d} \mathfrak{T}\right] \geq 0.
\end{equation} 

Using the expression $ \frac{d}{dt}\mathfrak{E}=(\mathbf{P} ^{\rm cons}+\mathbf{P} ^{\rm fr}) ^{\flat_g} : \nabla ^g \dot{\varphi} - \operatorname{DIV} \mathbf{J} _Q+ \rho _{\rm ref}R$ for the internal energy balance in material representation and replacing the heat power supply by using the entropy inequality $\dot S+ \operatorname{DIV}\mathbf{J} _ S \geq \frac{1}{\mathfrak{T}} \rho _{\rm ref}R$, (or, in Clausius form $\dot S\geq - \operatorname{DIV}\left( \frac{1}{\mathfrak{T}  } \mathbf{J} _ Q\right) +\frac{1}{\mathfrak{T}  } \rho _{\rm ref}R$, where $\mathbf{J} _ Q=\mathfrak{T} \mathbf{J} _ S$), we obtain a reformulation of \eqref{Clausius_Duhem_I} known as the {\bfi Clausius-Duhem inequality} (see, \cite{Lavenda1978}), namely
\begin{equation}\label{Clausius_Duhem_II} 
\dot S- \frac{1}{\mathfrak{T}}\dot{\mathfrak{E}}+ \frac{1}{\mathfrak{T}}(\mathbf{P} ^{\rm cons}+\mathbf{P} ^{\rm fr})^{\flat_g} : \nabla ^g \dot  \varphi - \frac{1}{\mathfrak{T}^2 }\mathbf{J} _Q \cdot \mathbf{d} \mathfrak{T}  \geq 0,
\end{equation} 
written here in material representation, which is a useful rewriting only if $R\neq 0$.
}
\end{remark}

\subsubsection{Spatial representation}\label{subsec_NSF_spatial} 

From a mathematical point of view, the passage from the material to the spatial description follows from the material relabeling symmetry through a Lagrangian reduction process. In this section, we shall implement such a Lagrangian reduction in the constrained variational formalism developed above in \S\ref{Mat_NSF}. 

One first observes that both the phenomenological and variational constraints \eqref{KC_NSF} and \eqref{VC_NSF} possesses the needed $ \operatorname{Diff}( \mathcal{B} )$-invariance to implement this reduction. The material Lagrangian is given by \eqref{general_Lagrangian} with internal energy \eqref{fluid_energy_pot}, which is also $ \operatorname{Diff}( \mathcal{B} )$-invariant. The corresponding spatial (or reduced) Lagrangian is provided by
\begin{equation}\label{spat_lagr_rho_s} 
\ell( \mathbf{v}, \rho , s)=\int_ \mathcal{S} \frac{1}{2}\rho  | \mathbf{v} | _g ^2 \mu _g -\int_ \mathcal{S} \varepsilon ( \rho , s) \mu _g.
\end{equation} 
Therefore, by using the relabeling symmetry, the reduced variables
\begin{equation}\label{reduced_variables_NSF} 
\mathbf{v} =\dot \varphi \circ \varphi ^{-1} , \quad \rho = \rho _{\rm ref} \circ \varphi ^{-1} J_ \varphi ^{-1} , \quad s =S\circ \varphi ^{-1} J_ \varphi ^{-1} ,\quad  \sigma  = \Sigma \circ \varphi ^{-1} J_ \varphi ^{-1} , \quad \gamma  = \Gamma  \circ \varphi ^{-1},
\end{equation} 
and the properties of the Piola transformation, the reduction of the variational formalism \eqref{VP_NSF} yields
\begin{equation}\label{VP_NSF_spatial}
\delta \int_{t _1 }^{ t _2 }\left( \ell( \mathbf{v} , \rho , s)+\int_ \mathcal{\mathcal{S} } (s- \sigma )( \partial _t \gamma + \mathbf{d} \gamma\cdot \mathbf{v} ) \mu _g \right) dt=0,
\end{equation}
together with the reduced phenomenological and variational constraints \eqref{KC_NSF} and \eqref{VC_NSF}, respectively, given by
\begin{equation}\label{KC_NSF_spatial}
\frac{\delta \ell}{\delta s} (\partial _t \sigma + \operatorname{div}( \sigma \mathbf{v} )) =-(\boldsymbol{\sigma} ^{\rm fr})^{\flat_g}: \nabla  \mathbf{v} +\mathbf{j} _S \cdot \mathbf{d} ( \partial _t \gamma + \mathbf{d} \gamma \cdot \mathbf{v} )-\rho r,
\end{equation} 
\begin{equation}\label{VC_NSF_spatial}
\frac{\delta \ell}{\delta s} (\delta \sigma + \operatorname{div}( \sigma \boldsymbol{\zeta}  ))=-(\boldsymbol{\sigma} ^{\rm fr})^{\flat_g}: \nabla \boldsymbol{\zeta}+\mathbf{j} _S \cdot \mathbf{d} (\delta  \gamma + \mathbf{d} \gamma \cdot \boldsymbol{\zeta} ),
\end{equation}
where $ \boldsymbol{\zeta} = \delta \varphi \circ \varphi ^{-1} $ and $\delta \gamma$ are arbitrary curves vanishing at the endpoints, the entropy flux $ \mathbf{j} _S$ and the friction stress $\boldsymbol{\sigma} ^{\rm fr}$ in spatial representation are defined as the inverse Piola transformation of $ \mathbf{J} _S$ and $ \mathbf{P} ^{\rm fr}$, that is,
\begin{equation}\label{def_EntFlux_FricStress}
\mathbf{j} _S:= (\varphi _\ast \mathbf{J} _S) J_ \varphi ^{-1}\quad\text{and}\quad \boldsymbol{\sigma} ^{\rm fr} (x)( \alpha _x ,  \beta_x):= J_\varphi^{-1} \mathbf{P}^{\rm fr} (X)(\alpha _x , T ^\ast _X\varphi (\beta _x)),
\end{equation}
and $r(x):=R( \varphi ^{-1} (x))$. From the definition of the spatial variables $ \mathbf{v} $ and $ \rho $, the variations $ \delta \mathbf{v} $ and $ \delta \rho $ are subjects to the constraints
\[
\delta \mathbf{v} = \partial _t \boldsymbol{\zeta}+[ \mathbf{v} , \boldsymbol{\zeta} ], \quad \delta \rho =- \operatorname{div}( \rho \boldsymbol{\zeta} ).
\]

Taking the variations of the action integral \eqref{VP_NSF_spatial} and using the divergence theorem with the boundary condition $ \mathbf{v}|_{\partial \mathcal{S}}=0$, we have
\begin{align*} 
&\int_{ t _0 }^{ t _1 }\int_ \mathcal{S} \left( \frac{\delta \ell}{\delta \mathbf{v} } \cdot (\partial _t \boldsymbol{\zeta} + [ \boldsymbol{\zeta} , \mathbf{v} ]) -\frac{\delta \ell}{\delta \rho  }\operatorname{div}( \rho \boldsymbol{\zeta} )  +\frac{\delta \ell}{\delta s  }\delta s- [ \partial _t (s- \sigma )+ \operatorname{div}((s- \sigma ) \mathbf{v} ) ]\delta  \gamma \right.\\
&\qquad \qquad \qquad \qquad \quad \left.\phantom{\int_{ t _0 }^{ t _1 }} +( \delta s- \delta \sigma )( \partial _t \gamma + \mathbf{d} \gamma \cdot \mathbf{v} )+(s- \sigma ) \mathbf{d} \gamma \cdot (\partial _t \boldsymbol{\zeta} + [ \boldsymbol{\zeta} , \mathbf{v} ])\right) \mu _g=0.
\end{align*} 
Using  the divergence theorem with the boundary condition $ \boldsymbol{\zeta} |_{ \partial \mathcal{S} }=0$, the variational constraint \eqref{VC_NSF_spatial}, the fact that $ \frac{ \delta \ell }{\delta  s}\neq 0$, and collecting the terms associated to the variations $\boldsymbol{\zeta} $, $ \delta\gamma $, $ \delta s$, we obtain
\begin{align} 
\boldsymbol{\zeta}  :&\;\;( \partial _t + \operatorname{ad}^*_ \mathbf{v} )\left( \frac{\delta \ell}{\delta \mathbf{v} }+( s- \sigma ) \mathbf{d} \gamma \right) -\rho  \,\mathbf{d} \frac{\delta \ell}{\delta \rho }+ \sigma \mathbf{d} ( \partial _t \gamma + \mathbf{d} \gamma \cdot \mathbf{v} ) \label{zeta_NSF_spatial}\\
& \;\; + \operatorname{div}\left( (\partial _t \gamma + \mathbf{d}\gamma \cdot \mathbf{v} ) \left( \frac{\delta \ell}{\delta s} \right)  ^{-1}\!\!\!\boldsymbol{\sigma} ^{\rm fr} \right) - \operatorname{div}\left( (\partial _t \gamma + \mathbf{d}\gamma \cdot \mathbf{v} ) \left( \frac{\delta \ell}{\delta s} \right)  ^{-1}\!\!\!\mathbf{j} _S \right) \cdot \mathbf{d} \gamma =0, \nonumber  \\
\delta \gamma  :&\;\;\partial _t ( s- \sigma )+ \operatorname{div}((s- \sigma ) \mathbf{v} ) =\operatorname{div}\left( (\partial _t \gamma + \mathbf{d}\gamma \cdot \mathbf{v} ) \left( \frac{\delta \ell}{\delta s} \right)  ^{-1}\mathbf{j} _S\right) ,\label{delta_gamma_NSF_spatial}\\
\delta s :&\;\; \partial _t \gamma + \mathbf{d} \gamma \cdot \mathbf{v} =-\frac{\delta \ell}{\delta s}.\label{delta_s_NSF_spatial}
\end{align}
where we assumed $ \delta \gamma |_{ \partial \mathcal{S} }=0$ and where $ \operatorname{ad}^*_ \mathbf{v} \mathbf{m} $ (the coadjoint operator) denotes the dual operation 
to the Lie bracket and reads $ \operatorname{ad}^*_ \mathbf{v} \mathbf{m} =  \mathbf{v} \cdot \nabla \mathbf{m} + \nabla \mathbf{v} ^\mathsf{T} \mathbf{m} + \mathbf{m} \operatorname{div} \mathbf{v} $. 
\color{black}
Using \eqref{delta_s_NSF_spatial} and \eqref{delta_gamma_NSF_spatial} in the phenomenological constraint \eqref{KC_NSF_spatial} leads to
\begin{equation*}
T(\partial _t s+ \operatorname{div}(s\mathbf{v}) +\operatorname{div} \mathbf{j} _S)=(\boldsymbol{\sigma} ^{\rm fr})^{\flat_g}: \nabla  \mathbf{v} -\mathbf{j} _S \cdot \mathbf{d} T \cdot \mathbf{v} +\rho r,
\end{equation*}
where $T:=-\frac{\partial \ell}{\partial s}$ is the temperature in spatial representation.

If $ \delta \gamma |_{ \partial \mathcal{S} }$ is free, then we obtain the condition
\begin{equation}\label{adiabatically_closed_cond} 
\mathbf{j} _S \cdot \mathbf{n} ^{\flat _g}=0 \quad\text{on}\quad \partial \mathcal{S},
\end{equation} 
where $\mathbf{n}$ is the outward pointing unit normal vector field on $ \partial \mathcal{S} $ relative to $g$, that is, there is no transport of heat to the exterior. In this case, it is also consistent with the assumption that there exist no heat sources coming from the exterior, namely, $\rho r=0$, in which case the fluid is adiabatically closed.  

\medskip

Our results are summarized in the following box.
\begin{framed}
\noindent\underline{\textsf{Variational formalism for the Navier-Stokes-Fourier system -- spatial representation:}}\phantom{$\frac{1}{\int_{A_A}}$}\\
In spatial representation, the Navier-Stokes-Fourier equations are given by
\begin{equation}\label{NSF_spatial} 
\left\{
\begin{array}{l}
\vspace{0.2cm}\rho (\partial _t \mathbf{v} + \mathbf{v}  \cdot \nabla \mathbf{v})=- \operatorname{grad} p +\operatorname{div} \boldsymbol{\sigma} ^{\rm fr}, \quad p=  \frac{\partial \varepsilon}{\partial \rho }\rho + \frac{\partial \varepsilon }{\partial s}s - \varepsilon, \\[2mm]
\vspace{0.2cm}\partial _t \rho + \operatorname{div}( \rho  \mathbf{v} )=0,\\[2mm]
T( \partial _t s+ \operatorname{div}( s \mathbf{v} )+ \operatorname{div} \mathbf{j} _S )= ( \boldsymbol{\sigma}  ^{\rm fr} )^{\flat_g} : \nabla  \mathbf{v}  - \mathbf{j} _S \cdot \mathbf{d} T +  \rho r,\quad T= \frac{\partial \varepsilon }{\partial s}.
\end{array}
\right.
\end{equation}
These equations arise from the variational condition  \eqref{VP_NSF_spatial}, where $ \mathbf{v}, \rho , s, \gamma , \sigma $ satisfy the nonholonomic constraint \eqref{KC_NSF_spatial} and for variations $ \delta \mathbf{v} = \partial _t \boldsymbol{\zeta} + [ \boldsymbol{\zeta} , \mathbf{v} ]$, $ \delta \rho  =- \operatorname{div}( \rho \boldsymbol{\zeta} )$, $ \delta s$, $ \delta \gamma $, $ \delta \sigma $, such that $ \boldsymbol{\zeta} $, $ \delta \sigma $ and $ \delta \gamma $ verify the variational constraint \eqref{VC_NSF_spatial} and where we impose $ \delta \gamma |_{ \partial \mathcal{S} }=0$. Direct computations using \eqref{delta_s_NSF_spatial} and \eqref{delta_gamma_NSF_spatial} in \eqref{zeta_NSF_spatial} and in the nonholonomic constraint \eqref{KC_NSF_spatial} leads to \eqref{NSF_spatial}. If the constraint $ \delta \gamma  |_{ \partial \mathcal{S} }=0$ is not imposed, then the variational formalism yields the condition $\mathbf{j} _S \cdot \mathbf{n} ^{\flat_g}=0$ and if in addition $\rho\, r=0$, then the fluid is adiabatically closed.
\end{framed}
 
\begin{remark}[Interpretation of $ \sigma $]\label{interpretation_sigma}{\rm The Lagrangian time derivative of the variables $ \sigma $ corresponds to the total entropy production density of the system. The entropy balance equation \eqref{delta_gamma_NSF_spatial} is consistent with equation (10) of \S3.1 in  \cite{deGrootMazur1969} (note that their notation of the entropy production ``$\sigma$'' corresponds to our notation ``$i$''). Indeed, from \eqref{delta_gamma_NSF_spatial} we have 
\[
\partial _t \sigma + \operatorname{div}( \sigma \mathbf{v} )=  \frac{1}{T} \left(( \boldsymbol{\sigma}  ^{\rm fr} )^{\flat_g}: \nabla \mathbf{v}   - \mathbf{j} _S \cdot \mathbf{d} T+\rho r\right)=i+ \frac{r \rho }{T},
\]
where $i$ is the internal entropy production. As in Remark \ref{Interpretation_Sigma}, in order to complete the system \eqref{NSF_spatial}, it is necessary to specify the phenomenological expressions for $ \boldsymbol{\sigma} ^{\rm fr}$ and $ \mathbf{j} _S$ in accordance with the second law of thermodynamics. 
The second law, applied locally, requires that these expressions are such that the internal entropy production density is positive, i.e., $i  \geq 0$. See also Remark \ref{TP_NSF} for the phenomenological expressions. Concerning the total entropy $\mathsf{S}(t)= \int_ \mathcal{S} s(t,x) \mu _g(x)$, we have
\begin{equation}\label{total_entropy_multi_comp} 
\frac{d}{dt} \mathsf{S}(t) =- \int_{ \partial \mathcal{S} }\mathbf{j} _S \cdot \mathbf{n} ^{\flat_g}\mu _g^\partial +\int_ \mathcal{S}  i \mu _g +\int_ \mathcal{S}\frac{1}{T}r \rho \mu _g \geq - \int_{ \partial \mathcal{S} }\frac{\mathbf{j} _Q \cdot \mathbf{n} ^{\flat_g}}{T}\mu _g^\partial+\int_ \mathcal{S}\frac{r \rho }{T}\mu _g.
\end{equation} 
}
\end{remark}

\begin{remark}[Energy balance]{\rm For completeness, we write here the energy balance in an intrinsic expression.
The total energy density $e:= \frac{1}{2} \rho | \mathbf{v} | ^2_g  + \varepsilon ( \rho , s)$ satisfies the equation
\[
\partial _t e+ \operatorname{div}( e \mathbf{v} ) =\operatorname{div}(\boldsymbol{\sigma} ^{\flat_g}\cdot \mathbf{v} )- \operatorname{div} \mathbf{j} _Q +\rho r,
\]
(compare with \eqref{energy_density_behavior}), where $ \boldsymbol{\sigma}: =-p g^\sharp+ \boldsymbol{\sigma} ^{\rm fr}$, and $ \mathbf{j} _Q= T \mathbf{j} _S$ is the heat flux, so that the total energy $\mathsf{E}(t)=\int_ \mathcal{S} e(t,x) \mu _g(x) $ verifies the balance law
\[
\frac{d}{dt} \mathsf{E}(t)= -\int_{\partial \mathcal{S} } \mathbf{j} _Q \cdot \mathbf{n} ^{\flat_g}\mu _g^\partial +\int_ \mathcal{S} \rho r\mu _g =P^{\rm ext}_H(t),
\]
where we recall that $\mu _g $ denotes the volume form associated to the Riemannian metric $g$ and $ \mu _g ^ \partial $ denotes the volume form induced by $ \mu _g $ on the boundary $ \partial \mathcal{S} $.
We consistently have $\frac{d}{dt} E= 0$ in the adiabatically closed case $\mathbf{j} _Q \cdot \mathbf{n}^{\flat_g}=0$ (see \eqref{adiabatically_closed_cond}) and $ \rho r=0$.}
\end{remark}

\begin{remark}[General Lagrangian]{\rm To obtain \eqref{NSF_spatial}, we have used the explicit expression \eqref{spat_lagr_rho_s} of the Lagrangian of an ideal compressible fluid. For a general Lagrangian $\ell( \mathbf{v}, \rho , s)$, we obtain the system
\begin{equation*}\label{general_NSF_spatial} 
\left\{
\begin{array}{l}
\displaystyle\vspace{0.2cm}\left( \partial _t + \operatorname{ad}^*_ \mathbf{v} \right) \frac{\delta \ell}{\delta \mathbf{v} }  =\rho\,\mathbf{d}  \frac{\delta \ell}{\delta \rho }+s \,\mathbf{d} \frac{\delta \ell}{\delta s}   +\operatorname{div} \boldsymbol{\sigma} ^{\rm fr},\qquad \partial _t \rho + \operatorname{div}( \rho  \mathbf{v} )=0,\\[2mm]
\displaystyle\frac{\delta \ell}{\delta s} ( \partial _t s+ \operatorname{div}( s \mathbf{v} )+ \operatorname{div} \mathbf{j} _S )= -( \boldsymbol{\sigma}  ^{\rm fr} )^{\flat_g} :\nabla  \mathbf{v}  - \mathbf{j} _S \cdot \mathbf{d} \frac{\delta \ell}{\delta s}- r \rho .
\end{array}
\right.
\end{equation*}}
\end{remark}

\begin{remark}[Thermodynamic phenomenology]\label{TP_NSF}{\rm As mentioned earlier, in order to determine from \eqref{NSF_spatial} the time evolution of all the fields, it is necessary to provide phenomenological expressions for $ \boldsymbol{\sigma} ^{\rm fr}$ and $ \mathbf{j} _S$ (which are examples of the so-called \textit{fluxes}) in terms of $ \operatorname{Def} \mathbf{v} $ and $ \mathbf{d} T$ (which are examples of the so-called \textit{affinities}) that are compatible with the second law $i \geq 0$. In the present example, we have the well-known relations

\begin{equation*}\label{friction_stress_NSF} 
\boldsymbol{\sigma} ^{\rm fr}=2 \mu  (\operatorname{Def} \mathbf{v})^{\sharp _g }+ \left( \zeta - \frac{2}{3}\mu \right)(\operatorname{div} \mathbf{v} )  g^\sharp\quad\text{and}\quad T\mathbf{j} _S^{\flat_g}= - \kappa \mathbf{d}  T \;\; \text{(Fourier law)},
\end{equation*} 
where $ \mu \geq 0 $ is the first coefficient of viscosity (shear viscosity), $ \zeta \geq 0$ is the second coefficient of viscosity (bulk viscosity), and $ \kappa \geq 0$ is the conductivity.}
\end{remark}

\begin{remark}{\rm
Another variational approach to the Navier-Stokes-Fourier system has been developed by \cite{FuFu2012}, in which the internal conversion due to frictional forces from the mechanical to the heat power in dissipated systems is written as nonholonomic constraints in the integral form over the fluid domain. However, this variational approach does not involve the variables $ \Gamma $ nor $ \Sigma $. Therefore, in their setting there is no systematic way to pass from a phenomenological constraint to a variational constraint, although it is quite essential when one consider several irreversible processes such as matter transfer and chemical reactions in addition to viscosity (see \S\ref{sec_multi_comp}). 

In our approach, we have established a variational formalism that enables to incorporate nonlinear nonholonomic constraints associated with nonequilibrium thermodynamics into the Lagrangian setting by using a generalized Lagrange-d'Alembert principle, and we have shown a systematic way to pass from the phenomenological constraint to the variational constraint  by simply replacing the rate of each of the irreversible processes (e.g., $ \dot \Gamma $ here) by the associated variations (e.g., $ \delta \Gamma $ here). Furthermore, we have also shown a systematic procedure to pass from the material representation to the spatial representation through a Lagrangian reduction procedure based on relabeling symmetry.
}
\end{remark}

\subsection{Variational formalism for  multicomponent reacting viscous fluids}\label{sec_multi_comp} 

In this section, we will extend the variational formalism for nonequilibrium thermodynamics to the case of a multicomponent fluid, which is subject to the irreversible processes associated to {\it viscosity, heat transport, internal matter transport as well as chemical reactions}. We will consider a fluid of $K$ chemical components $A=1,...,K$ amongst which $r$ chemical reactions $a=1,...,r$ are possible. 

\subsubsection{Material representation}\label{mat_repr_muticomp} 

We shall show that, despite the complexity of the system of equations for the nonequilibrium thermodynamics of a multicomponent reacting fluid, the Lagrangian variational formalism in material representation has still the same structure described in the previous examples. Again, the variational formalism in spatial representation is much more involved, however, it will be clearly explained by a reduction process associated to the relabeling symmetry applied to the material variational formalism.
\medskip

Recall from example 2 in \S\ref{VP_no_therm_mat}, that in absence of all the irreversible processes, the equations for a fluid with $K$ chemical components $A=1,...,K$, follow from Hamilton's principle applied to a Lagrangian of the form $L_{g, \{N_{A\,\rm ref}\}, S_{\rm ref}}( \varphi , \dot \varphi )$, where $N_{A\,\rm ref}(X)$ denotes the number density of the substance $A=1,...,K$ in the reference configuration and $S_{\rm ref}(X)$ is the entropy density in the reference configuration. To these fixed fields are associated the continuity equations $ \partial _t n_A + \operatorname{div}( n_A \mathbf{v} )=0$ and $ \partial _t s+ \operatorname{div}(s \mathbf{v} )=0$ in the spatial representation. 

When irreversible processes are included, not only the entropy continuity equation is modified but also the continuity equations $\partial _t n_A + \operatorname{div}( n_A \mathbf{v} )=0$ for the substances $A=1,...,K$ are modified, in which an additional current density and source density occur due to $r$ chemical reactions $a=1,...,r$ amongst the $K$ substances. As a consequence, the number densities $N_{A\,\rm ref}(X)$ are no more fixed even in the material representation, but they become {\it dynamic} variables $N_A(t,X)$.

Accordingly, for the present case, we will make the following change in the notation of the Lagrangian:
\[
L_{g, \{N_{A\,\rm ref}\}, S_{\rm ref}}( \varphi , \dot \varphi ) \leadsto L_{g}( \varphi , \dot \varphi ,\{N_A\}, S).
\]
The occurrence of the variables $N_A$ must be accompanied with their corresponding conjugate variables $W ^A $ whose variation $ \delta W ^A $ would ensure $ \frac{d}{dt} N_A=0$ in the reversible case. This would amount to add the term $\int_ \mathcal{B} N_A \dot W^A \mu _{G_{\rm ref}}$ in the action functional. In the irreversible case, however, the equation $ \frac{d}{dt} N_A=0$ has to be broken because of the additional current density and source density occurring due to $r$ chemical reactions $a=1,...,r$ amongst the $K$ substances. This can be done by regarding $\dot W ^A $ as a \textit{rate} of  $W ^A $ associated to the irreversible processes of internal matter transfer, similarly to the velocity $\dot \Gamma $ associated to heat transfer in the preceding section. We assume that there is no matter transfer with the exterior. Based on these comments, we can now establish an appropriate variational formalism.

\paragraph{Variational formalism for the nonequilibrium thermodynamics of a reacting multicomponent fluid.} The continuum version of the variational formalism in Definition \ref{LdA_def_discrete_Systems} adapted to the continuum case and to the setting of chemical reactions in \S\ref{DCR} (second version) is given by
\begin{align} \label{VP_multi_comp} 
&\delta \int_{t _1 }^{ t _2 }\left(L_g ( \varphi , \dot \varphi , \{N_A\},S)+ \int_ \mathcal{B}N_A \dot W^A \mu _{G_{\rm ref}} + \int_ \mathcal{B} ( S- \Sigma )\dot \Gamma \mu _{G_{\rm ref}}\right) dt=0\nonumber\\[2mm]
&\hspace{8.5cm}\textsc{Variational Condition}
\end{align}
with nonholonomic phenomenological and variational constraints
\begin{align} 
\frac{\partial \mathfrak{L}}{\partial S} \dot \Sigma &= -( \mathbf{P} ^{\rm fr})^{\flat_g}: \nabla  \dot \varphi +\mathbf{J} _S \cdot\mathbf{d} \dot \Gamma + \mathbf{J} _A \cdot \mathbf{d} \dot W ^A + J _a \dot \nu ^a - \rho _{\rm ref}R,  &\textsc{Phen. Constr.} \label{KC_multi_comp}\\[2mm]
\dot\nu ^a &= \nu ^a_{A} \dot W ^A, \qquad a=1,...,r,\label{chemical_constraint_K}  &\textsc{Chem. Constr.}\\[2mm]
\frac{\partial \mathfrak{L}}{\partial S} \delta  \Sigma &= -( \mathbf{P} ^{\rm fr})^{\flat_g}: \nabla  \delta  \varphi + \mathbf{J} _S\cdot\mathbf{d}  \delta  \Gamma +\mathbf{J} _A \cdot \mathbf{d} \delta  W ^A  + J_a \delta  \nu ^a, &\textsc{Var. Constr.}\label{VC_multi_comp} \\[2mm]
\delta \nu ^a &= \nu ^a_A \delta  W ^A , \qquad a=1,...,r,\label{chemical_constraint_V}&\textsc{Var. Constr.}
\end{align}
where $ \nu ^a_{A}:= \nu ^{''a}_{A}- \nu ^{'a}_{A}$ as in \S\ref{DCR}, and verify $\nu ^a_{A}M^A=0$ (Lavoisier law) meaning that mass is conserved in each separate chemical reaction $a=1,...,r$. In \eqref{KC_multi_comp} and \eqref{VC_multi_comp}, $\mathbf{P} ^{\rm fr}(t,X)$ is the Piola-Kirchhoff friction stress tensor, $ \mathbf{J} _S(t,X)$ is the entropy flux density, $ \mathbf{J} _A(t,X)$ is the diffusive flux density of substance $A$, $J_a(t,X)$ is the chemical reaction rate of reaction $a$, and $\rho _{\rm ref}(X)R(t,X)$ is the heat power supply density, all of which are expressed in material representation.
In all these formulas, Einstein's summation convention is used. The variations $\delta \varphi $, $ \delta W ^A $, and $ \delta \Gamma $ are assumed to vanish at the endpoint $(t=t _1 , t _2 )$. Total mass is conserved by assuming that the fluxes $ \mathbf{J} _A$ verify
\begin{equation*}\label{total_mass_conservation} 
M^A \mathbf{J} _A(t,X)=0, \quad \text{for all $A=1,...,N$.} 
\end{equation*} 

In a similar way to the simpler case of \S\ref{VHCF}, the introduction of the variables $ \Gamma $, $ \Sigma $, and $ W ^A $ allows us to propose a variational formalism with the same clear and physically meaningful structure:

\begin{itemize}
\item The variational condition \eqref{VP_multi_comp} is an extension of Hamilton's principle \eqref{HP_principle} for fluid dynamics in material representation. We note that the whole expression under the temporal integral can be interpreted as a Lagrangian $\bar L$ defined on the tangent bundle of the configuration space $Q= \operatorname{Emb}( \mathcal{B} , \mathcal{S} ) \times  [\mathcal{F} ( \mathcal{B} ) \times \mathcal{F} ( \mathcal{B} ) ^\ast ]^K\times \mathcal{F} ( \mathcal{B} ) \times \mathcal{F} ( \mathcal{B} ) ^\ast \times \mathcal{F} ( \mathcal{B} ) ^\ast$, i.e., 
\[
\bar L=\bar L ( \Psi , \dot \Psi ): T Q \rightarrow \mathbb{R},
\]
where $\Psi =( \varphi ,\{ W ^A \}, \{N _A \}, \Gamma , \Sigma , S) \in Q$, $\mathcal{F} ( \mathcal{B} )$ is the space of real valued functions on $ \mathcal{B} $, $ \mathcal{F} ( \mathcal{B} ) ^\ast $ is its dual space of densities on $ \mathcal{B} $, and $[\cdots]^{K}$ denotes the direct product of $k$ copies of $[\cdots]$.

\item  The phenomenological constraint \eqref{KC_multi_comp} is the expression of the thermodynamic power density associated to all the irreversible processes involved: viscosity, heat transport, matter transport, chemical reactions, all of which are characterized by the thermodynamic fluxes $J_{\alpha}$ (as if it were  forces in mechanics) acting on the thermodynamic affinities $X^{\alpha}$ regarded as the rates $\dot{\gamma}^{\alpha}$ (as if they were velocities in mechanics); namely $ \mathbf{P}^{\rm fr}$ "acting" on $ \frac{d}{dt}  \varphi $ and $ \mathbf{J} _S $ "acting" on $\frac{d}{dt} \Gamma $, $ \mathbf{J} _A$ "acting" on $\frac{d}{dt}  W ^A $ and $J _a $ "acting" on $\frac{d}{dt}  \nu ^a $, resulting in a \textit{power} or \textit{rate of work} density.

\item The variational constraint \eqref{VC_multi_comp} can be obtained by replacing velocities by virtual displacements associated to each of the irreversible processes, thereby expressing the variational constraint as a sum of virtual thermodynamic work densities. This interpretation is possible thanks to the introduction of the variables $ \Gamma(t,X)$, $ W ^A (t,X)$, and $ \nu_a (t,X)$ whose time derivative will be identified with the temperature $ \mathfrak{T}(t,X)$, the chemical potential $\Upsilon  ^A (t,X)$, and the affinity $\Lambda  _a(t,X) $, respectively, all written in material representation.
\end{itemize}

\paragraph{The equations of motion in material coordinates.}
Since we assumed no-slip boundary conditions, we have $ \delta \varphi |_{ \partial \mathcal{B} }=0$ and by computing the variations in \eqref{VP_multi_comp}, we obtain
\begin{align*} 
&\int_{t _1 }^{ t _2 } \int_ \mathcal{B} \left[ \left( \frac{\partial \mathfrak{L} }{\partial \varphi }- \operatorname{DIV} \frac{\partial \mathfrak{L} }{\partial T_X \varphi }- \frac{D}{Dt}\frac{\partial \mathfrak{L} }{\partial \dot \varphi }\right) \delta \varphi  + \frac{\partial \mathfrak{L}}{\partial N _A }\delta N _A + \frac{\partial \mathfrak{L}}{\partial S}\delta S \right .\\
& \qquad \qquad \left. \phantom{\int_{t _1 }^{ t _2 } \int_ \mathcal{B}} +\delta N _A \dot W ^A  - \dot N _A \delta W ^A -  (\dot S- \dot \Sigma ) \delta \Gamma + ( \delta S- \delta \Sigma )\dot \Gamma \right] \mu _{ G_{\rm ref}}dt=0,
\end{align*}
where we recall that $ \mathfrak{L}$ denotes the Lagrangian density as in \eqref{general_Lagrangian}. 

Using the variational constraints \eqref{VC_multi_comp}, \eqref{chemical_constraint_V} and the fact that $ \frac{\partial \mathfrak{L} }{\partial S}\neq 0$, and collecting the terms associated to the variations $ \delta  \varphi $, $ \delta W ^A $, $ \delta N_A$, $ \delta \Gamma$, and $ \delta S$, one obtains
\begin{align} 
\delta \varphi :&\;\;\rho _{\rm ref} \frac{D \mathbf{V} }{Dt}= \operatorname{DIV} \left(  \mathbf{P} ^{\rm cons}- \dot\Gamma\left( \frac{\partial \mathfrak{L} }{\partial S}\right) ^{-1} \mathbf{P} ^{\rm fr} \right) + \rho _{\rm ref} \mathbf{B} ^{\rm cons},\nonumber\\[2mm]\delta N_A: & \;\; \dot W ^A=-\frac{\partial \mathfrak{L} }{\partial N_A},\label{delta_N_multi_comp} \\[2mm]
\delta W^A: & \;\; \dot N_A=\operatorname{DIV}\left( \dot\Gamma\left( \frac{\partial \mathfrak{L} }{\partial S}\right) ^{-1} \mathbf{J} _A \right) - \dot\Gamma\left( \frac{\partial \mathfrak{L} }{\partial S}\right) ^{-1} J _a \nu ^a_{A},\nonumber\\[2mm]
\delta \Gamma  :&\;\; \dot S =\operatorname{DIV}\left( \dot\Gamma\left( \frac{\partial \mathfrak{L} }{\partial S}\right) ^{-1} \mathbf{J} _S \right)  +\dot \Sigma, \label{delta_Gamma_multi_comp}\\
\delta  S: & \;\;\dot \Gamma = - \frac{\partial \mathfrak{L} }{\partial S},\label{delta_S_multi_comp}
\end{align}  \color{black} 
where $\mathbf{V} $, $\mathbf{P} ^{\rm cons}$ and $ \mathbf{B}^{\rm cons} $ are defined as before. It is instructive to compare these conditions with those obtained for the chemical reactions in the discrete case in \eqref{conditions_second_version}. Using the relations \eqref{delta_N_multi_comp}, \eqref{delta_Gamma_multi_comp}, \eqref{delta_S_multi_comp}, and the chemical constraint \eqref{chemical_constraint_K} in the phenomenological constraint \eqref{KC_multi_comp} yields 
\[
\mathfrak{T}  (\dot S+  \operatorname{DIV} \mathbf{J} _S)  = ( \mathbf{P} ^{\rm fr})^{\flat_g}: \nabla ^g \dot \varphi - \mathbf{J} _S \cdot \mathbf{d}  \mathfrak{T} -\mathbf{J} _A \cdot \mathbf{d} \Upsilon ^A   +J _a\Lambda ^a +\rho _{\rm ref}R,
\]
where we recall that $ \mathfrak{T}  := - \frac{\partial \mathfrak{L} }{\partial S}$ is the temperature, $ \Upsilon ^A:= - \frac{\partial \mathfrak{L}}{\partial N _A }$ is the chemical potential, and $ \Lambda ^a:=- \nu ^a_{A} \Upsilon ^A $ is the affinity, all in material representation.
\medskip

If we impose the variation of $ \Gamma $ and $ W ^A$ to vanish at the boundary, i.e., $ \delta \Gamma |_{ \partial \mathcal{B} }= \delta W ^A |_{ \partial \mathcal{B} }=0$, then there are no supplementary boundary conditions arising from the variational formalism. If, however, $ \delta \Gamma $, respectively, $ \delta W ^A $ has no constraints on the boundary, then it implies the condition
\[
\mathbf{J} _S \cdot \mathbf{N}^{\flat_G} =0,\quad\text{respectively,}\quad \mathbf{J} _A \cdot \mathbf{N}^{\flat_G} =0\quad\text{on}\quad \partial \mathcal{B} ,
\]
where $\mathbf{N}$ is the outward pointing unit normal vector field on $ \partial \mathcal{B} $ relative to $G_{\rm ref}$, that is, there is no transfer of heat or matter to the exterior. In this case, if $ \rho _{\rm ref}R=0$, the fluid is \textit{adiabatically closed}. 
\medskip

Our results are summarized in the box below. 

\begin{framed}
\noindent\underline{\textsf{Variational formalism for multicomponent reacting fluids -- material representation:}}\\
In material representation, the evolution equations are given by\phantom{$\int_A^B$}
\begin{equation}\label{box_multi_comp} 
\left\{
\begin{array}{l}
\vspace{0.2cm}\displaystyle\rho _{\rm ref} \frac{D \mathbf{V} }{Dt}= \operatorname{DIV}( \mathbf{P} ^{\rm cons}+ \mathbf{P} ^{\rm fr})+ \rho _{\rm ref} \mathbf{B} ^{\rm cons}, \\[2mm]
\vspace{0.2cm}\displaystyle\dot N_A+ \operatorname{DIV} \mathbf{J} _A = J _a \nu ^a_{A},\\[2mm]
\mathfrak{T}  (\dot S+  \operatorname{DIV} \mathbf{J} _S)  = ( \mathbf{P} ^{\rm fr})^{\flat_g}: \nabla  ^g\mathbf{V}  - \mathbf{J} _S \cdot \mathbf{d}  \mathfrak{T} -\mathbf{J} _A \cdot \mathbf{d} \Upsilon ^A   +J _a\Lambda ^a +\rho _{\rm ref}R,
\end{array}
\right.
\end{equation}
with no-slip boundary conditions $ \mathbf{V} |_{ \partial \mathcal{B} }=0$. These equations are obtained from the variational condition for nonequilibrium thermodynamics  \eqref{VP_multi_comp} with the phenomenological and chemical constraints \eqref{KC_multi_comp} and \eqref{chemical_constraint_K}, together with their associated variational constraints \eqref{VC_multi_comp} and \eqref{chemical_constraint_V}, where $ \delta \Gamma |_{ \partial \mathcal{B} }=\delta W ^A  |_{ \partial \mathcal{B} }=0$. If the constraint $\delta W ^A |_{ \partial \mathcal{B} }=0$ and $\delta \Gamma |_{ \partial \mathcal{B} }=0$ are not imposed, then the variational formalism yields $ \mathbf{J} _S \cdot \mathbf{N}^{\flat_G}=0$ and $ \mathbf{J} _A \cdot \mathbf{N}^{\flat_G}=0$ and if, in addition, $\rho _{\rm ref}R=0$, then the fluid is adiabatically closed.
\end{framed}

From the second equation in \eqref{box_multi_comp} and the mass conservation conditions $M^A \mathbf{J} _A=0$ and $ M^A \nu^a _{A}=0$, it follows that the \textit{total mass density} $\mathfrak{R}:= M^AN_A$ is conserved as $\dot{\mathfrak{R}}=0$, where $M^A$ is the molar mass of the chemical component $A$.

External stress and body forces can be easily included in the variational picture, by augmenting \eqref{VP_multi_comp}  with the appropriate virtual work term (second term in \eqref{Forced_HP_elasticity}). In this case, the first equation above reads
\[
\rho _{\rm ref} \frac{D \mathbf{V} }{Dt}= \operatorname{DIV}( \mathbf{P} ^{\rm cons}+ \mathbf{P} ^{\rm fr}+ \mathbf{P} ^{\rm ext})+ \rho _{\rm ref} (\mathbf{B} ^{\rm cons}+ \mathbf{B} ^{\rm ext}).
\]

\begin{remark}[Interpretation of $ \Sigma $]
\normalfont
Recall from \eqref{delta_Gamma_multi_comp} that the entropy balance equation is written as
\[
\dot S=- \operatorname{DIV} \mathbf{J} _S+\dot \Sigma,
\]
where the total (internal and external) entropy production of the system is given by
\[
\dot \Sigma=  \frac{1}{\mathfrak{T}}\left[   ( \mathbf{P} ^{\rm fr})^{\flat_g}: \nabla ^g \dot \varphi - \mathbf{J} _S \cdot \mathbf{d}  \mathfrak{T} -\mathbf{J} _A \cdot \mathbf{d} \Upsilon ^A   +J _a\Lambda ^a +\rho _{\rm ref}R\right].
\]
In order to complete the system \eqref{box_multi_comp} it is necessary to specify the phenomenological expressions for $ \mathbf{P} ^{\rm fr}$,  $\mathbf{J} _S$, $ \mathbf{J} _A$, and $J ^a$ in accordance with the second law of thermodynamics. The second law, applied locally, imposes that these expressions are such that the internal entropy production density $I$ be positive 
\begin{equation}\label{Clausius_Duhem_I_multi_comp} 
I=\frac{1}{\mathfrak{T}}\left[  ( \mathbf{P} ^{\rm fr})^{\flat_g}: \nabla^g  \dot \varphi - \mathbf{J} _S \cdot \mathbf{d} \mathfrak{T}-\mathbf{J} _A \cdot \mathbf{d} \Upsilon ^A   +J _a\Lambda ^a \right] \geq 0.
\end{equation} 
Using the expression $ \frac{d}{dt}\mathfrak{E}=(\mathbf{P} ^{\rm cons}+\mathbf{P} ^{\rm fr}) ^{\flat_g} : \nabla^g \dot \varphi - \operatorname{DIV} (\mathbf{J} _Q+ \mathbf{J} _A \Upsilon ^A )+ \rho _{\rm ref}R$ for the internal energy balance in material representation and replacing the heat power supply by using the entropy inequality $\dot S+ \operatorname{DIV}J_S \geq \frac{1}{\mathfrak{T}} \rho _{\rm ref}R$, (or, in Clausius form $\dot S\geq - \operatorname{DIV}\left( \frac{1}{\mathfrak{T}  } \mathbf{J} _ Q\right) +\frac{1}{\mathfrak{T}  } \rho _{\rm ref}R$), we obtain a reformulation of \eqref{Clausius_Duhem_I_multi_comp} that generalizes the Clausius-Duhem inequality \eqref{Clausius_Duhem_II} to the case of the multicomponent reacting fluid, namely,
\begin{equation*}\label{Clausius_Duhem_II_multi_comp} 
\dot S- \frac{1}{\mathfrak{T}}\dot{\mathfrak{E}}+ \frac{1}{\mathfrak{T}}(\mathbf{P} ^{\rm cons}+\mathbf{P} ^{\rm fr})^{\flat_g} : \nabla ^g \dot \varphi - \frac{1}{\mathfrak{T}^2 }\mathbf{J} _Q \cdot \mathbf{d} \mathfrak{T} - \frac{1}{\mathfrak{T} } \operatorname{DIV}( \mathbf{J} _A \Upsilon ^A )   \geq 0.
\end{equation*} 
\end{remark}

\subsubsection{Spatial representation}\label{multi_comp_spatial} 

In a similar way to \S\ref{subsec_NSF_spatial}, the passage from the material to the spatial description follows from the material relabeling symmetry through a Lagrangian reduction procedure in material variables described in \S\ref{mat_repr_muticomp}. In spatial coordinates, the Lagrangian of the multicomponent fluid reads
\begin{equation*}\label{spat_lagr_nA_s} 
\ell( \mathbf{v}, \{n_A\} , s)=\int_ \mathcal{S} \frac{1}{2}\rho  | \mathbf{v} | _g ^2 \mu _g -\int_ \mathcal{S} \varepsilon ( \{n_A\} , s) \mu _g,
\end{equation*}
where we recall that $ \rho := \sum_{A=1}^N M^A n_A$, is the total mass density.
In addition to the reduced variables in \eqref{reduced_variables_NSF}, we need to define
\[
n_A=N_A \circ \varphi ^{-1} J_ \varphi ^{-1} , \quad w^A =W^A\circ \varphi ^{-1}  ,\quad  \upsilon  ^a = \nu ^a \circ \varphi ^{-1}. 
\]
Under these changes of variables, the variational formalism in material representation \eqref{VP_multi_comp} reads\begin{equation}\label{VP_multi_comp_spatial}
\delta \int_{t _1 }^{ t _2 }\left( \ell( \mathbf{v} , \rho , s)+\int_ \mathcal{S} ( \partial _t w ^A +   \mathbf{d}  w^A \cdot \mathbf{v}) n _A  \mu _g +\int_ \mathcal{\mathcal{S} } (s- \sigma )( \partial _t \gamma + \mathbf{d} \gamma\cdot \mathbf{v} ) \mu _g  \right) dt=0
\end{equation}
in spatial representation.
Similarly, the material phenomenological and chemical constraints together with their associated variational constraints \eqref{KC_multi_comp}, \eqref{chemical_constraint_K}, \eqref{VC_multi_comp}, \eqref{chemical_constraint_V} read
\begin{equation}\label{KC_multi_comp_spatial}
\begin{aligned} 
\frac{\delta \ell}{\delta s}(\partial _t \sigma + \operatorname{div}( \sigma \mathbf{v} )) &=-(\boldsymbol{\sigma} ^{\rm fr})^{\flat_g}: \nabla ^g \mathbf{v} +\mathbf{j} _S \cdot \mathbf{d} ( \partial _t \gamma + \mathbf{d} \gamma \cdot \mathbf{v} )\\
&  + \mathbf{j} _A \cdot \mathbf{d} ( \partial _t w ^A +\mathbf{d} w ^A \cdot \mathbf{v} )+ j _a ( \partial _t \upsilon ^a+ \mathbf{d} \upsilon ^a \cdot \mathbf{v} )- \rho r,
\end{aligned}
\end{equation} 
\begin{equation}\label{KC_chemistry_spatial}
\hspace{-5.7cm}
\partial _t \upsilon ^a+ \mathbf{d} \upsilon ^a \cdot \mathbf{v}= \nu^a  _{A}(\partial _t w ^A +\mathbf{d} w ^A \cdot \mathbf{v} ),
\end{equation} 
\begin{equation}\label{VC_multi_comp_spatial}
\begin{aligned} 
\frac{\delta \ell}{\delta s}(\delta \sigma + \operatorname{div}( \sigma \boldsymbol{\zeta}  )) =&-(\boldsymbol{\sigma} ^{\rm fr})^{\flat_g}: \nabla ^g\boldsymbol{\zeta}  +\mathbf{j} _S \cdot \mathbf{d} (\delta \gamma + \mathbf{d} \gamma \cdot\boldsymbol{\zeta}  ),\\
&  + \mathbf{j} _A \cdot \mathbf{d} (\delta  w ^A + \mathbf{d} w ^A \cdot\boldsymbol{\zeta} ) + j _a ( \delta \upsilon ^a+ \mathbf{d} \upsilon ^a \cdot \boldsymbol{\zeta}  )- \rho r,
\end{aligned}
\end{equation}
\vspace{2mm}
\begin{equation}\label{VC_chemistry_spatial}
\hspace{-5.8cm}
\delta \upsilon ^a+ \mathbf{d} \upsilon ^a \cdot\boldsymbol{\zeta} = \nu^a _{A}(\delta  w ^A +\mathbf{d} w ^A \cdot\boldsymbol{\zeta}  )
\end{equation}
in spatial representation. Here, $ \mathbf{j} _S$ and $\boldsymbol{\sigma} ^{\rm fr}$ in spatial representation are defined as in \eqref{def_EntFlux_FricStress}, while $ \mathbf{j} _A$, $j_a$ and $r$ are respectively defined by
\[
\mathbf{j} _A:= (\varphi _\ast \mathbf{J} _A) J_ \varphi ^{-1}, \quad j _a := J_a \circ \varphi ^{-1} J_ \varphi ^{-1} \quad\text{and}\quad r:= R \circ \varphi ^{-1}.
\]
As in \S\ref{subsec_NSF_spatial} above, we choose variations of the form $\delta \mathbf{v} = \partial _t \boldsymbol{\zeta}+[ \mathbf{v} , \boldsymbol{\zeta}] $, where $ \boldsymbol{\zeta} = \delta \varphi \circ \varphi ^{-1} $ is an arbitrary curve vanishing at the endpoints, and the variations of the variables $n_A$, $w^A$, $s$, $ \sigma $, and $ \gamma $ verify the variational constraints, and $\delta w^{A}$ and $\delta \gamma$ vanish at endpoints.

Taking the variations of the action integral \eqref{VP_multi_comp_spatial} and using the divergence theorem with the boundary condition $ \mathbf{v}|_{\partial \mathcal{S}}=0$, we have
\begin{align*} 
&\int_{ t _1 }^{ t _2 }\int_ \mathcal{S} \left( \frac{\delta \ell}{\delta \mathbf{v} } \cdot (\partial _t \boldsymbol{\zeta} + [ \boldsymbol{\zeta} , \mathbf{v} ]) + \frac{\delta \ell}{\delta n_A  }\delta n_A  +\frac{\delta \ell}{\delta s  }\delta s-( \partial _t n _A + \operatorname{div}( n _A \mathbf{v} )) \delta w ^A \right.\\[2mm]
& \qquad + \delta n _A ( \partial _t w ^A + \mathbf{d} w ^A \cdot \mathbf{v} )+ n _A \mathbf{d} w ^A \cdot (\partial _t \boldsymbol{\zeta} + [ \boldsymbol{\zeta} , \mathbf{v} ])- [ \partial _t (s- \sigma )+ \operatorname{div}((s- \sigma ) \mathbf{v} ) ]\delta  \gamma\\[2mm]
& \left .\phantom{\int_{ t _0 }^{ t _1 }}+( \delta s- \delta \sigma )( \partial _t \gamma + \mathbf{d} \gamma \cdot \mathbf{v} )+(s- \sigma ) \mathbf{d} \gamma \cdot (\partial _t \boldsymbol{\zeta} + [ \boldsymbol{\zeta} , \mathbf{v} ])\right) \mu _gdt =0.
\end{align*}
Using  the divergence theorem with the boundary condition $ \boldsymbol{\zeta} |_{ \partial \mathcal{S} }=0$, the variational constraints \eqref{VC_multi_comp_spatial} and \eqref{VC_chemistry_spatial}, the fact that $ \frac{ \delta \ell }{\delta  s}\neq 0$, and collecting the terms associated to the variations $\boldsymbol{\zeta} $, $ \delta n_A$, $ \delta w^A$, $ \delta\gamma $, $ \delta s$, we obtain
\begin{align} 
\boldsymbol{\zeta}  :&\;\;( \partial _t + \operatorname{ad}^*_ \mathbf{v} )\left( \frac{\delta \ell}{\delta \mathbf{v} }+( s- \sigma ) \mathbf{d} \gamma \right) -\rho  \,\mathbf{d} \frac{\delta \ell}{\delta \rho }+ \sigma \mathbf{d} ( \partial _t \gamma + \mathbf{d} \gamma \cdot \mathbf{v} ) \label{zeta_multi_comp_spatial} \\
& \;\; + \operatorname{div}\left( (\partial _t \gamma + \mathbf{d}\gamma \cdot \mathbf{v} ) \left( \frac{\delta \ell}{\delta s} \right)  ^{-1}\!\!\!\boldsymbol{\sigma} ^{\rm fr} \right) - \operatorname{div}\left( (\partial _t \gamma + \mathbf{d}\gamma \cdot \mathbf{v} ) \left( \frac{\delta \ell}{\delta s} \right)  ^{-1}\!\!\!\mathbf{j} _S \right) \cdot \mathbf{d} \gamma =0, \nonumber \\[2mm]
\delta n _A :& \;\; \partial _t w ^A + \mathbf{d} w ^A \cdot \mathbf{v} =-\frac{\delta \ell}{\delta n _A } ,\label{delta_n_multi_comp_spatial} \\[2mm] 
\delta w _A :& \;\; \partial _t n _A + \operatorname{div}( n _A \mathbf{v} )  =\operatorname{div}\left( (\partial _t \gamma + \mathbf{d}\gamma \cdot \mathbf{v} ) \left( \frac{\delta \ell}{\delta s} \right)  ^{-1} \mathbf{j} _A\right) -(\partial _t \gamma + \mathbf{d}\gamma \cdot \mathbf{v} ) \left( \frac{\delta \ell}{\delta s} \right)  ^{-1} j _a \nu ^a_{A} ,\label{delta_w_multi_comp_spatial} \\[2mm]  
\delta \gamma  :&\;\;\partial _t ( s- \sigma )+ \operatorname{div}((s- \sigma ) \mathbf{v} ) =\operatorname{div}\left( (\partial _t \gamma + \mathbf{d}\gamma \cdot \mathbf{v} ) \left( \frac{\delta \ell}{\delta s} \right)  ^{-1}\mathbf{j} _S\right),\label{delta_gamma_multi_comp_spatial} \\
\delta s :&\;\; \partial _t \gamma + \mathbf{d} \gamma \cdot \mathbf{v} =-\frac{\delta \ell}{\delta s}, \label{delta_s_multi_comp_spatial}  
\end{align}\color{black} 
where we assumed $ \delta \gamma |_{ \partial \mathcal{S} }=0$ and $ \delta w ^A |_{ \partial \mathcal{S} }=0$. Using \eqref{delta_s_multi_comp_spatial} and \eqref{delta_gamma_multi_comp_spatial} in the phenomenological constraint \eqref{KC_multi_comp_spatial} leads to
\begin{equation*}
T( \partial _t s+ \operatorname{div}( s \mathbf{v} )+ \operatorname{div} \mathbf{j} _S )= ( \boldsymbol{\sigma}  ^{\rm fr} )^{\flat_g} : \operatorname{Def} \mathbf{v}  - \mathbf{j} _S \cdot \mathbf{d} T - \mathbf{j} _A\cdot \mathbf{d} \mu ^A + j _a \mathcal{A} ^a + \rho r,
\end{equation*}
where $T:=-\frac{\delta  \ell}{\delta  s}= \frac{\partial \varepsilon }{\partial s}$ is the temperature, $ \mu^A:= - \frac{\partial \mathfrak{L}}{\partial n_A }= \frac{\partial \varepsilon }{\partial n _A}$ is the chemical potential, and $ \mathcal{A} ^a:=- \nu ^a_{A} \mu ^A $ is the affinity, all in spatial representation.
\medskip

If $ \delta \gamma |_{ \partial \mathcal{S} }$ and $ \delta w ^A |_{ \partial \mathcal{S} }$ are free, then we obtain the conditions:
\begin{equation}\label{adiabatically_closed_cond_multi_comp} 
\mathbf{j} _S \cdot \mathbf{n} ^{\flat _g}=0 \quad\text{and}\quad\mathbf{j} _A \cdot \mathbf{n} ^{\flat _g}=0 \quad\text{on}\quad \partial \mathcal{S},
\end{equation} 
where $\mathbf{n}$ is the outward pointing unit normal vector field on $ \partial \mathcal{S} $ relative to $g$. In this case, if $r\rho=0$, the fluid is adiabatically closed.
\medskip

Our results are summarized in the following box.
\begin{framed}
\noindent\underline{\textsf{Variational formalism for multicomponent reacting fluids -- spatial representation:}}\phantom{$\frac{1}{\int_{A_A}}$}\\
In spatial representation, the evolution equation are given by
\begin{equation}\label{multi_comp_spatial} 
\!\!\!\!\!\!\!\left\{
\begin{array}{l}
\vspace{0.2cm}\rho (\partial _t \mathbf{v} + \mathbf{v} \cdot  \nabla  \mathbf{v})=- \operatorname{grad} p +\operatorname{div} \boldsymbol{\sigma} ^{\rm fr}, \quad p=  \frac{\partial \varepsilon}{\partial n _A  }n _A  + \frac{\partial \varepsilon }{\partial s}s - \varepsilon, \\[2mm]
\vspace{0.2cm}\partial _t n _A  + \operatorname{div}( n _A   \mathbf{v} )+ \operatorname{div} \mathbf{j} _A  =j_a \nu _{A}^{a},\\[2mm]
T( \partial _t s+ \operatorname{div}( s \mathbf{v} )+ \operatorname{div} \mathbf{j} _S )=( \boldsymbol{\sigma}  ^{\rm fr} )^{\flat_g} : \operatorname{Def} \mathbf{v}  - \mathbf{j} _S \cdot \mathbf{d} T - \mathbf{j} _A\cdot \mathbf{d} \mu ^A + j _a \mathcal{A} ^a + \rho r,
\end{array}
\right.
\end{equation}
where $T=-\frac{\delta  \ell}{\delta  s}= \frac{\partial \varepsilon }{\partial s}$, $ \mu^A= - \frac{\partial \mathfrak{L}}{\partial n_A }= \frac{\partial \varepsilon }{\partial n _A}$, and $ \mathcal{A} ^a=- \nu ^a_{A} \mu ^A $. 
These equations arise from the variational condition  \eqref{VP_multi_comp_spatial}, where $ \mathbf{v}, n _A , w ^A  , s, \gamma , \sigma , \upsilon ^a $ satisfy the phenomenological and chemical constraints \eqref{KC_multi_comp_spatial}-\eqref{KC_chemistry_spatial} and for variations of the form  $ \delta \mathbf{v} = \partial _t \boldsymbol{\zeta} + [ \boldsymbol{\zeta} , \mathbf{v} ]$, $ \delta n _A $, $ \delta w ^A $, $ \delta s$, $ \delta \gamma $, $ \delta \sigma $, $ \delta \upsilon ^a $ such that $ \boldsymbol{\zeta} $, $ \delta \sigma $, $ \delta w ^A $, $ \delta \upsilon ^a $, $ \delta \gamma $ verify the variational constraints \eqref{VC_multi_comp_spatial}-\eqref{VC_chemistry_spatial} and with $ \delta \gamma |_{ \partial \mathcal{S} }=\delta w ^A |_{ \partial \mathcal{S} }=0$. Direct computations using \eqref{delta_s_multi_comp_spatial} and \eqref{delta_n_multi_comp_spatial}--\eqref{delta_gamma_multi_comp_spatial} in \eqref{zeta_multi_comp_spatial} and in the phenomenological constraint \eqref{KC_multi_comp_spatial} leads to \eqref{multi_comp_spatial}. If the constraint $ \delta \gamma  |_{ \partial \mathcal{S} }=0$ and $\delta w ^A |_{ \partial \mathcal{S} }=0$ are not imposed, then the variational formalism yields $\mathbf{j} _S \cdot \mathbf{n} ^{\flat_g}=0$ and $ \mathbf{j} _A \cdot \mathbf{n} ^{\flat_g}=0$ and if, in addition, $\rho\, r=0$, then the fluid is adiabatically closed.
\end{framed}

\begin{remark}[Interpretation of $ \sigma $]{\rm As before, the Lagrangian time derivative of the variable $ \sigma $ corresponds to the total (internal and external) entropy production density of the system and it follows from \eqref{delta_gamma_multi_comp_spatial} that 
\[
\partial _t \sigma + \operatorname{div}( \sigma \mathbf{v} )=  \frac{1}{T} \left( ( \boldsymbol{\sigma}  ^{\rm fr} )^{\flat_g} : \nabla \mathbf{v}   - \mathbf{j} _S \cdot \mathbf{d} T- \mathbf{j} _A\cdot \mathbf{d} \mu ^A + j _a \mathcal{A} ^a +\rho r\right)=i+ \frac{r \rho }{T},
\]
where $i$ is the internal entropy production density and hence, of course, ${r \rho }/{T},$ is the external entropy production density.
Since the second law of thermodynamics implies $i\geq 0$, for the total entropy, the relation \eqref{total_entropy_multi_comp} is kept unchanged, except the fact that the expression of $i$ is different. 
}
\end{remark}

\begin{remark}[Energy balance]{\rm The total energy density $e:= \frac{1}{2} \rho | \mathbf{v} | ^2_g  + \varepsilon ( \{n _A \} , s)$ satisfies the equation
\[
\partial _t e+ \operatorname{div}( e \mathbf{v} ) =\operatorname{div}(\boldsymbol{\sigma} ^{\flat_g}\cdot \mathbf{v} )- \operatorname{div} \mathbf{j} _Q-\operatorname{div} (\mathbf{j} _A \mu ^A ) +\rho r,
\]
so that the total energy $\mathsf{E}(t)=\int_ \mathcal{S} e(t,x) \mu _g(x) $ verifies the balance law
\[
\frac{d}{dt} \mathsf{E}(t)= -\int_{\partial \mathcal{S} } \mathbf{j} _Q \cdot \mathbf{n} ^{\flat_g}\mu _g^\partial  -\int_{\partial \mathcal{S} }\mu ^A  \mathbf{j} _A \cdot \mathbf{n} ^{\flat_g}\mu _g^\partial +\int_ \mathcal{S} \rho r\mu _g =P^{\rm ext}_H(t)+P^{\rm ext}_M(t).
\]
We consistently have $\frac{d}{dt} \mathsf{E}= 0$ in the adiabatically closed case, i.e.,  $\mathbf{j} _Q \cdot \mathbf{n}^{\flat_g}=\mathbf{J} _A \cdot \mathbf{n} ^{\flat_g}=0$ (see \eqref{adiabatically_closed_cond_multi_comp}) and $ \rho r=0$.}
\end{remark} 

\begin{remark}[General Lagrangian]{\rm For a general Lagrangian $\ell( \mathbf{v} , \{n_A\},s)$, the variational formalism yields the system
\begin{equation*}
\!\!\!\!\left\{
\begin{array}{l}
\displaystyle\vspace{0.2cm}\left( \partial _t + \operatorname{ad}^*_ \mathbf{v} \right) \frac{\delta \ell}{\delta \mathbf{v} }  =\rho\,\mathbf{d}  \frac{\delta \ell}{\delta \rho }+s \,\mathbf{d} \frac{\delta \ell}{\delta s}   +\operatorname{div} \boldsymbol{\sigma} ^{\rm fr},\qquad \partial _t n _A  + \operatorname{div}( n _A   \mathbf{v} )+ \operatorname{div} \mathbf{j} _A  =j _a \nu^a _{A},\\
\displaystyle\frac{\delta \ell}{\delta s} ( \partial _t s+ \operatorname{div}( s \mathbf{v} )+ \operatorname{div} \mathbf{j} _S )= -( \boldsymbol{\sigma}  ^{\rm fr} )^{\flat_g} :\nabla  \mathbf{v}  - \mathbf{j} _S \cdot \mathbf{d} \frac{\delta \ell}{\delta s}- \mathbf{j} _A \cdot \mathbf{d} \frac{\delta \ell}{\delta n _A }-j _a \nu ^a_{A}\frac{\delta \ell}{\delta n _A }+r \rho .
\end{array}
\right.
\end{equation*}}
\end{remark} 

\begin{remark}[Thermodynamic phenomenology and Onsager's relations]\label{TP_NSF}{\rm The system of equations \eqref{multi_comp_spatial} needs to be supplemented with phenomenological expressions for the \textit{thermodynamic fluxes} $J _\alpha $ (i.e., $ \boldsymbol{\sigma} ^{\rm fr}$, $ \mathbf{j} _S$, $ \mathbf{j} _A $, and $j _a $) in terms of the \textit{thermodynamic affinities} $X^\alpha $ (i.e., $ \operatorname{Def} \mathbf{v} $, $ \mathbf{d} T$, $ \mathbf{d} \mu ^A $, $ \mathcal{A} ^a=- \nu ^a_{A} \mu ^A $) compatible with the second law $I = J_ \alpha X ^\alpha \geq 0$, where $I$ is the internal entropy production density.
It is empirically accepted  that for a large class of irreversible processes and under a wide range of experimental conditions, the thermodynamic fluxes $ J_ \alpha $ are linear functions of the thermodynamic affinities $ X^\alpha $, i.e., $J_ \alpha = \mathcal{L} _{ \alpha \beta } X^\beta $, where the transport coefficients $ \mathcal{L} _ { \alpha \beta }(...)$ are state functions that must be determined by experiments or if possible by kinetic theory.
Besides defining a positive quadratic form, the coefficients $ \mathcal{L} _ { \alpha \beta }(...)$ must also satisfy {\it Onsager's reciprocal relations} (\cite{Onsager1931}) due to the microscopic time reversibility and the {\it Curie principle} associated to material invariance (see, for instance, \cite{deGrootMazur1969}, \cite{KoPr1998}, \cite{Woods1975}). 
In the case of the multicomponent fluid, writing $ (\boldsymbol{\sigma} ^{\rm fr})^{(0)}=\boldsymbol{\sigma} ^{\rm fr}- \frac{1}{3}( \operatorname{Tr}\boldsymbol{\sigma} ^{\rm fr} )g^\sharp$ and, similarly, $ (\operatorname{Def} \mathbf{v}) ^{(0)}=  \operatorname{Def} \mathbf{v} - \frac{1}{3}(\operatorname{div} \mathbf{v} )\delta $, we have the following phenomenological linear relations
\[
-\begin{bmatrix}
\mathbf{j} _S\\
\mathbf{j} _A
\end{bmatrix}= 
\begin{bmatrix}
\mathcal{L} _{SS} & \mathcal{L} _{SB}\\
\mathcal{L} _{AS}&\mathcal{L} _{AB}
\end{bmatrix}\begin{bmatrix}
\mathbf{d} T\\
\mathbf{d} \mu ^B 
\end{bmatrix}\!, \;
\begin{bmatrix}
\operatorname{Tr}\boldsymbol{\sigma} ^{\rm fr} \\
j _a 
\end{bmatrix}= 
\begin{bmatrix}
\mathcal{L} _{00} & \mathcal{L} _{0b}\\
\mathcal{L} _{a0}&\mathcal{L} _{ab}
\end{bmatrix}\begin{bmatrix}
\frac{1}{3} \operatorname{div} \mathbf{v}  \\
\mathcal{A} ^b
\end{bmatrix}\!, \;
(\boldsymbol{\sigma} ^{\rm fr}) ^{(0)}= 2 \mu (\operatorname{Def}\mathbf{v} ^{\sharp_g})^{(0)}\!,
\]
where all the coefficients may depend on $(s, \{n_A\})$. The first linear relation describes the vectorial phenomena of heat conduction (Fourier law), diffusion (Fick law) and their cross effects (Soret and Dufour effects), while the second relation describes the scalar processes of bulk viscosity and chemistry and their possible cross-phenomena. The associated friction stress reads
\[
\boldsymbol{\sigma} ^{\rm fr}= 2 \mu \operatorname{Def} \mathbf{v} + \left( \frac{1}{9} \mathcal{L} _{00} - \frac{2}{3} \mu \right) (\operatorname{div} \mathbf{v} ) g^\sharp+ \frac{1}{3} \mathcal{L} _{0b}\mathcal{A} ^b g^\sharp,
\]
(compare with \eqref{friction_stress_NSF}). The condition $M^A \mathbf{j} _A=0$ is satisfied if $M^A\mathcal{L} _{AS}=M^A \mathcal{L} _{AB}=0$.
\medskip

All these phenomenological considerations take place in the phenomenological constraint \eqref{KC_multi_comp_spatial} or \eqref{KC_multi_comp} and the associated variational constraints \eqref{VC_multi_comp_spatial} or \eqref{VC_multi_comp}, but they are not involved in the variational condition \eqref{VP_multi_comp_spatial} or \eqref{VP_multi_comp}. Note that our variational formalism holds independently on the linear character of the phenomenological laws.}
\end{remark}

\begin{remark}[Analogy between thermodynamics and mechanics]\label{forces_VS_velocity}{\rm In our approach, we have rewritten the thermodynamic power $J_ \alpha X^{\alpha} $ associated to an irreversible process $ \alpha$ by introducing the \textit{thermodynamic displacement} $ \Lambda ^\alpha $ such that $X^{\alpha} =\dot \Lambda ^{\alpha}$. In this way, the power $J_ \alpha X^{\alpha} = J_ \alpha \dot \Lambda ^{\alpha} $ takes a similar form with the mechanical power $ \left\langle F^{\rm fr},\dot q \right\rangle $. 
By employing this analogy with mechanics, the {\it thermodynamic affinities} $X ^\alpha= \dot \Lambda ^\alpha  $, such as $ \dot \varphi $, $\dot \Gamma $, $\dot W ^A $, $\dot \nu ^a $, may be interpreted as {\it velocities} , while the {\it thermodynamic fluxes} $J_ \alpha $, such as $ \mathbf{P}^{\rm fr}$, $ \mathbf{J} _S $, $ \mathbf{J} _A$, $J _a $, as \textit{friction forces} paired with these velocities. 
}
\end{remark}

\section{Conclusions}

In this paper,  we have established a Lagrangian variational formalism for discrete and continuum systems including nonequilibrium thermodynamics.  From a mathematical point of view, this variational formalism is an extension of the standard Lagrange-d'Alembert principle used in nonholonomic mechanics. In this extension, the nonholonomic constraint is nonlinear and is given by the expression of the entropy production associated to all the irreversible processes involved. 

To achieve this goal, we have employed an analogy between thermodynamics and mechanics, in which the thermodynamic affinities are regarded as a rate of the thermodynamic displacement $ \Lambda  $, (i.e., $X^ \alpha =\dot\Lambda  ^\alpha$), associated to each irreversible process. This allows us to formulate the associated variational constraint in a systematic way, namely, by replacing all the velocities by their corresponding virtual displacement $ \delta \Lambda  ^\alpha $ and by removing the effect of the exterior of the system: 
\[
J_ \alpha \dot \Lambda  ^\alpha +P^{\rm ext} \;\;\leadsto \;\; J_ \alpha  \delta \Lambda  ^\alpha.
\]
 
Our variational formalism has thus a clear and systematic structure that appears to be common for the macroscopic description of the nonequilibrium thermodynamics of physical systems. 
In particular, it applies to both discrete systems (i.e., systems with finite degrees of freedom) and continuum systems (i.e., systems with infinite degrees of freedom). It has been illustrated with examples from classical mechanics, electric circuits, chemical reactions, matter transfer, and multicomponent reacting viscous fluids.
\medskip

Note that the examples that we consider in this paper are restricted to closed systems; namely, we have not shown examples of systems that include external matter transfer, although our variational formalism is also applicable to such open systems typically appearing in biology, mechanical engines, etc. The application to more complex systems including open systems will be explored as a future work.

\paragraph{Acknowledgements.} The authors thank C. Gruber for extremely helpful discussions. F.G.B. is partially supported by the ANR project GEOMFLUID, ANR-14-CE23-0002-01; H.Y. is partially supported by JSPS (Grant-in-Aid 26400408), JST (CREST), Waseda University (SR 2014B-162, SR 2015B-183), the IRSES project ``Geomech'' (246981) within the 7th European Community Framework Programme, and the MEXT ``Top Global University Project'' at Waseda University.


\begin{thebibliography}{xx}
\bibitem[Appell(1911)]{Ap1911}
Appell, P [1911], Sur les liaisons exprim\'ees par des relations non lin\'eaires entre les vitesses, \textit{C.R. Acad. Sci. Paris}, \textbf{152}, 1197--1199.

\bibitem[Arnold(1988)]{Arnold}
Arnold, V.~I.[1988], {\it Dynamical Systems: Vol III}. Springer-Verlag, New York.


\bibitem[Biot(1975)]{Bi1975}
Biot, M. A. [1975],  A virtual dissipation principle and Lagrangian equations in non-linear irreversible thermodynamics,  \textit{Acad. Roy. Belg. Bull. Cl. Sci.} \textbf{5} (61), 6--30.

\bibitem[Biot(1984)]{Bi1984}
 Biot, M. A. [1984], New variational-Lagrangian irreversible thermodynamics with application to viscous flow, reaction-diffusion, and solid mechanics. \textit{Adv. Appl. Mech.} \textbf{24}, 1--91. 

\bibitem[Bloch(2003)]{Bl2003}
Bloch, A.~M. [2003], {\it Nonholonomic Mechanics and Control}, volume~24 of {\it Interdisciplinary Applied Mathematics}, Springer-Verlag, New York. With the collaboration of J. Baillieul, P. Crouch and J. Marsden, and with scientific input from P. S. Krishnaprasad, R. M. Murray and D. Zenkov.

\bibitem[Cendra, Ibort, de Le\'on, and Mart\'in de Diego(2004)]{CeIbdLdD2004}
Cendra, H., A. Ibort, M. de Le\'on, and D. Mart\'in de Diego [2004], A generalization of Chetaev's principle for a class of higher order nonholonomic constraints, \textit{J. Math. Phys.} \textbf{45}, 2785.


\bibitem[Chetaev(1934)]{Ch1934}
Chetaev, N.~G. [1934], On Gauss principle, \textit{Izv. Fiz-Mat. Obsc. Kazan Univ.},
\textbf{7}, 68--71.

\bibitem[Chua and McPherson(1974)]{ChMc1974}
Chua, L. O. and J. D. McPherson [1974], Explicit topological formulation of Lagrangian and Hamiltonian equations for nonlinear networks. 
\textit{IEEE Trans. Circuits Syst.}, \textbf{21}, 277--286.

\bibitem[de Groot and Mazur(1969)]{deGrootMazur1969}
de Groot, S.~R and P. Mazur [1969], {\it Nonequilibrium Thermodynamics}, North-Holland.






\bibitem[Ferrari and Gruber(2010)]{FeGr2010}
Ferrari, C. and C. Gruber [2010], Friction force: from mechanics to thermodynamics, \textit{Europ. J. Phys.} \textbf{31}(5), 1159--1175.


\bibitem[Fukagawa and Fujitani(2012)]{FuFu2012}
Fukagawa, H. and Y. Fujitani [2012], A variational principle for dissipative fluid dynamics, \textit{Prog. Theor. Phys.} \textbf{127}(5), 921--935.



\bibitem[Gay-Balmaz, Marsden, and Ratiu(2012)]{GBMaRa2012}
Gay-Balmaz, F., J.~E. Marsden, and T.~S. Ratiu [2012], Reduced variational formulations in free boundary continuum mechanics. {\it J. Nonlinear Sc.} \textbf{22}, 553--597.

\bibitem[Gay-Balmaz and Yoshimura(2015)]{GBYo2015}
Gay-Balmaz, F. and H. Yoshimura [2015], Dirac reduction for nonholonomic mechanical systems and semidirect products, {\it Advances in Applied Mathematics}, \textbf{63}, 131--213.


\bibitem[Gibbs(1902)]{Gibbs1902}
Gibbs, J.~W. [1902], {\it Collected Works}, Scribner, New-York.


\bibitem[Glansdorff and Prigogine(1971)]{GlPr1971}
Glansdorff, P. and I. Prigogine [1971], {\it Thermodynamic Theory of Structure, Stability, and Fluctuations}, Wiley-Interscience.

\bibitem[Green and Naghdi(1991)]{GrNa1991}
Green, A. E. and P. M. Naghdi [1991], A re-examination of the basic postulates of thermomechanics, {\it Proc. R. Soc. London.}  Series A: {\it Mathematical, Physical and Engineering Sciences}, {\bf 432}(1885), 171--194.

\bibitem[Green and Naghdi(1993)]{GrNa1993}
Green, A. E. and P. M. Naghdi [1993], Thermoelasticily without energy dissipation, {\it Journal of Elasticity}, {\bf 31}, 189--208.





\bibitem[Gruber(1997)]{Gr1997}
Gruber, C. [1997], \textit{Thermodynamique et M\'ecanique Statistique}, Institut de physique th\'{e}orique, EPFL. 

\bibitem[Gruber(1999)]{Gr1999}
Gruber, C. [1999], Thermodynamics of systems with internal adiabatic constraints: time evolution of the adiabatic piston, \textit{Eur. J. Phys.} \textbf{20}, 259--266. 
\bibitem[Gruber and Brechet(2011)]{GrBr2011}
Gruber, C. and S.~D. Brechet [2011], Lagrange equation coupled to a thermal equation: mechanics as a consequence of thermodynamics, \textit{Entropy} \textbf{13}, 367--378. 


\bibitem[Gyarmati(1970)]{Gyarmati1970}
Gyarmati, I. [1970], \textit{Nonequilibrium Thermodynamics: Field Theory and
Variational Principles}, Springer-Verlag, New York.

\harvarditem[Holm, Marsden and Ratiu]{Holm, Marsden and Ratiu}{1998}{HMR1998}
Holm, D. D., J. E. Marsden and T. S. Ratiu [1998], The Euler-Poincar\'e equations and semidirect products with applications to continuum theories, {\it Adv. in Math.} \textbf{137}, 1--81.

\bibitem[Ichiyanagi(1994)]{Ichiyanagi1994}
Ichiyanagi M. [1994], Variational principles in irreversible processes, \textit{Phys. Rep.} \textbf{243}, 125--182.


\bibitem[Jacobs and Yoshimura(2014)]{JaYo2014}
Jacobs, H. and H. Yoshimura [2014], Tensor products of Dirac structures and interconnection in Lagrangian mechanics, 
\textit{J. Geo. Mech.} \textbf{6}(1), 67--98.

\bibitem[Jimenez and Yoshimura(2015)]{JiYo2015}
Jimenez, F. and H. Yoshimura [2015], Dirac structures in vakonomic mechanics, \textit{J. Geo. and Phys.} \textbf{94}, 158--178.



\bibitem[Kondepudi and Prigogine(1998)]{KoPr1998}
Kondepudi, D. and I. Prigogine [1998], \textit{Modern Thermodynamics}, John Wiley \& Sons.


\bibitem[Lavenda(1978)]{Lavenda1978}
Lavenda, B.~H. [1978], \textit{Thermodynamics of Irreversible Processes}, Macmillan, London.


\bibitem[Lew, Marsden, Ortiz, and West(2003)]{LeMaOrWe2003}
Lew, A., Marsden, J.~E., Ortiz, M., and West, M. [2003] Asynchronous variational integrators, \textit{Arch. Rational Mech. Anal.}, 
\textbf{167}(2), 85--146.



\bibitem[Marle(1998)]{Ma1998}
Marle, C.-M. [1998], Various approaches to conservative and nonconservative non-holonomic systems, \textit{Rep. Math. Phys.} \textbf{42}, 1/2, 211--229.

\bibitem[Marsden and Hughes(1983)]{MaHu1983}
Marsden, J.~E. and T.~J.~R. Hughes [1983], \textit{Mathematical Foundations of Elasticity} (Prentice Hall, New York, 1983) (reprinted by
Dover, New York, 1994).

\bibitem[Marsden and West(2001)]{MaWe2001}
Marsden, J.~E. and West, M. [2001], Discrete mechanics and variational integrators, \textit{Acta Numer.}, \textbf{10}, 357--514.

\bibitem[Mata and Lew(2011)]{MaLe2011} 
Mata, P. and Lew,~A.~J. [2011], Variational time integrators for finite-dimensional thermo-elasto-dynamics without heat conduction. \textit{Internat. J. Numer. Methods Engrg.} \textbf{88}(1), 1--30. 



\bibitem[Onsager(1931)]{Onsager1931}
Onsager, L. [1931], Reciprocal relations in irreversible processes I, \textit{Phys. Rev.} \textbf{37}, 405--426; Reciprocal relations in irreversible processes II, \textit{Phys. Rev.} \textbf{38}, 2265--2279.  

\bibitem[Onsager and Machlup(1953)]{OnMa1953}
Onsager, L. and S. Machlup [1953], Fluctuations and irreversible processes, \textit{Phys. Rev.} \textbf{91}, 1505--1512.

\bibitem[Machlup  and Onsager(1953)]{MaOn1953}
Onsager, L. and S. Machlup [1953], Fluctuations and irreversible processes II. Systems with kinetic energy. \textit{Phys. Rev.} \textbf{91}, 1512--1515.


\bibitem[Oster, Perelson, Katchalsky(1973)]{OsPeKa1973}
Oster, G.~F., A.~S. Perelson, A. Katchalsky [1973], Network thermodynamics: dynamic modelling of biophysical systems, \textit{Quarterly Reviews of Biophysics}, \textbf{6}(1), 1--134.

\bibitem[Pironneau(1983)]{Pi1983}
Pironneau, Y. [1983], Sur les liaisons non holonomes non lin\'eaires,
d\'eplacements virtuels \`a travail nul, conditions de Chetaev, Proceedings
of the IUTAM-ISIMM Symposium on Modern Developments in Analytical
Mechanics, Torino 1982, \textit{Atti della Acad. della sc. di Torino}, \textbf{117}, 671--686.

\bibitem[Podio-Guidugli(2009)]{Po2009}
Podio-Guidugli, P. [2009], A virtual power format for thermomechanics,  
\textit{Continuum Mechanics and Thermodynamics}, \textbf{20}(8), 479--487.

\bibitem[Prigogine(1947)]{Prigogine1947}
Prigogine [1947], Etude thermodynamique des ph\'enom\`enes irr\'eversibles, Thesis, Paris: Dunod and Li\`ege: Desoer.




\bibitem[Simo, Marsden, and Krishnaprasad(1988)]{SiMaKr1988}
Simo, J.~C., J.~E. Marsden and P.~S. Krishnaprasad [1988], The Hamiltonian structure of nonlinear elasticity: The material, spatial and convective representations of solids, rods and plates, \textit{Arch. Rational Mech. Anal.}, \textbf{104}, 125--183.

\bibitem[Stueckelberg and Scheurer(1974)]{StSc1974}
Stueckelberg, E.~C.~G. and P.~B. Scheurer [1974], \textit{Thermocin\'etique ph\'enom\'enologique galil\'eenne}, Birkh\"auser, 1974.

\bibitem[Truesdell(1969)]{Truesdell1969}
Truesdell, C. [1969], \textit{Rational Thermodynamics}, McGraw-Hill, New-York.




\bibitem[von Helmholtz(1884)]{He1884}
von Helmholtz, H. [1884], Studien zur Statik monocyklischer Systeme.
\textit{Sitzungsberichte der K\"{o}niglich Preussischen Akademie der Wissenschaften
zu Berlin}, 159--177.


\bibitem[Woods(1975)]{Woods1975}
Woods, L.~C. [1975], \textit{The Thermodynamics of Fluid Systems}, Clarendon Press Oxford 1975.



\bibitem[Yoshimura and Marsden(2006a)]{YoMa2006a} Yoshimura, H. and J.~E. Marsden [2006a], Dirac structures in Lagrangian mechanics. Part I: Implicit Lagrangian systems, \textit{J. Geom. and Phys.} \textbf{57}, 133--156.
  
\bibitem[Yoshimura and Marsden(2006b)]{YoMa2006b} Yoshimura, H. and J.~E. Marsden [2006b], Dirac structures in Lagrangian mechanics. Part II: Variational structures, \textit{J. Geom. and Phys.} \textbf{57}, 209--250.

\bibitem[Yoshimura and Marsden(2006c)]{YoMa2006c}
Yoshimura, H. and J.~E. Marsden [2006c], Dirac structures and implicit Lagrangian systems in electric networks, \textit{Proc. of the 17th International Symposium on Mathematical Theory of Networks and Systems}, Paper WeA08.5, pages 1--6, July 24-28, 2006, Kyoto.





\bibitem[Ziegler(1968)]{Ziegler1968}
Ziegler, H. [1968],  A possible generalization of Onsager's theory, in H. Barkus and L.I. Sedov (eds.), \textit{Irreversible Aspects of Continuum Mechanics}, Springer, New York.


\end{thebibliography}
\end{document}